%% file: EPO6_arxiv_M.tex
\documentclass[prb,singlecolumn,a4paper,amsmath,amssymb]{revtex4}

\usepackage{geometry}
\usepackage{amsfonts}
\usepackage{latexsym}
\usepackage{graphicx}
\usepackage[usenames]{color}
\usepackage[utf8]{inputenc}
\usepackage[english,bulgarian,russian]{babel}
\usepackage{hyperref}
\usepackage{caption}
\usepackage{morefloats}
\usepackage{epstopdf}
\usepackage{systeme}
\usepackage{listings}
\usepackage{array}
\usepackage{circuitikz}
\usepackage{tikz}

\newcommand{\be}{\begin{equation}}
\newcommand{\ee}{\end{equation}}

\newcommand{\kb}{k_\mathrm{_B}}

\begin{document}

\title{Measurement of the electron charge $q_e$ using Schottky noise. 
Problem of the 6-th Experimental Physics Olympiad. Sofia 8 December 2018}


\author{Todor~M.~Mishonov, Emil~G.~Petkov, Aleksander~A.~Stefanov, Aleksander~P.~Petkov, Viktor~I.~Danchev, Zehra~O.~Abdrahim, Zlatan~D.~Dimitrov}
\email[E-mail: ]{mishonov@bgphysics.eu}
\affiliation{Physics Faculty,
``St.~Kliment Ohridski'' University at Sofia,\\
5 James Bourchier Blvd., BG-1164 Sofia, Bulgaria}
\author{Iglika~M.~Dimitrova}
\affiliation{Faculty of Chemical Technologies, Department of Physical Chemistry,
University of Chemical Technology and Metallurgy,
8 Kliment Ohridski Blvd., BG-1756 Sofia, Bulgaria}
\author{Riste Popeski-Dimovski}
\email[E-mail: ]{ristepd@gmail.com}
\affiliation{Institute of Physics, Faculty of Natural Sciences and Mathematics, ``Ss. Cyril and Methodius'' University, Skopje, R. Macedonia}
\author{Marina Poposka}
\email[E-mail: ]{marinapoposka@gmail.com}
\affiliation{High school SOU Gimnazija ``Mirche Acev'',
96a Prilepski braniteli Str., MKD-7500 Prilep, R.~Macedonia}
\author{Sla\dj ana Nikoli\'c}
\affiliation{Milan D. Mili\'cevi\'c School, 27a Borivoja Stevanovi\'ca Str., 
RS-11000 Belgrade, Serbia}
\author{Slavoljub Miti\'c}
\affiliation{Svetozar Markovi\'c Gymnasium, 1 Branka Radi\'cevi\'ca Str., RS-18106 Ni\v{s}, Serbia}
\author{Vassil~N.~Gourev}
 \affiliation{Department of Atomic Physics, Physics Faculty,\\
 ``St.~Kliment Ohridski'' University at Sofia,\\
5 James Bourchier Blvd., BG-1164 Sofia, Bulgaria}
\author{Ria Rosenauer, Felix Schwarzfischer}
\affiliation{Karls-Gymnasium, T\"ubinger Stra\ss e 38, 70178 Stuttgart, Germany}
\author{Vasil~G.~Yordanov, Albert~M.~Varonov}
\affiliation{Laboratory for fundamental constants measurement,\\
 ``St.~Kliment Ohridski'' University at Sofia, Bulgaria}
\email[E-mail: ]{avaronov@phys.uni-sofia.bg}

\begin{abstract}
Several consecutive experiments are described with a printed circuit board PCB set-up, especially designed for these experiments. 
Doing the consecutive experimental tasks opens up possibility to determine the value of electron charge 
$q_e.$
The fluctuations of the voltage $U(t)$ should be measured for different illuminations of a 
photodiode.
The voltage is amplified 1~million times $Y=10^6$. 
The amplified voltage $YU(t)$ is applied to the device, which gives the result of the value of 
the time averaged square of the voltage
$U_\mathrm{S}=\left<(Y U(t))^2\right>/U_0$.
This voltage $U_\mathrm{S}$ is measured with a multimeter.
The series of measurements gives the possibility to determine the 
$q_e$ using the well known Schottky formula for the spectral density of the current noise
$(I^2)_f=2q_e\left<I\right>.$
For the junior high school students, the basic problem is to analyze the analog squaring.
Students' work is separated and graded in four categories S, M, L, XL divided by age of students.
For the last XL categories, the tasks contain problems oriented to physics university education program and include theoretical research of the PCB set-up as an engineering device.
This is the problem of EPO6, December 2018 ``Day of the Charge'' considered. 
EPO6 is organized by Sofia branch of Union of physicists in Bulgaria in cooperation with Faculty of physics of Sofia University and Society of Physicists of Republic of Macedonia. 
\end{abstract}
\captionsetup{labelfont={normalsize},textfont={small},justification=centerlast}
\maketitle

\section*{Pre-Introduction}

The first version of this work has demonstrated that an educational experimental set-up can easily be done. In the second version the elaborated set-up is described, in which: 1) the electronic circuit is so stable that we do not need screening boxes and BNC cables between them, 2) the photo diode is cooled by ice water, which is of crucial importance for the accuracy of determination of electron charge, 3) the set-up was reproduced in more than 150 copies, which was distributed in many distant countries Kazakhstan, Germany, Canada etc., 4)  the set-up and the text was given to the participants of the 6$^\mathrm{th}$ Experimental Physics Olympiad, 5) the problem together with the solution is given as a didactical guide in many languages: Bulgarian, Macedonian, Serbian, Russian, German.

The described experiment and experimental set-up is much simpler than the classical experiment with charged drops. The spectral density of a noise is already a notion accessible for teenagers by smart-phones and can be included in the high school education. In such a way we are witnesses of an emerging worldwide system and a realistic possibility to establish a cheap solution in the market of educational experiment for measurement of electron charge. We give perhaps the simplest and cheapest solution. The description of the experimental set-up with the corresponding theory is given as a research paper in Eur. Phys. Journal~\cite{qe-EPJ}.

Here it is better to cite an article from The Relaxation Times (Newsletter of Teachspin. Inc) Vol. III, No. 8 May 2010: The Millikan oil drop experiment has had its run; it’s time to give it a rest! (At least, that seemed to be the consensus of the Topical Conference in Ann Arbor last summer). There are other ways to measure the charge of the electron, that use modern instrumentation that teach students highly useful as well as transferable skill, and that will challenge students theoretical understanding of an important yet often neglected physical phenomena. Out of the ``stuff'' we usually want to get rid of electrical noise, our students can carry out 1-2\% measurement of both the charge of the electron and the Boltzmann constant.

\section{Introduction}

From its very beginning, the Experimental Physics Olympiad (EPO) is worldwide known;
all Olympiad problems have been published in Internet~\cite{EPO1,EPO2,EPO3,EPO4,EPO5} and from the very beginning there were 120 participants.
In the last years high-school students from 7 countries participated 
and the distance between the most distant cities is more than 4~Mm.

Let us describe the main differences between EPO and other similar competitions.
\begin{itemize}
\item Each participant in EPO receives as a gift from the organizers the set-up, 
which one worked with.
So, after the Olympiad has finished, even bad performed participant is able to repeat the experiment and reach the level of the champion.
In this way, the Olympiad directly affects the teaching level in the whole world.
After the end of the school year, the set-up remains in the school, where the participant has studied.
\item Each of the problems is original and is connected to fundamental physics or the  understanding of the operation of a technical patent.
\item The Olympic idea is realized in EPO in its initial from 
and everyone willing to participate from around the world can do that.
There is no limit in the participants number.
On the other hand, the similarity with other Olympiads is that the problems are direct illustration of the study material and alongside with other similar competitions mitigates the secondary education degradation, which is a world tendency.
\item One and the same experimental set-up is given to all participants but the tasks are different for the different age groups, the same as the swimming pool water is equally wet for all age groups in a swimming competition.
\end{itemize}

We will briefly mention the problems of former 5 EPOs: 
1) The setup of EPO1 was actually a student version of the American patent for auto-zero
and chopper stabilized direct current amplifiers.~\cite{EPO1}
2) The problem of EPO2~\cite{EPO2} was to measure Planck constant by diffraction of a LED light by a compact disk.
3) A contemporary realization of the assigned to NASA patent for the use of negative impedance converter for generation of voltage oscillations was the set-up of EPO3.~\cite{EPO3}
4) EPO4~\cite{EPO4} was devoted to the fundamental physics -- to determine the speed of light by measuring electric and magnetic forces.
The innovative element was the application of the catastrophe theory in the analysis of the stability of a pendulum.
5) The theme of the EPO5 was to measure the Boltzmann constant $\kb$ following the Einstein
idea of study thermal fluctuations of electric voltage of a capacitor.

The present EPO6 follows the tradition theme of the Olympiad to be the measurement of some 
fundamental constant. 
Inspired by a lecture on fluctuations given by Einstein
Walter Schottky concluded that electron charge also can be determined by
study of the voltage noise.
Now, a century after, due to appearance of low noise operational amplifiers 
the Schottky idea can be implemented as a high-school problem.

In short, the established traditions is a balance between contemporary working technical inventions and fundamental physics (\textit{Allah akbar}).

\section{Experimental set-up}
Your set-up should consist of the following items:
\begin{enumerate}
\item A printed circuit board (PCB) with a double board connector with a photodiode and lithium battery holder.
\item Four 9~V batteries and two double connection clips for them.
\item Three 1.5~V batteries and a holder for them with a potentiometer soldered to it.
\item A plastic bag containing the following items:
\begin{itemize}
\item Two operational amplifiers on connector boards with a green label each with a number written on.
\item A multiplier on connector boards with an orange label.
\item A cell lithium battery.
\end{itemize}
\item An optical fiber cable with a white cylinder at one of its ends.
\item A plastic cup with a straw glued across it.
\end{enumerate}
You should carry additionally 2 multimeters with their connecting probes and a calculator.
\section{Initial easy tasks. S}
\begin{enumerate}
\item Set the multimeters to measure voltage (as voltmeters).
\item Measure the voltages of the four 9~V batteries with maximal accuracy (use 20~V range of the voltmeter) and write them down.
Do the same with the three 1.5~V batteries (using the 2000~mV range of the voltmeter).
\item Place the three 1.5~V batteries in their holder.
Connect the potentiometer \emph{potentiometrically} to one of your multimeters.
Rotate the axis of the potentiometer.
Measure the interval of voltages which you get.
This will be the voltage source for the next tasks.
Reversing the polarity changes the voltage sign.
\item Connect the four 9~V batteries to the two doubled connection clips by buckling the electrodes.
\section{Analogue squaring. M}
\label{s:M}
\item \textbf{Attention! From this moment on there is a possibility to burn the 
integral circuits, if you connect the batteries improperly.} 
Orient the set-up you work with so that both wires ``OUT'' and ``COM'' at the edge to be on the right side, take a look at Fig.~\ref{Fig:Board}.
\item Carefully connect the 9~V battery connection clip to the right hand side voltage supply 3-pin male connectors on the PCB, label to label (both labels facing each other).
\textbf{Work carefully -- if you make an error with the polarity you may burn the multiplier and operational amplifier.}
\item 
Connect the middle contact of the potentiometer 
with the input wire at the top of the set-up with a label ``IN''. 
Use the wire with ``crocodile'' connectors. 
\item Connect the other electrode of the potentiometer to the wire comes from the ``ground'' of the circuit with a sign ``COM''. 
\item 
Connect the first voltmeter V$_1$ with voltage $U_1$ to the potentiometer and ``IN''--``COM'' inputs of the circuit parallelly.
\item When you rotate the potentiometer axis, the voltage $U_1$ should change approximately between 0 and the total battery voltage 4.5~V.
\item Connect  the second voltmeter V$_2$, that shows voltage $U_2$, between the output wire of 
the circuit ``OUT'' and the common point ``COM''.
This way $U_2$ is the voltage between ``OUT'' and ``COM'' points.
\item Check the  whether the ``COM'' electrodes of both multimeters and the set-up are properly connected.
\item The 4 double connector pins with an orange label below them on the right hand side of the PCB are labeled ``AD633'' in Fig.~\ref{Fig:Board}.
Connect the multiplier to ``AD633'' so that \textbf{the orange label on the multiplier faces the orange label on the board}.
\item Rotate the potentiometer axis, wait 1 minute and write down the voltages $U_1$ and $U_2$.
Switch the polarity of the voltage source and repeat the measurements of the input voltage $U_1$ and the output voltage $U_2$.
Arrange the results in a table with columns:
number of measurement $i$, $U_1$ and $U_2$. 
\item Represent the results graphically where $U_1$ is set on abscissa (horizontal coordinate) and $U_2$ is set on ordinate (vertical coordinate) on a millimeter paper.
\item Add a column $(U_1)^2$ to the table and represent the results graphically, $(U_1)^2$ -- abscissa (horizontal axis) and $U_2$ -- ordinate (vertical axis) on a millimeter paper.
Draw a straight line that passes most closely to the experimental points.
This fitting (approximating) line is described by the equation
$U_2= (U_1)^2/U_0 + \mathrm{const}.$
Select two points on the straight line, measure the differences on abscissa 
$\Delta(U_1^2)$, and on ordinate $\Delta(U_2)$ 
and determine the slope
$U_0=\Delta(U_1^2)/\Delta(U_2)$.
This parameter $U_0$ with dimension of voltage is essential for the determination of the
electron charge $q_e$ as it is described in the next section. \label{it:u0}

\section{Voltage fluctoscopy. Determination of Electron charge $q_e$. L} 

\begin{figure}[h]
\centering
\includegraphics[scale=0.3]{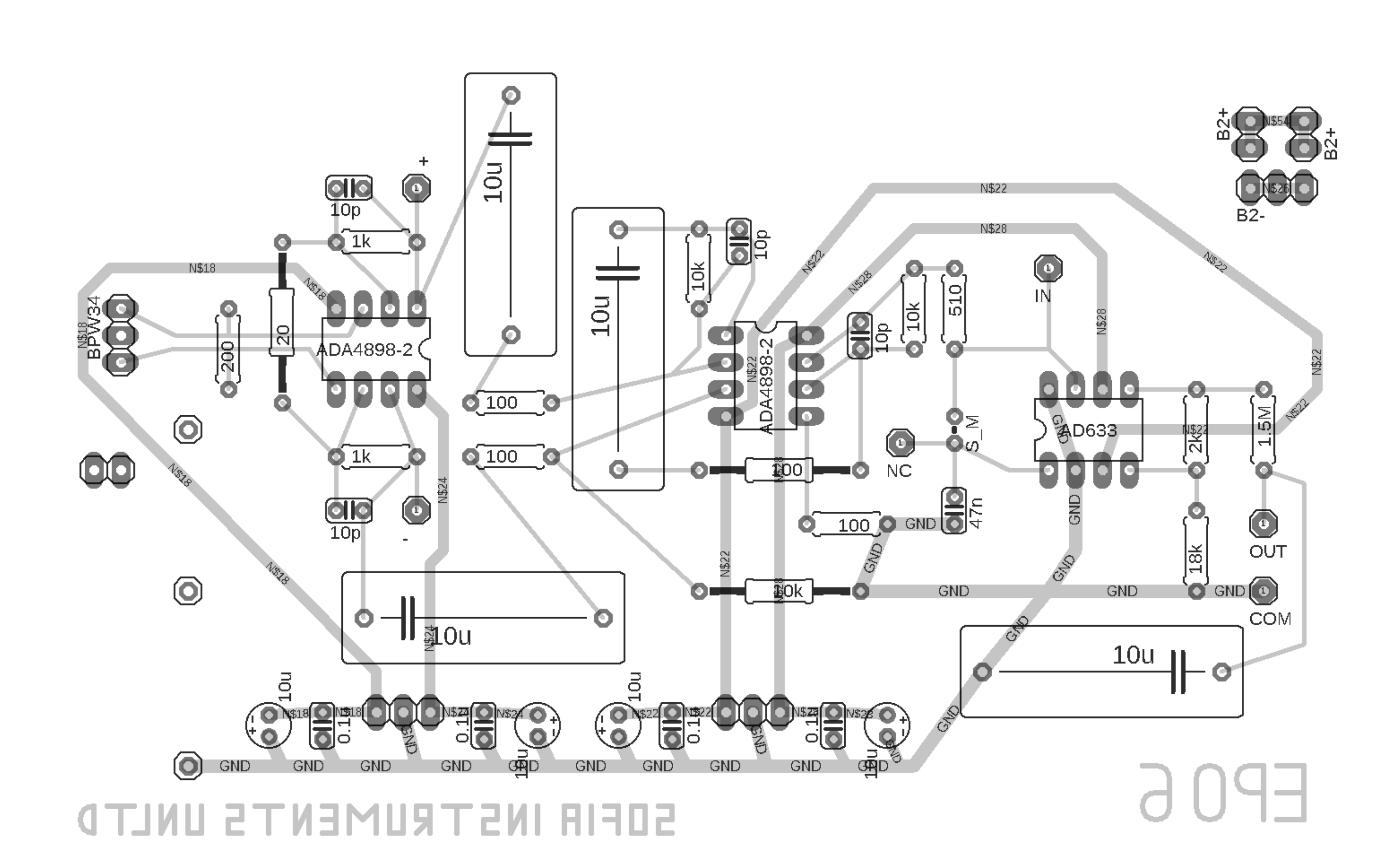}
\caption{A diagram of the printed circuit board of the experimental set-up.
An incadescent light bulb with a white cylinder attached on it not shown here is soldered between the ``B2-'' and ``B2+'' pins (not named on the PCB) to the top right of the board.
The two unnamed 3-pins located at the bottom of the board are the voltage supply pins.}
\label{Fig:Board}
\end{figure}
\item Put the 3 1.5~V batteries into their battery holder. 
Connect with two crocodiles the potentiometer two outer electrodes to the lamp ``B2-'' and one of ``B2+'' pins in the upper right corner of the set-up shown in Fig.~\ref{Fig:Board}.
Here polarity does not matter, upon proper connection the lamp should light up. 
\item Connect with a ``crocodile'' cable the ``($-$)'' cable of set-up with ``COM'' point of voltmeter V$_1$.
\item Analogously connect with another ``crocodile'' cable the ``($+$)'' electrode of set-up with 
``$\mathrm{V\,\Omega\, m\!A}$'' input of the voltmeter V$_1$.
In such way voltmeter V$_1$ shows the potential difference $U_{\pm}$ between the ``($+$)''  and the ``($-$)'' points and this voltage is proportional to the average photo-current of the photo diode.
\item Place the 3~V cell lithium battery in its battery holder on the double board connector with the photo-diode soldered,
\textbf{the ``($+$)''  side of the battery should be connected with the ``($+$)''  side of the battery holder.}
\item The two 4 double connector pins on the right hand side and center of the PCB with green labels next to them are labeled ``ADA4898-2'' places shown in Fig.~\ref{Fig:Board}.
Connect the two operational amplifiers to the two ``ADA4898-2'' so that \textbf{the green label of each operational amplifier faces the green label on the board. Be very careful, an opposite connection may result in an amplifier damage.}
\item Carefully connect the board connector with the soldered on it photodiode and lithium battery holder in its place ``BPW34'' in Fig.~\ref{Fig:Board} of the PCB (printable circuit board) as it is drawn with marker on the PCB.
Put very carefully the fiber optic cable in the black straw in the plastic cup. 
The white cylinder should enter the straw from the longer side of the straw.
Check whether the fiber cable can move through straw without significant friction.
\item Put the other end of the fiber cable to the white cylinder attached to the light bulb.
\item Put the photo diode in the straw from its shorter side by placing the plastic cup be next to the photo diode board connector, which is connected to the PCB.
\item Connect one of the 9~V battery clip to the left 3-pin supply voltage on the PCB. \\
\textbf{Be careful! The orange labels of the batteries voltage clips should match with the orange labels of the set-up.}
Switch the voltmeter to 2000~mV range and move the light-guide with white cylinder through the straw. This voltage is $U_\pm$ and should change up to 1000~mV.
In such a way, moving the light-guide you change the photo-current and measure the photo voltage $U_\pm$ created by the photo-current passing through the resistor $R$.
If you can not reach the value of at least 700~mV or there is no voltage at all, ask for help the quaestors in the auditorium. 
This is an important preliminary part of the measurement of the experiment.
\textbf{Do not try moving the white cylinder attached to the light bulb as you can easily break the latter.}
\item Ask for ice and water to be placed in your plastic cup with straw in it and place the fiber cable at the end of the straw far away from the photodiode.
\item After placing water and ice in the plastic cup, observe the cup for any leaks.
If present, carefully remove the cup and ask for its change.
\item Connect  the ``COM'' output of the PCB to the ``COM'' input of your voltmeter V$_2$ (voltmeter V$_2$ must be the same voltmeter that you used in Sec.~\ref{s:M}).
\item Connect  the ``OUT'' output of the PCB with the ``$\mathrm{V\,\Omega\, m\!A}$'' input of the voltmeter V$_2$.
Now, voltmeter V$_2$ measures the voltage $U_\mathrm{S}$ which is proportional to the current noise of the photo-diode.
\item Connect the other 9~V battery clip to the right 3-pin supply voltage on the PCB. \\
\textbf{Be careful! The orange labels of the batteries voltage clips should match with the orange labels of the set-up.}
\item Ask the queastors for a scotch tape and stick to the table all 6 cables of the experimental set-up in the following way:
\begin{itemize}
\item First place all four 9~V batteries below the PCB.
\item Stick the ``+'' and ``-'' cables above the PCB, ensuring they are not in contact.
\item Stick the ``IN'' and ``NC'' cables above the PCB and away from the ``+'' and ``-'' cables.
\item Stick the ``OUT'' and ``COM'' cables to the right hand side of the PCB, ensuring they are not in contact.
\end{itemize}
Now your experimental set-up should resemble a spider.
\item Look at the measured voltage of voltmeter V$_2$, it should stabilize around several tens of mV.
\item Begin measuring $U_\pm$ and $U_\mathrm{S}$: change $U_\pm$ by carefully moving the fiber cable in the straw a little bit closer to the photo diode.
The change of $U_\pm$ between each successive measurement should be at least 100~mV.
Patiently wait for $U_\mathrm{S}$ to stabilize, wait for at least 2 minutes.
\emph{If any disturbance changes significantly the voltmeter reading, again wait patiently for stabilization.}
Arrange the results in a table with columns:
number of measurement $i$, $U_\pm$ and $U_\mathrm{S}$. 
\item Represent the results graphically where $U_\pm$ is set on abscissa (horizontal coordinate) and $U_\mathrm{S}$ is set on ordinate (vertical coordinate) on a millimeter paper (if you need additional millimeter paper, the queastors will give you).
Draw the straight line which best explains the linear dependence
$U_\mathrm{S}=k U_\pm +\mathrm{const}$.
Choose 2 points on the line and determine the slope
$k=\Delta U_\mathrm{S}/ \Delta U_\pm$,
where $\Delta$ means difference.
This mathematical procedure is called a linear regression.
For our set-up $k\simeq 10^{-3}.$
\item Finally, determine the electron charge $q_e$
using the formula
\begin{equation}
q_e=2 k \frac{y_1}{Y^2} \frac{R_\mathrm{_L}}{R} C_\mathrm{_L}U_0,
\label{electron_charge}
\end{equation}
where $Y=1.01\times 10^{6}$ is the total amplification of our amplifier,
and $y_1=101$ is the amplification of the first step of the multiplier.
You can find the value of $C_\mathrm{_L}$ written on a label placed at the bottom right hand corner of the PCB ($C \equiv C_\mathrm{_L}$), $R_\mathrm{_L}=510~\Omega$ and $R=200~\Omega$. \\
Congratulations! You have just measured a fundamental constant
and you are in a good company! :)
\item Your measurement could be a little bit more precise by including the non-ideal effects of the operational amplifiers using he formula
\begin{equation}
q_e=2 k \frac{y_1}{(1-\varepsilon)Y^2} \frac{R_\mathrm{_L}}{R} 
C_\mathrm{_L}U_0.
\label{electron_charge_mod}
\end{equation}
The small correction $\varepsilon(C_\mathrm{_L})$ as a function of the capacitance $C_\mathrm{_L}$ is graphically presented in Fig.~\ref{Fig:Eps}.
\begin{figure}[h]
\centering
\includegraphics[scale=0.8]{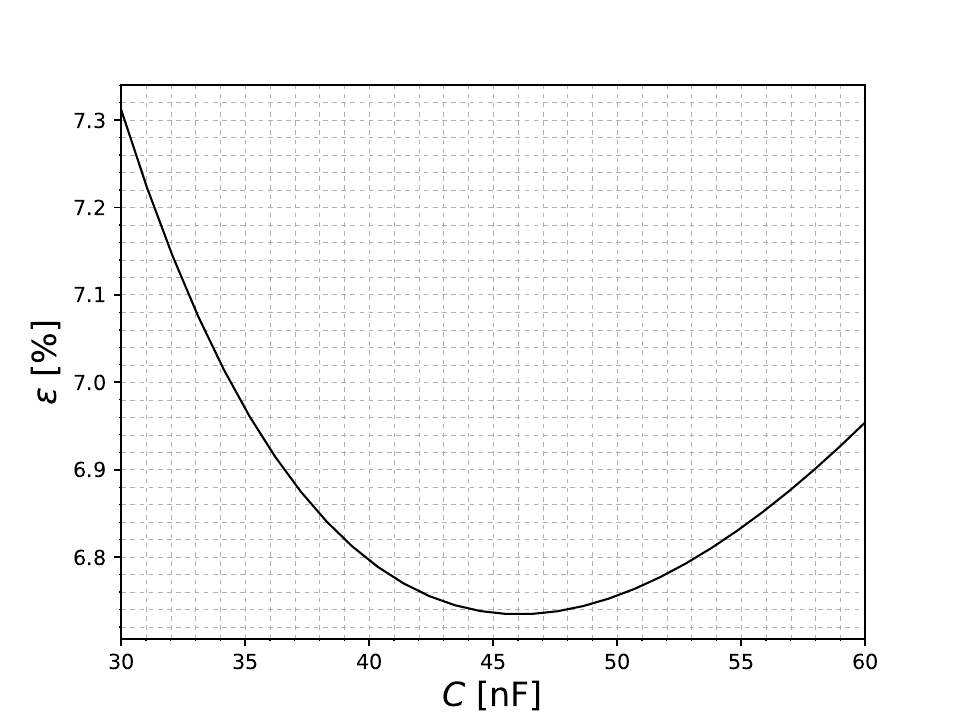}
\caption{Error $\varepsilon$ in percent in determination of $q_e$ as a function of the  
capacitor $C_\mathrm{_L} \equiv C$.}
\label{Fig:Eps}
\end{figure}
Find the error $\varepsilon$ from Fig.~\ref{Fig:Eps} according to your $C_\mathrm{_L}$ value written on the PCB and do not forget to divide $\varepsilon$ by 100 before calculating $q_e$ from Eq.~(\ref{electron_charge_mod}).

\end{enumerate}

A century ago this method for determination of electron charge $q_e$
by measurement of the shot noise 
was suggested by 
Walter Schottky.\cite{Schottky:1918}
At this time Walter Schottky was working with Max Plank 
and was inspired by a lecture of Albert Einstein on electric fluctuations.
Cerebrating the century anniversary, we have implemented again the Schottky idea
as university educational experiment.~\cite{qe_EPJ}
A year later we present a hundred times multiplication of the set-up for high school students.
After 101 years new science enters in the high school education
and physics is still the contemporary culture.

\section{Homework problem. XL}
\begin{figure}[t]
\begin{minipage}[t]{0.31\linewidth}
\includegraphics[scale=0.28]{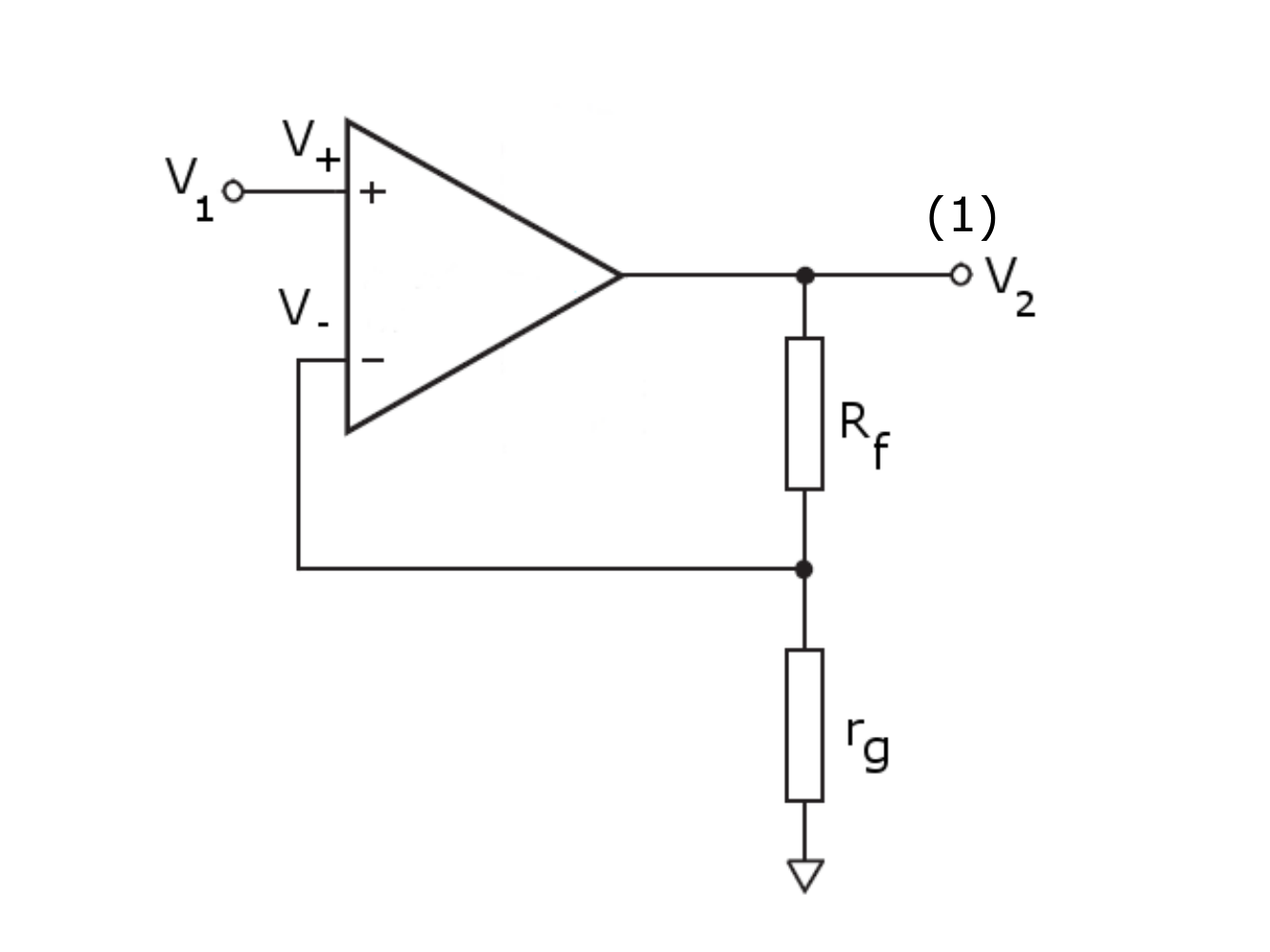}
\caption{Non-inverting amplifier}
\label{Non-inverting amplifier}
\end{minipage}
\begin{minipage}[t]{0.31\linewidth}
\includegraphics[scale=0.28]{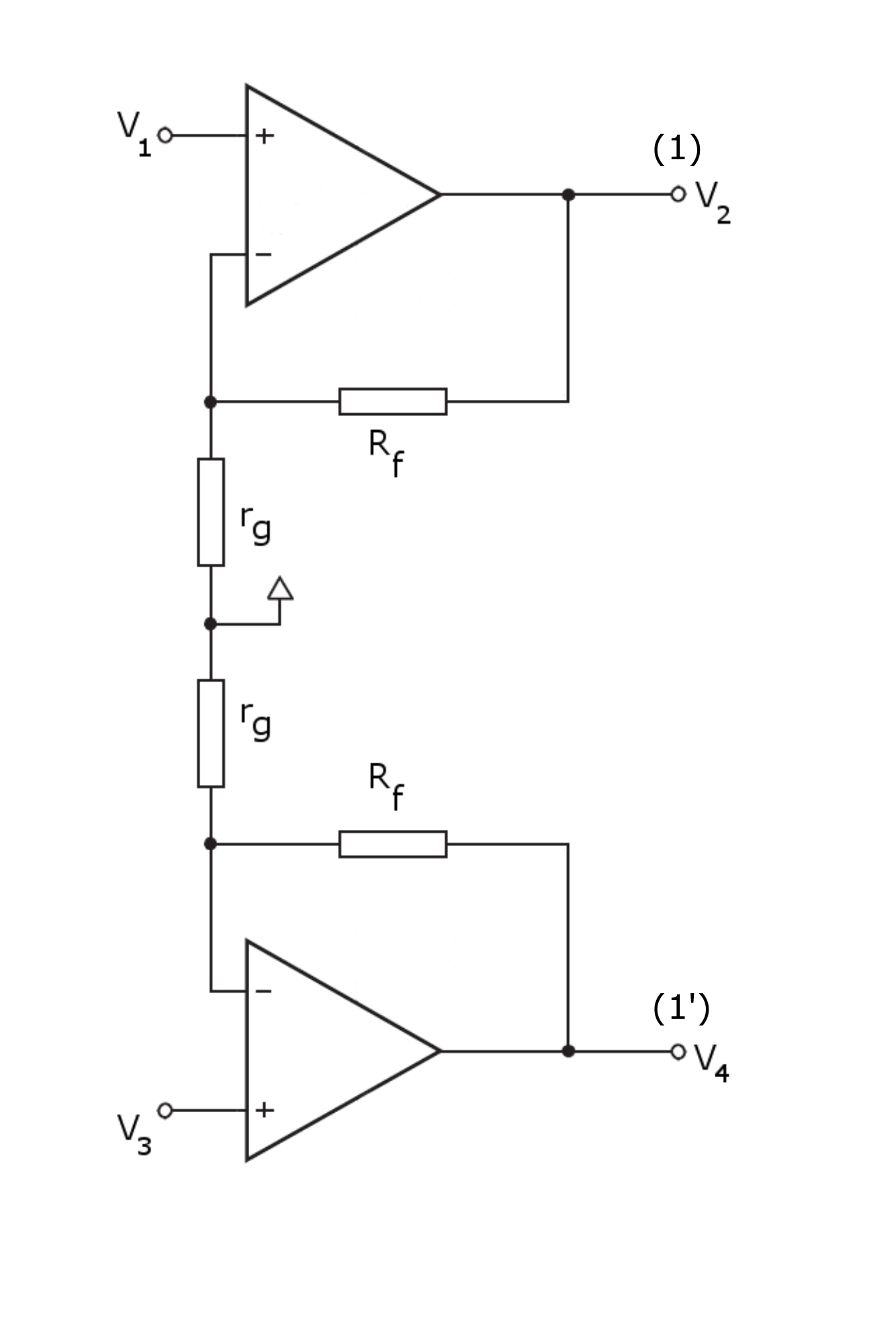}
\caption{Buffer}
\label{Buffer}
\end{minipage}
\begin{minipage}[t]{0.36\linewidth}
\includegraphics[scale=0.28]{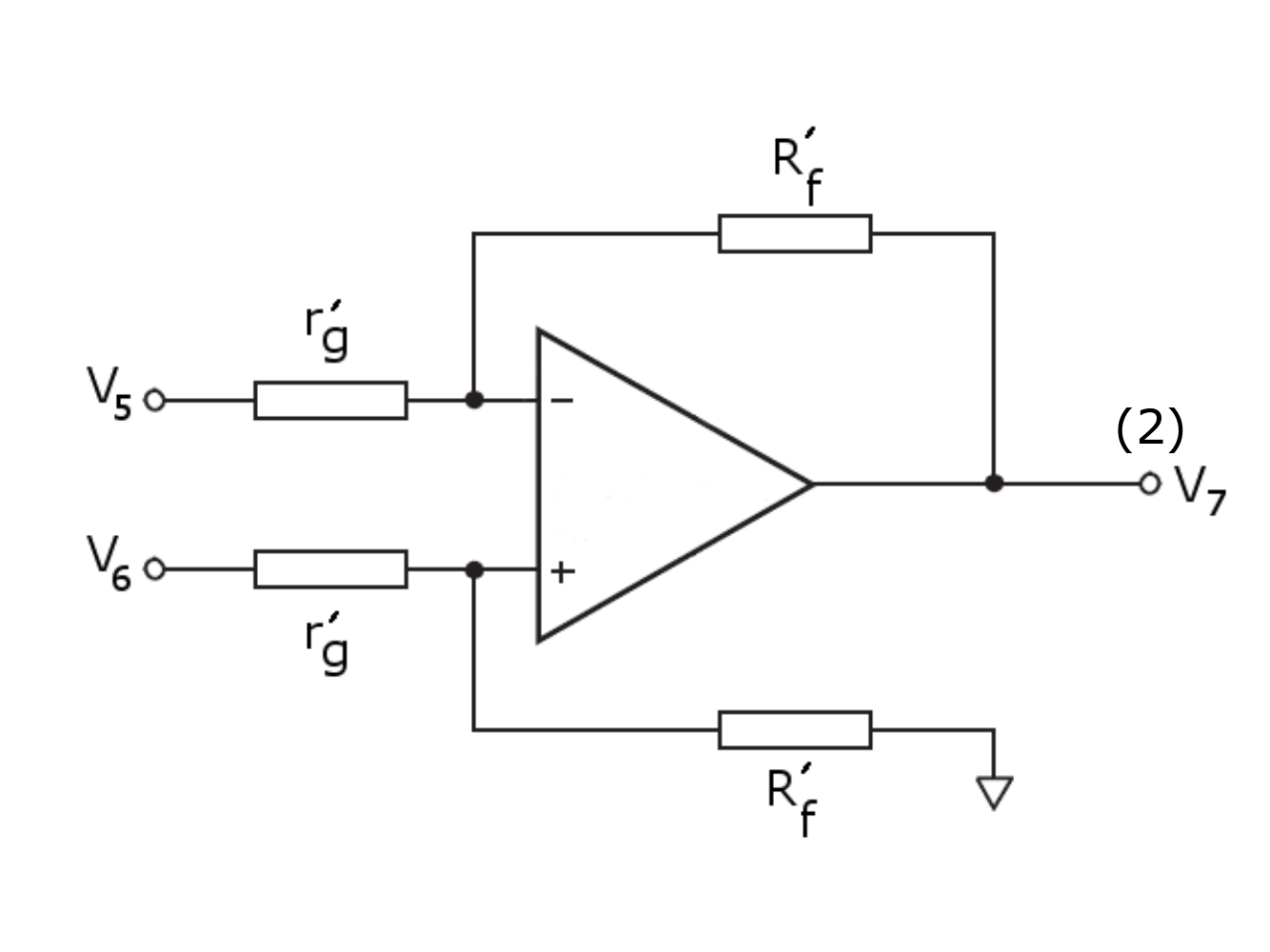}
\caption{Differential amplifier}
\label{Differential amplifier}
\end{minipage}
\begin{minipage}[c]{0.4\linewidth}
\includegraphics[scale=0.28]{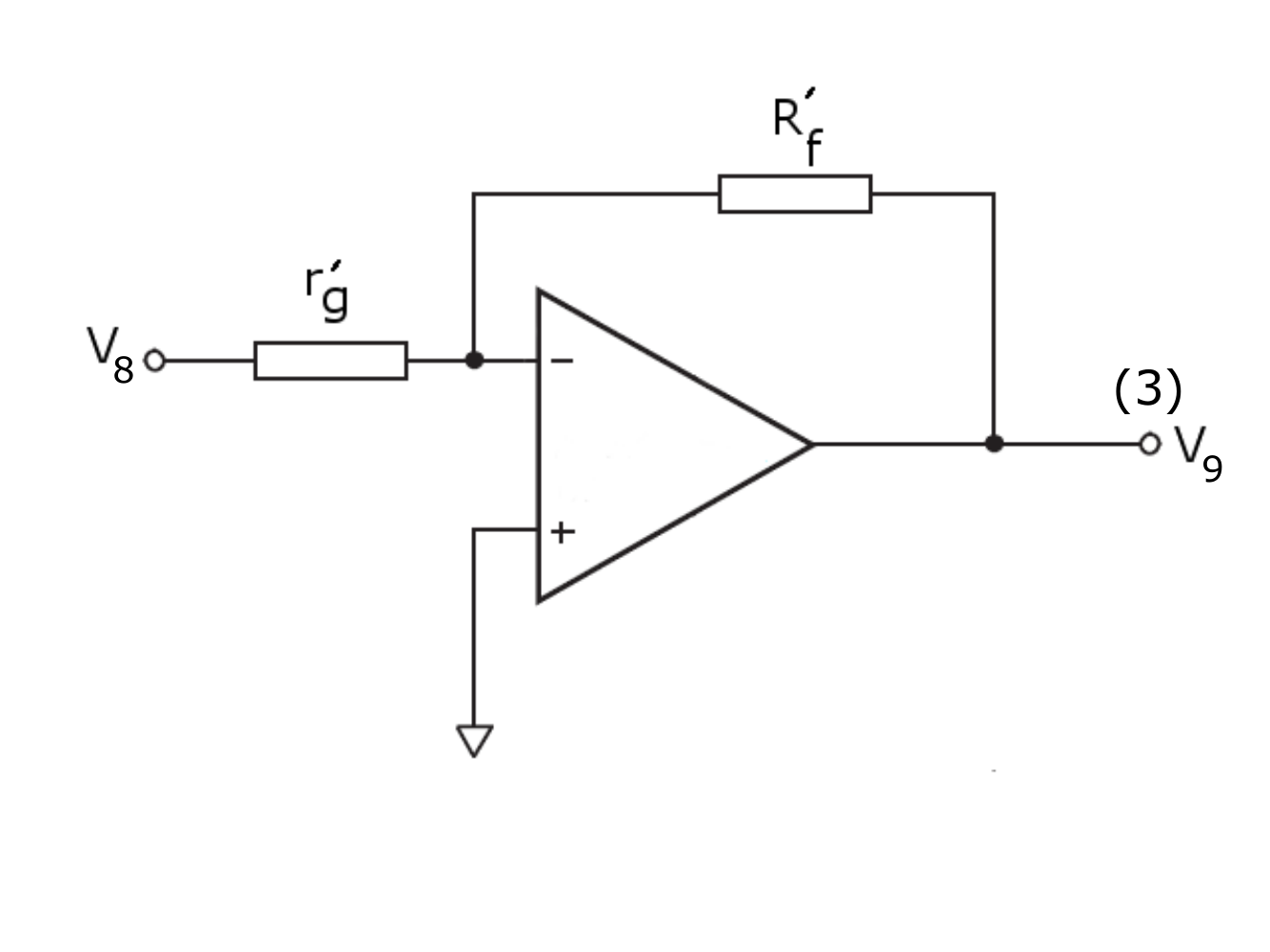}
\caption{Inverting amplifier}
\label{Inverting amplifier}
\end{minipage}
\begin{minipage}[c]{0.57\linewidth}
\includegraphics[scale=0.28]{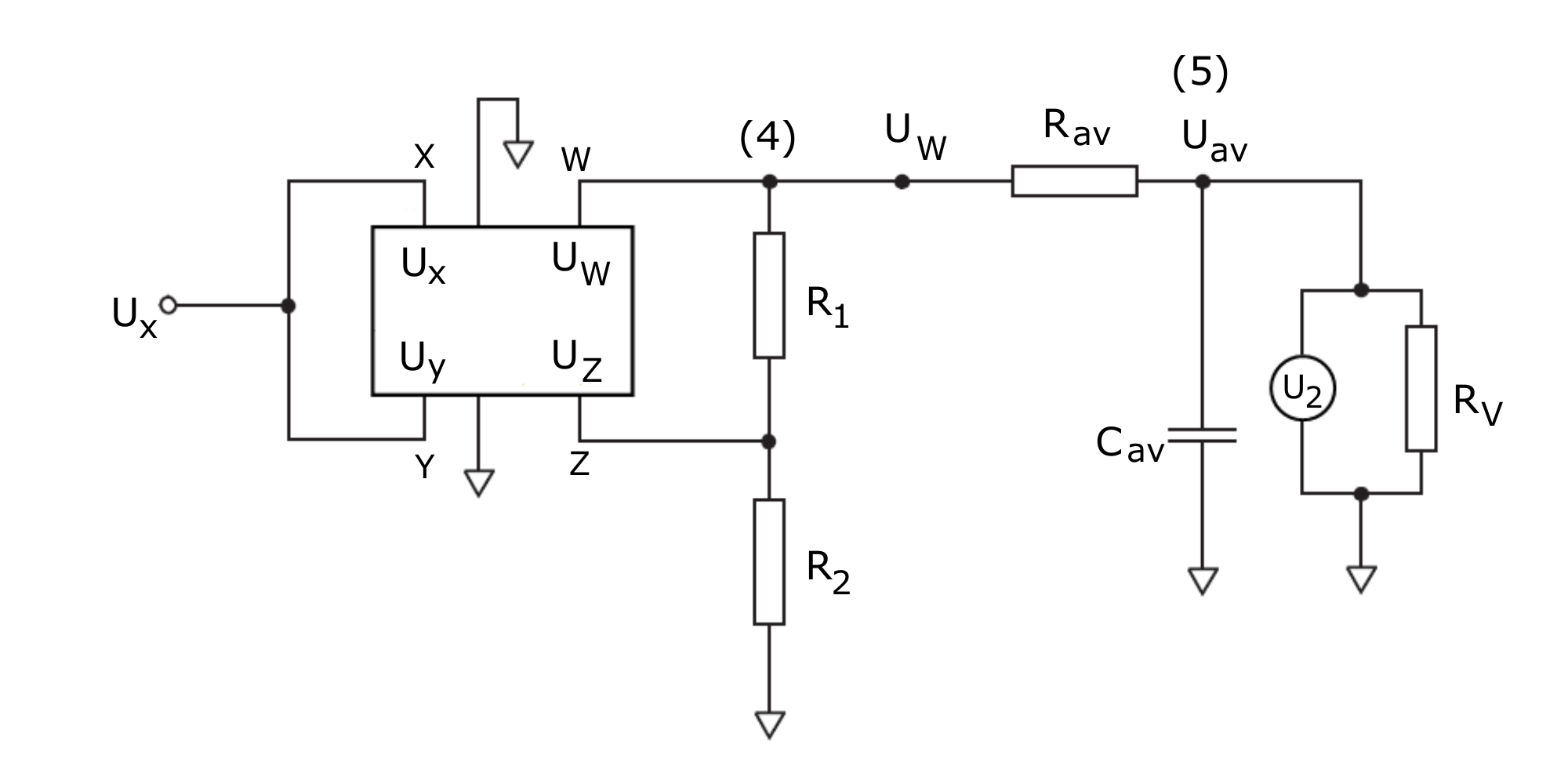}
\caption{Analog multiplier}
\label{Analog multiplier}
\end{minipage}
\end{figure}

Derive the formulas Eq.~(\ref{electron_charge}) 
and Eq.~(\ref{electron_charge_mod}) analyzing the circuit and the tabulated values of the parameters from Table~\ref{tbl:values}. 
Calculate the amplifications $y_1$ of the buffer and total amplification 
of all steps of the amplifier $Y$. 
The different details of the circuit are depicted in
Figs.~\ref{Non-inverting amplifier},\ref{Buffer}, 
\ref{Differential amplifier},
\ref{Inverting amplifier},
\ref{Analog multiplier}.

\begin{center}
\begin{table}[h]
\begin{tabular}{| c | r |}
		\hline
		&  \\ [-1em]
		Circuit element  & Value  \\ \tableline
			&  \\ [-1em]
			$R$ &200~$\Omega$ \\
			$r_\mathrm{_G}$ & 20~$\Omega$ \\
			$R_\mathrm{F}$ &  1~k$\Omega$  \\
			$C_\mathrm{F}$ &  10~pF  \\ 
			$C_\mathrm{G}$ & 10~$\mu$F \\
			$R_\mathrm{G}$ &  100~$\Omega$  \\ 
			$R_\mathrm{F}^\prime$ & 10~k$\Omega$ \\
			$C_\mathrm{F}^\prime$ & 10~pF \\
			$R_\mathrm{_L}$ & 510~$\Omega$ \\
			$R_1$ &  2~k$\Omega$  \\ 
			$R_2$ & 18~k$\Omega$  \\
			$R_\mathrm{av}$ & 1.5~M$\Omega$ \\
			$C_\mathrm{av}$ & 10~$\mu$F \\
			$R_\mathrm{_V}$ & $\approx 1~\mathrm{M} \Omega$ \\
			$V_\mathrm{CC}$ & +9~V \\
			$V_\mathrm{EE}$ & -9~V \\
\tableline
\end{tabular}
	\caption{Table of the numerical values of the circuit elements.}
	\label{tbl:values}
\end{table}
\end{center}

The solution to which must be send during the night after the Olympiad, by sunrise, at the e-mail of the Olympiad epo@bgphysics.eu.
The best solution will win the Sommerfeld price with a monetary equivalent of DM137. 

\section{Problems for further work}
The experimental set-up is actually a lock-in voltmeter and with it you can measure 
AC signal smaller than $\mathrm{\mu V}$.
You can write us how to use ``NC'' output in this case.

\clearpage
\section{Solution of the problems}

\subsection{Initial easy tasks. S}
\begin{enumerate}
\item Set the multimeter for voltage measurement.
\item The measured voltages of the four 9~V batteries are $U_{\mathrm{B}1}=9.28$~V, $U_{\mathrm{B}2}=9.45$~V, $U_{\mathrm{B}3}=9.39$~V, $U_{\mathrm{B}4}=9.59$~V and the measured voltages of the three 1.5~V AAA type batteries are $U_{\mathrm{B}5}$=1.578~V, $U_{\mathrm{B}6}$=1.578~V and $U_{\mathrm{B}7}$=1.578~V.
\item The potentiometer has three terminals, the outer two of which are soldered to the battery holder.
A potentiometer connectivity means connection to the middle and one of the outer terminals,
$U_\mathrm{min}=0.03$~V, $U_\mathrm{max}=4.28$~V and therefore the voltage interval is 
$U \epsilon [0.03,4.28]$~V.
\item Simply connect the 9~V batteries to their connection clips.

\subsection{Analogue squaring. M}

\item Orient the experimental set-up according to the instructions.

\item Connect the 9~V battery clip to the described in the instructions voltage supply of the experimental set-up.

\item Connect the potentiometer to the first input wire of the PCB according to the instructions.

\item Connect the potentiometer to the second wire of the PCB according to the instructions.

\item Connect the voltmeter to the potentiometer according to the instructions.

\item A check for proper connectivity.

\item Connect the second voltmeter to the output of the experimental set-up according to the instructions.

\item A check for proper connection of both voltmeters, the ``COM'' electrodes of which should be connected.

\item Connect the AD633 multiplier to the PCB according to the instructions.

\item The results from the measurements are presented in the three columns designated i, $U_1$ and $U_2$ of Table~\ref{tbl:mult}.
\begin{table}[h]
\begin{tabular}{ c  r  r  r }
		\tableline \tableline
		&  \\ [-1em]
		i & $U_1$ [V] & \hspace{2.5pt} $U_2$ [V] & \hspace{5pt} $U_1^2$ [V$^2$] \\ \tableline 
			&  \\ [-1em] 
			1	&	0.01	&	0.000	&	0.000 \\
			2	&	0.25	&	0.021	&	0.063 \\
			3	&	0.49   &	0.093	&	0.240 \\
			4	&	0.75   &	0.220	&	0.562 \\
			5	&	1.00   &	0.397	&	1.000 \\
			6	&	1.24   &	0.615	&	1.538	\\
			7	&	1.50	&	0.905	&	2.250	\\
			8	&	1.76    &	1.238	&	3.098	\\
			9	&	2.00    &	1.603	&	4.000	\\
			10	&	2.26	&	2.050	&	5.108	\\
			11	&	2.50	&	2.510	&	6.250	\\
			12	&	2.59	&	2.680	&	6.708	\\
			13	&	0.00	&	0.000	&	0.000	\\
			14	&	-0.12	&	0.002	&	0.014	\\			
			15	&	-0.37	&	0.052	&   0.137	\\
			16	&	-0.64	&	0.161	&	0.410	\\
			17	&	-0.87	&	0.300	&	0.757	\\
			18 &	-1.10  &	0.486	&	1.210	\\
			19 &	-1.35	&	0.730	&	1.823	\\
			20 &	-1.60	&	1.031	&	2.560	\\
			21	&	-1.85	&	1.377	&	3.423	\\
			22	&	-2.10	&	1.782	&	4.410	\\
			23	&	-2.34	&	2.200	&	5.476 	\\
			24	&	-2.54	&	2.600	&	6.452 	\\		
\tableline \tableline
\end{tabular}
	\caption{Results from the measurements in task 14 and the data processing in task 16 (last column).}
	\label{tbl:mult}
\end{table}

\item The results from the second and third columns of Table~\ref{tbl:mult} are represented graphically in Fig.~\ref{fig:par}.
\begin{figure}[h]
\begin{minipage}[t]{0.44\linewidth}
\centering
\includegraphics[scale=0.76]{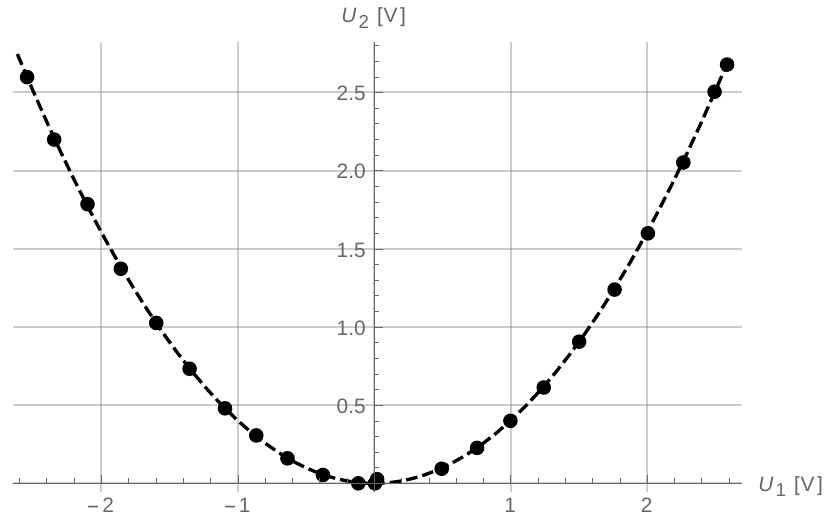}
\caption{The graphically represented dependency $U_2$ of $U_1$ from Table~\ref{tbl:mult}.}
\label{fig:par}
\end{minipage}
\hfill
\begin{minipage}[t]{0.52\linewidth}
\centering
\includegraphics[scale=0.3]{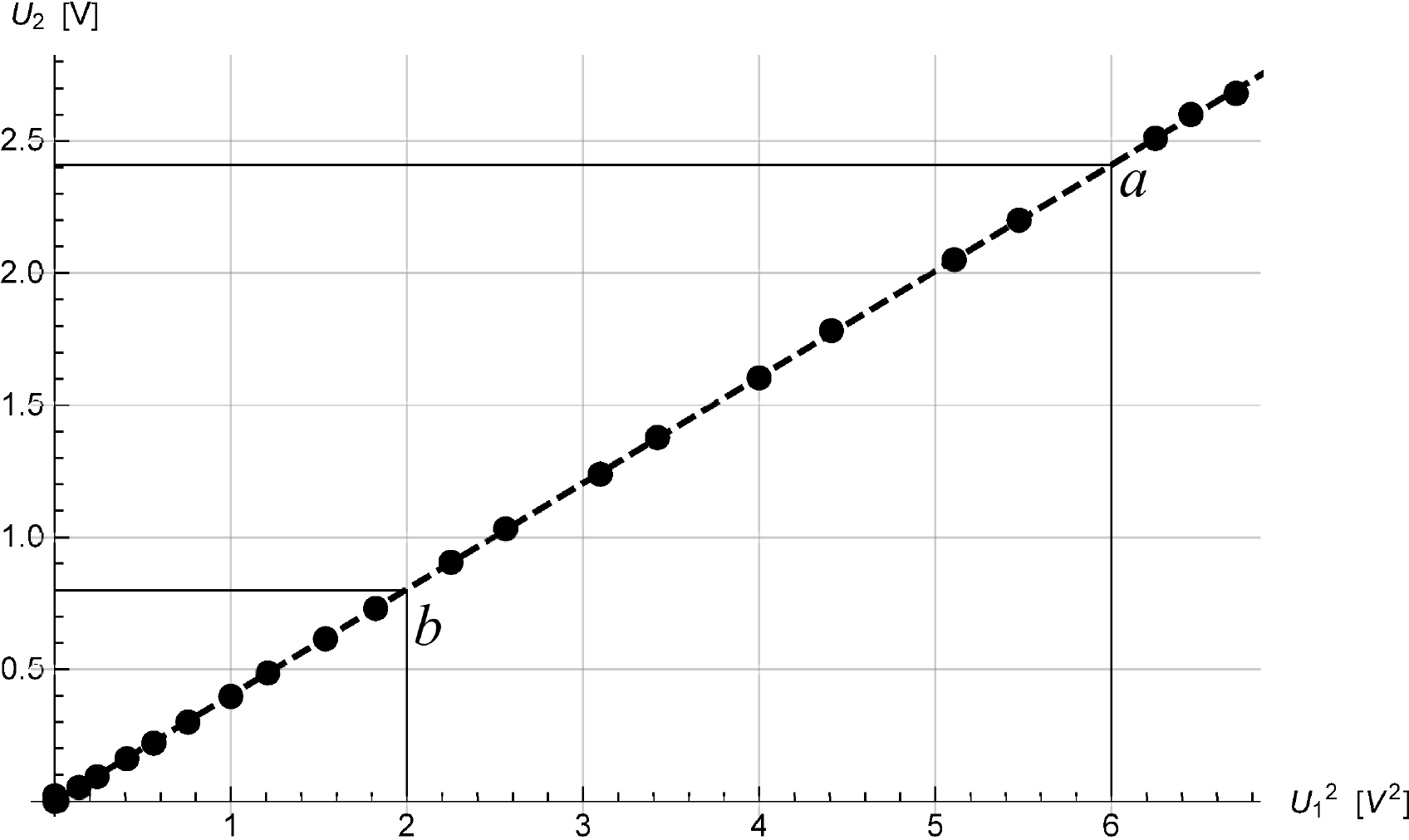}
\caption{The graphically represented dependency $U_2$ of $(U_1)^2$ from Table~\ref{tbl:mult}.}
\label{fig:lin}
\end{minipage}
\end{figure}
\item The last column of Table~\ref{tbl:mult} and the results presented in Fig.~\ref{fig:lin}.
We choose points (a) and (b) in Fig.~\ref{fig:lin}, so that in abscissa $(U_1)_a^2=6$~V$^2$ and $(U_1)_b^2=2$~V$^2$.
Next we find their corresponding values in ordinate $(U_2)_a=2.4$~V and, $(U_2)_b=0.8$~V by drawing a line parallel to the abscissa towards the ordinate and the value of the intersection point on the ordinate is the corresponding $(U_2)$.
Finally, we calculate
\be
U_0=\frac{\Delta(U_1^2)}{\Delta(U_2)}=
\frac{(U_1)_a^2-(U_1)_b^2}{(U_2)_a-(U_2)_b}=\frac{6-2}{2.4-0.8}=2.5~\mathrm{V}.
\label{eq:u0}
\ee
The value of $U_0$ may be different depending on the used voltmeter for the measurement of $U_2$, a short discussion is presented in App.~\ref{sec:LI}.

\subsection{Voltage fluctoscopy. Determination of electron charge $q_e$. L} 

\item Connect the batteries to the lamp on the PCB according to the instructions.

\item Connect the voltmeter ``COM'' input according to the instructions.

\item Connect the voltmeter ``$\mathrm{V\,\Omega\, m\!A}$'' input according to the instructions.

\item Connect the 3~V cell lithium battery to the double board connector according to the instructions.

\item Connect the two ``ADA4898-2'' operational amplifiers to the PCB board according to the instructions.

\item Connect the board connector to the PCB board and put one end of the fiber optic cable in the black straw attached to the plastic cup according to the instructions.

\item Connect the other end of the fiber optic cable to the light bulb according to the instructions.

\item Place the plastic cup on the board and insert the photodiode in the black straw according to the instructions.

\item Connect the power voltage of the first ``ADA4898-2'' operational amplifier according to the instructions.
Check for proper connectivity by ensuring voltage readings of the connected voltmeter.

\item Get ice and water pull the fiber cable at the end of the straw, as explained.

\item Verify that no leaks from the plastic cup are present.

\item Connect the ``COM'' input of the second voltmeter V$_2$ to the PCB according to the instructions.

\item Connect  the ``$\mathrm{V\,\Omega\, m\!A}$'' input of voltmeter V$_2$ to the PCB according to the instructions.

\item Power on the right ``ADA4898-2'' operational amplifier by connecting the 9~V battery clip to the PCB according to the instructions.

\item Arrange and stick the batteries and all 6 cables according to the instructions.
Battery and cable placement and movement could lead to drastically changes in the experimental data during the experiment.

\item Verify that voltmeter V$_2$ has correct reading.

\item Start measuring according to the instructions.
The measured experimental data is shown in Table~\ref{tbl:meas}.
\begin{center}
\begin{table}[h]
\begin{tabular}{ c  r  r  }
		\tableline \tableline
		&  \\ [-1em]
		$i$  \hspace{1.5pt} & $U_\pm$ [mV] & \hspace{5pt} $U_\mathrm{S}$ [mV] \\ \tableline 
			&  \\ [-1em] 
			1 & 177 & 16.7 \\
			2 & 292 & 16.9 \\
			3 & 420 & 17.2 \\
			4 & 525 & 17.5 \\
			5 &	 661 & 17.8 \\
			6 & 771 & 17.9 \\ 
			7 & 899 & 18.2 \\
			8 & 1040 & 18.6 \\
			9 & 1200 & 18.9 \\
\tableline \tableline
\end{tabular}
\caption{Experimental results from the measurements of the photo voltage $U_\pm$ and the time averaged voltage fluctuations $U_\mathrm{S}$ of the photodiode.}
	\label{tbl:meas}
\end{table}
\end{center}

\item Fig.~\ref{fig:k} shows the graphical representation of the experimental data from Table~\ref{tbl:meas} and the straight line of the linear regression.
\begin{figure}[h]
\centering
\includegraphics[scale=0.5]{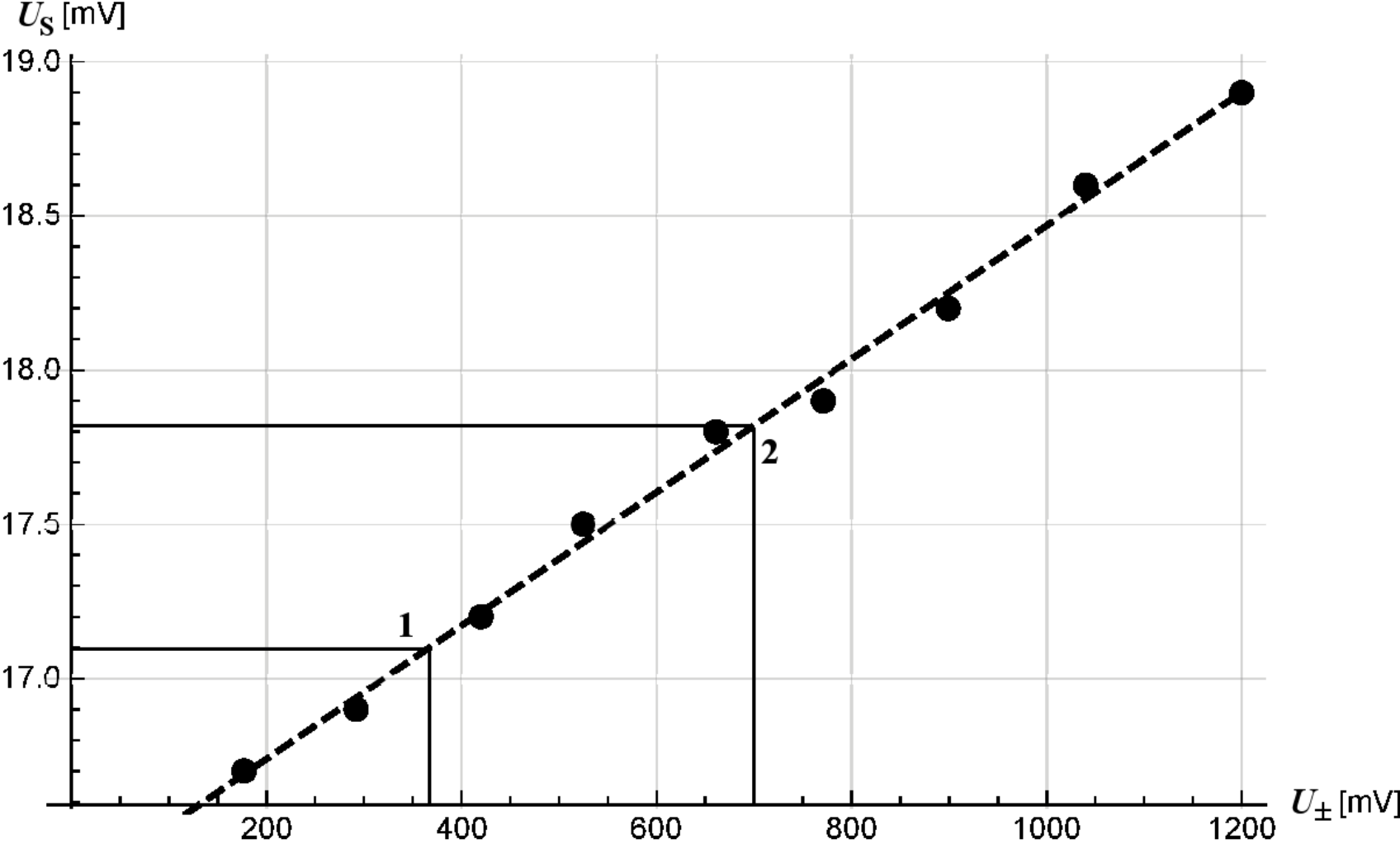}
\caption{Linear regression of the photo voltage $U_\pm$ and the time averaged voltage fluctuations $U_\mathrm{S}$ from Table~\ref{tbl:meas}.
These measurements were made on a warm mid-March St.~Theodor Day with the experimental set-up being placed on a window ledge due to central heating still on in the room.
The cooling of the plastic cup filled with ice and water is not sufficient to deal with spring temperature at mid latitudes (Sofia) and central heating therefore for warm and hot weather, additional cooling is necessary.
Placing the experimental set-up in a fridge is probably the easiest and most accessible solution.}
\label{fig:k}
\end{figure}
Choosing points (1) and (2) from the line of the linear regression in Fig.~\ref{fig:k}, we calculate the slope of the line
\be
k=\frac{\Delta U_\mathrm{S}}{\Delta U_\pm}=
\frac{(U_\mathrm{S})_2-(U_\mathrm{S})_1}{(U_\pm)_2-(U_\pm)_1}=
\frac{17.82-17.10}{700-370}=\frac{0.72}{330}=2.182 \times 10^{-3}.
\ee

\item From the PCB label $C_\mathrm{L}=42.7 \times 10^{-9}$~nF and therefore the electron charge
\be
q_e=2 k \frac{y_1}{Y^2} \frac{R_\mathrm{_L}}{R} C_\mathrm{_L}U_0 =
2 \,(2.182 \times 10^{-3}) \,\frac{101}{(1.01)^2 \times 10^{12}} \, \frac{510}{200} \, (42.7 \times 10^{-9}) \, (2.5) = 1.18 \times 10^{-19}~\mathrm{C}.
\ee

\item From Fig.~\ref{Fig:Eps} $\varepsilon$=6.76\% for $C_\mathrm{L}=42.7 \times 10^{-9}$~nF and therefore
\be
q_e=\frac{q_e}{(1-\varepsilon)} = \frac{1.18 \times 10^{-19}}{(1-0.0676)}=
1.26 \times 10^{-19}~\mathrm{C}.
\ee
The obtained value for the electron charge is around 20\% lower than its real value 1.6$\times 10^{-19}$~C, which is an excellent result for a high school experiment made by a high school student!

\end{enumerate}

\input{EPO6_bg_a}
\input{EPO6_mk_a}
\input{EPO6_sr_a}
\input{EPO6_ru_a}

\input{EPO6_de_a}

\clearpage

\renewcommand{\acknowledgmentsname}{Acknowledgment}
\acknowledgments

The authors would like to thank Peter Todorov for his invaluable support for EPO6 and the previous olympiads, to Mikhail Zhabitsky for the Russian translation, to Beka Nathan, Svetla Strashimirova, Evgenia Stoinovska, Nonka Bailova, Grigori Matein, Elton Shumka, Fisnik Hoxha, Orlin Georgiev, Kalin Angelov,
Bojidar Kovachev and Mladen Matev for the indispensable help in the EPO6 organization.
Diana from WMWare has fed with sandwiches the most clever children on the planet.
Pancho Cholakov from K1 Electronics suggested us to use ADA4898. 
Thank you also to Andrea Rosenauer for proofreading the German translation.
The spirit of the Olympiad was again unforgettable and unforgotten.
	

\renewcommand{\appendixname}{Appendix}
\clearpage
\appendix

\section{A photograph of the experimental set-up}
\renewcommand{\figurename}{Fig.}
\begin{figure}[h]
\centering
\includegraphics[scale=0.15]{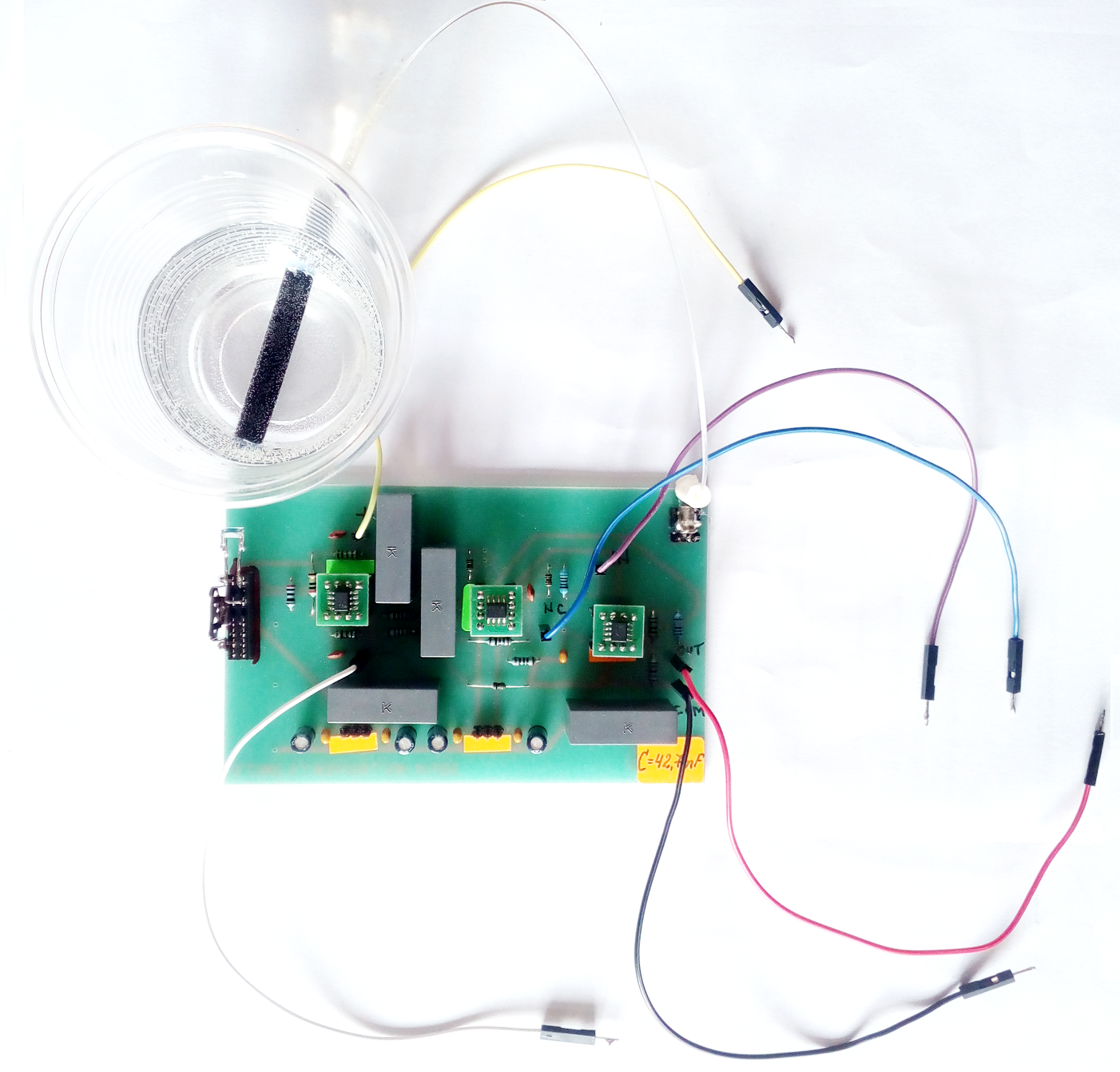}
\caption{A photograph of the experimental set-up.
The wires between which the difference $U_\mathrm{S}$ of the output voltage ``OUT'' and the ground ``COM'' are visible on the right.
The photo voltage $U_\pm$ is measured between the first pair of cables from left to right.
The cable ``NC'' can be used for further research.
The photodiode is soldered on the double board connector connected to the PCB on the left of the PCB below the plastic cup filled with water.
The lamp is at the upper right corner of the PCB with one end of the fiber optic cable inserted in it, while the other end is in the black straw glued across the plastic cup.
When measuring, the photodiode needs to be cooled.
For this purpose, the plastic cup is moved next to the double board connector, so that the photodiode fully enters in the black straw from the remaining free end of the latter (the other end has already a fiber optic cable in as showed in the picture).
The EPO6 participants do not need of this photo because they have the real set-up but for the other readers it is necessary.}
\label{Fig:Photo}
\end{figure}

\section{Lock-in voltmeter hidden in the experimental set-up}
\label{sec:LI}

Let us now take a look at the element named S\textunderscore M in Figs.~\ref{Fig:Board}.
S\textunderscore M is a switch, whose initial position is closed, that equals the potentials of both ``IN'' and ``NC'' inputs.
Or in other words, $U_X=U_Y$, as we have already considered in the operation of the non-linear element.

For the Olympiad problems the switch S\textunderscore M is not necessary and hence it has not been placed on the printed circuit board in Fig.~\ref{Fig:Board}.
Instead, both switch contacts are soldered together, which in practice means that the switch is permanently closed or short-circuited (i.e. missing).
The joint connecting both switch contacts can be easily removed by cutting with diagonal pliers (wire cutters) or unsoldering it and in this way the switch is opened and now $U_X \neq U_Y$, which means that the multiplier multiplies two different signals.

Let us now consider a different element in principle.
A weak signal $U_s(t)=U_s \cos(\omega t)$ is applied at the amplifier input, which is amplified $Y$ times, so that at the multiplier $X$ input voltage $U_X(t)=Y U_s(t)$ is applied.
In case of a removed connection between $X$ and $Y$ inputs, i.e. open switch S\textunderscore M in Fig.~\ref{Fig:Board}, at the ``NC'' labeled input of the board a known signal $U_r(t)=U_r\cos(\omega t)$ is applied.
This signal goes to the $Y$ multiplier input $U_Y(t)=U_s(t).$

The analog multiplier shown in Fig.~\ref{Analog multiplier} gives the following output voltage $U_W$ as a function of the input voltages $U_X$ and $U_Y$
\be
U_W=\frac{U_X U_Y}{U_\mathrm{m}}+U_Z,
\label{eq:mult}
\ee
where the voltage $U_Z$ is simply an offset voltage and for the used AD633 multiplier the voltage dimension constant $U_\mathrm{m}=10$~V.
The offset voltage can be easily expressed if we consider a current $I_W = U_W/(R_1+R_2)$ flowing from point (4) to ground in Fig.~\ref{Analog multiplier}.
Then the offset
\be
U_Z=I_W R_2 = \frac{R_2}{R_1+R_2} U_W
\ee
and substituting it in Eq.~\ref{eq:mult}, we obtain
\be
U_W=\frac{U_X U_Y}{U_\mathrm{m}} + \frac{R_2}{R_1+R_2} U_W
\ee
and after elementary rearrangements
\be
U_W=\frac{U_X U_Y}{U_\mathrm{m}} \frac{R_1+R_2}{R_1}.
\label{eq:multw}
\ee
Finally, the circuit ends with averaging filter with large time constant $R_\mathrm{av}C_\mathrm{av} = 15$~sec and after this filter we measure the time averaged voltage $U_V = \left<U_W \right>$.
The resistances $R_\mathrm{V}$ and $R_\mathrm{av}$ form a voltage divider, for which
\be
U_\mathrm{V}=\frac{R_\mathrm{V}}{R_\mathrm{av}+R_\mathrm{V}} \left < U_W \right >
\ee
and substituting here Eq.~\ref{eq:multw} for $U_W$, we obtain
\be
U_V = \frac{\left <U_X U_Y \right >}{U_\mathrm{m}} \frac{R_1+R_2}{R_1}
\frac{R_\mathrm{V}}{R_\mathrm{av}+R_\mathrm{V}} = \frac{\left < U_X U_Y \right >}{U_0},
\label{eq:uv}
\ee
where
\be
U_0 \equiv \frac{R_1}{R_1+R_2}\frac{R_\mathrm{av}+R_\mathrm{V}}{R_\mathrm{V}}U_\mathrm{m},
\ee
which is precisely what you have measured and calculated in task \ref{it:u0}.
It is worth noting that $R_\mathrm{V}$ is the internal resistance of your voltmeter (which is also evident from Fig.~\ref{Analog multiplier}) and therefore by calculating $U_0$, you actually calculate the internal resistance of your voltmeter.
That is why the value for $U_0$ calculated here in the solution Eq.~(\ref{eq:u0}) may be different from your result.

Substituting $U_X(t)=Y U_s(t)$ and $U_Y(t)=U_s(t)$ into Eq.~(\ref{eq:uv})
\be
U_V=\frac{\left < YU_s(t) U_r(t) \cos^2(\omega t) \right >}{2 U_0}
\ee
and accounting that $\left < \cos^2(\omega t) \right > = 1/2$, for the time averaged voltage measured by the voltmeter we obtain
\be
U_V=\frac{Y U_r}{2 U_0}U_s .
\ee
In this way, at the circuit output the voltmeter measures direct voltage $U_V$, in which the studied by us periodic signal with amplitude $U_s$ is present.
The signal $U_r$ that we have applied to the $Y$ input is called carrier or reference signal and a device, whose operation we have just described is also called a lock-in voltmeter (amplifier).

Lock-in amplifiers are used for small signals measurements even when through an oscilloscope the amplified signal is lost in the background of the external noise.
In this synchronous measurement of two sinusoidal signals with common frequency, the measured signal can be separated from the noise with spectral density
$(\mathcal{E}^2)_f$, if
\be
U_s>\sqrt{(\mathcal{E}^2)_f \,\Delta f_\mathrm{av}},
\quad \Delta f_\mathrm{av}=\frac1{2\pi R_\mathrm{av}C_\mathrm{av}},
\ee
i.e. at large enough time averaging
$\tau_\mathrm{av}=R_\mathrm{av}C_\mathrm{av}$.
Still the noise should not overload the amplifier
\be
U_m>\sqrt{Y^2\, (\mathcal{E}^2)_f\,B},
\quad B= \frac{f_0}{y_1}
\ee
and this sets the upper limit of the possible gain coefficient
\be
Y<\frac{U_m}{\sqrt{(\mathcal{E}^2)_f\,B}}.
\ee

\section{EPO6 Feedback}

\begin{itemize}
\item Ohrid e nasiot vsicki zaednicki grad.
\item Everything was awesome. Maybe you can add an adittional part in the olympiad like a tour around the city.
\item Добре ще е да се актуализира регламента и мотивацията за участие.
\item This year was my first time participating. I was really fascinated by the Olympiad, I didn’t expect it to be this good. Next year I’m going to participate again for sure. 
\item introduce medals please
\end{itemize}

\section{Electron Charge Value at Home}

\begin{itemize}
\item Да. Мои ученици го повториха: $2.2 \times 10^{-19}$.
\item Yes, I did. I got $1.2 \times 10^{-18}$, I probably had a calculation mistake because I was closer than last time.
\end{itemize}

\end{document}

%% file: EPO6_bg_a.tex
\clearpage
\renewcommand{\figurename}{Фигура}
\renewcommand{\tablename}{Таблица}
\begin{center}
\textbf{Измерване на заряда на електрона $q_e$ използвайки шумове на Шотки.
Задача от 6-тата Експериментална олимпиада по физика. София, 8 декември 2018}
\end{center}
\normalsize
\indent
\setlength{\leftskip}{1.2cm}
\setlength{\rightskip}{1.2cm}
Описани са няколко последователни с постановка с печатна електронна платка (PCB), специално изобретена за тези експерименти.
Извършването на последователни експериментални задачи дава възможността да се определи стойността на заряда на електрона $q_e.$
Флуктуациите на напрежението $U(t)$ трябва да бъдат измерени за различни осветености на фотодиод.
Напрежението е усилено 1~милион пъти $Y=10^6$. 
Усиленото напрежение $YU(t)$ е приложено на устройство, което дава като резултат стойността на осредненото вдигнато на квадрат напрежение 
$U_\mathrm{S}=\left<(Y U(t))^2\right>/U_0$.
Това напрежение $U_\mathrm{S}$ се измерва с мултицет.
Серията от измервания дава възможността да се определи $q_e$ използвайки добре известната формула на Шотки за спектралната плътност на токовия шум
$(I^2)_f=2q_e\left<I\right>.$
За по-малките ученици, основната задача е да се анализира аналоговото повдигане на квадрат.
Ученическите работи се разделят и оценяват в четири възрастово разделени категории S, M, L, XL.
За последната XL категория, задачите са ориентирани към университетската образователна програма по физика и включват теоретично изследване на експерименталната платка като инженерно устройство.
Това е задачата на ОЕФ6, декември 2018 ``Денят на заряда''.
ОЕФ6 се организира от Софийски клон на Съюза на физиците в България със съдействието на Софийски Университет и Съюза на физиците в Република Македония.

\setlength{\leftskip}{0cm}
\setlength{\rightskip}{0cm}

\section{Въведение}

Още в самото си начало, Олимпиадата по Експериментална Физика (ОЕФ) е световно известна:
всички задачи от Олимпиадата са публикувани в Интернет~\cite{EPO1,EPO2,EPO3,EPO4,EPO5} и от самото си начало, участниците бяха 120.
През последните години, участваха ученици от 7 страни и разстоянието между най-отдалечените градове е повече от 4~Mm.
Нека опишем основните разлики между ОЕФ и други подобни състезания.
\begin{itemize}
\item Всеки участник в ОЕФ получава като подарък от организаторите постановката, с която е работил.
По този начин, след като Олимпиадата е приключила, дори лошо представил се участник има възможност да повтори експеримента и да достигне нивото на шампиона.
По този начин Олимпиадата директно влияе на нивото на преподаване в целия свят.
След края на учебната година, постановката остава в училището, където участникът е учил.
\item Всяка от задачите е оригинална и е свързана с фундаментална физика или разбирането на действието на технически патент.
\item В ОЕФ6 е реализирана олимпийската идея в първоначалната си форма и всеки, който има желание да участва от целия свят, може да го направи.
Няма ограничение в броя на участниците.
От друга страна, сходството с други олимпиади е, че задачите са пряка илюстрация на учебния материал и заедно с други подобни състезания смекчава деградацията на гимназиалното образование, което е световна тенденция.
\item Една и съща експериментална постановка се дава на всички участници, но задачите за различните възрастови групи са различни, също както водата в плувния басейн е еднакво мокра за всички възрастови групи в едно състезание по плуване.
\end{itemize}

Накратко ще споменем задачите от предишните 5 ОЕФ:
1) Постановката на ОЕФ1 бе всъщност ученическа версия на американския патент на самонулиране на постоянно токови усилватели.~\cite{EPO1}
2) Задачата на втората ОЕФ2~\cite{EPO2} бе да се измери константата на Планк като се използва 
дифракция от компакт диск на светлина от светодиоди.
3) Една съвременна реализация на принадлежащия на НАСА патент 
за използване на отрицателното съпротивление за генериране на електрични трептения 
бе постановката на ОЕФ3~\cite{EPO3}.
4) Четвъртата ОЕФ~\cite{EPO4} бе посветена на фундаменталната физика
-- да се определи скоростта на светлината чрез измерване на електрични и магнитни сили.
Иновативния елемент бе приложението на теорията на катастрофите
към анализа на устойчивостта на махала.
5) Темата на ОЕФ5 бе да се измери константата на Болцман $\kb$ следвайки идеята на Айнщайн за изучаване на топлинните флуктуации на електрическото напрежение на кондензатор.

Настоящата ОЕФ6 следва традиционната тема на Олимпиадата да се измерва някоя фундаментална константа.
Вдъхновен от лекция за флуктуации на Айнщайн, Валтер Шотки разбрал, че зарядът не електрона също може да се определи от изучаването на шумове на напрежението.
Сега, век по-късно, поради появата на ниско-шумни операционни усилватели, идеята на Шотки може да бъде реализирана като гимназиална задача.

Накратко, установените традиции са баланс между съвременно работещи технически изобретения и фундаментална физика (\textit{Allah akbar}).

\section{Експериментална постановка}
Вашата експериментална постановка трябва да съдържа:
\begin{enumerate}
\item Печатна електронна платка (PCB) и двойна рейка, на която е запоен фотодиод и държач за литиева батерия.
\item Четири батерии по 9~V и с два двойни конектора.
\item Три батерии по 1.5~V и държач за тях с потенциометър запоен за него.
\item Найлонов плик съдържащ следното:
\begin{itemize}
\item Два операционни усилвателя, запоени върху адаптери, с зелен етикет и номер върху тях.
\item Умножител на напрежение , запоен върху адаптер, с оранжев етикет.
\item Литиева батерия.
\end{itemize}
\item Оптично влакно с бял цилиндър в единия край.
\item Пластмасова чаша с залепена пластмасова сламка.
\end{enumerate}
В допълнение, трябва да имате два мултиметъра, с нужните кабели за свързване, както и калкулатор.
\section{Прости начални задачи. S}
\begin{enumerate}
\item 
Включете мултиметъра да измерва напрежение (като волтметър)
\item 
Измерете напрежението на четирите 
9~V 
батерии с максимална точност (използвайте обхвата от 20~V) и запишете напреженията в таблица.
Направете същото с трите 1.5~V батерии, като сега използвате обхвата от 
2000~mV на волтметъра.
\item Поставете трите 1.5~V  батерии в държача за батерии.
Свържете потенциометъра  \emph{потенциометрично} към един от мултиметрите.
Въртете оста на потенциометъра.
Измерете и запишете интервала на напреженията които получавате.
Това ще бъде източника на напрежение за следващите задачи.
Изменението на полярността на свързването ще измени знака на напрежението.
\item Свържете четирите  9~V батерии към клипсовете като закопчаете електродите.
\section{Аналогово повдигане на квадрат. M}
\label{s:M_b}

\item \textbf{Внимание! От този момент нататък има възможност да изгорите интегралните схеми,
ако включите тях или батериите неправилно.} 
Ориентирайте постановката така, че жиците с етикети 
``OUT'' и ``COM'' на края да бъдат отдясно, вижте Фиг.~\ref{Fig:Board_b}.
\item 
Внимателно свържете рейката на 9~V батерия към десните стърчащи 3 електрода на постановката.
Етикет към етикет (двата етикета трябва да са един към друг.)
\textbf{
Работете внимателно -- ако сбъркате полярността ще изгорите операционния усилвател.
}
\item 
Свържете средния контакт на потенциометъра
с етикета с входната жичка с етикет ``IN''. 
Използвайте жиците с ``крокодили''.
\item Свържете някой от крайните електроди на потенциометъра 
към жицата  излизаща от другия електрод на потенциометъра 
към ``земята'' на постановката с етикет ``COM''. 
\item 
Свържете първия волтметър V$_1$ който ще показва напрежение $U_1$
към потенциометъра и ``IN''--``COM'' входовете на постановката.
Това е успоредно свързване.
\item Когато въртите оста на потенциометъра, напрежението  $U_1$
трябва да се изменя приблизително между 0 
и сумарното напрежение на батериите от
4.5~V.
\item 
Свържете втория волтметър V$_2$, , който ще показва напрежение $U_2$,
между изходящата жица на постановката отбелязана с ``OUT'' и общата точка отбелязана с  ``COM''.
Така  $U_2$ е напрежението между точките  ``OUT'' и ``COM'' .
\item 
Проверете дали ``COM'' електродите на двата мултиметъра са правилно свързани.
\item 
Четирите сдвоени рейки с оранжев етикет под тях в дясната част на платката са означени 
``AD633'' на Фиг.~\ref{Fig:Board_b}.
Свържете умножителя към ``AD633'', така че \textbf{оранжевият етикет на умножителя да е срещу оранжевия етикет на платката}.
\item 
Завъртате малко оста на потенциометъра, чакате 1 минута и записвате в таблица напреженията
$U_1$ и $U_2$.
Обърнете полярността на източника на напрежение и повторете измерванията на
входното напрежение $U_1$ и изходното напрежение $U_2$.
Подредете резултатите в таблица с колони:
номер на измерването $i$, $U_1$ и $U_2$. 
\item Представете резултатите графично като  $U_1$ 
е по абсцисата (хоризонталната координата),
а $U_2$ е по ордината (вертикалната координата) на милиметровата хартия.
\item 
Към таблицата добавете колона  $(U_1)^2$ като пресметнете квадрата на напрежението.
Представете и този резултат графично на милиметрова хартия:
$(U_1)^2$ -- абсциса (хоризонтална ос)
а $U_2$ -- ордината (вертикална ос).
Начертайте права линия която най-добре преминава в близост до експерименталните точки.
Това линейно приближение (фитиране) се описва с уравнението
$U_2= (U_1)^2/U_0 + \mathrm{const}.$
Изберете две точки от правата линия и измерете разликата по абсцисата 
$\Delta(U_1^2)$, и по ординатата $\Delta(U_2)$.
Чрез тяхното отношение определете наклона 
$U_0=\Delta(U_1^2)/\Delta(U_2)$.
Този параметър $U_0$  с размерност на напрежения е съществен за определянето на заряда на електрона  $q_e$, както е описано в следващата глава.

\section{Флуктоскопия на напрежението. Определяне на заряда на електрона $q_e$} 

\begin{figure}[h]
\centering
\includegraphics[scale=0.3]{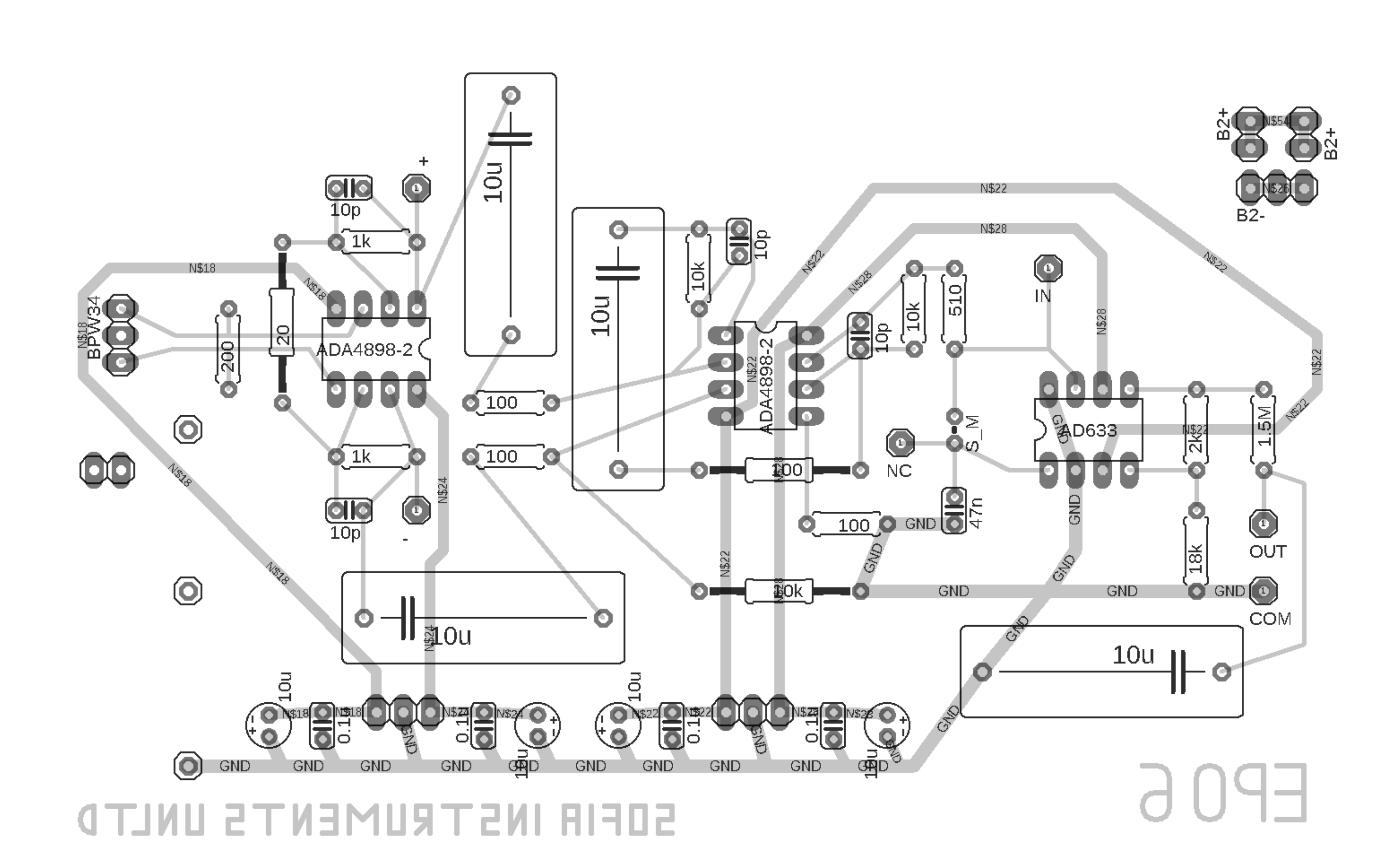}
\caption{Диаграма на печатната електронна платка реализираща експерименталната постановка. 
Тук е показана крушка с нажежаема жичка с бял цилиндър закачен за нея, която е запоена за пиновете ``B2-'' и ``B2+'' (не са отбелязани на платката), намиращи се в дясната страна на платката. За двата безименни 3-пина намиращи се в дъното на платката се закача захранването.
}
\label{Fig:Board_b}
\end{figure}
\item Поставете трите батерии по 1.5 V в съответния държач. Свържете, чрез два кабела тип ``крокодил'' двата външни електрода на потенциометъра към  на лампата ``B2-''  и единия от пиновете ``B2+'' в десния горен краи на постановката показана на Фиг.~\ref{Fig:Board_b}.
Тук поляритетът е без значение и при правилно свързване лампата трябва да свети.
\item Свържете чрез кабел тип ``крокодил'' ``COM'' входа на волтметъра V$_1$ с кабела ``($-$)'' от постановката.
\item Аналогично, свържете с друг ``крокодил'' кабела ``($+$)'' от постановката с входа ``$\mathrm{V\,\Omega\, m\!A}$'' на волтметъра V$_1$.
По този начин волтметърът  V$_1$ показва потенциалната разлика $U_{\pm}$  между точките ``($+$)'' и ``($-$)'', като това напрежение е пропорционално на средния фото-ток на фото диода.
\item Поставете литиевата батерия от 3~V в съответния държач, намиращ се на двойната рейка, на която е запоен фото-диода, \textbf{``($+$)'' на батерията да е към ``($+$)'' на държача.}
\item
Пиновете за операционните усилватели, с име ``ADA4898-2'', намиращи се в центъра и в дясната страна на платката и имащи зелени етикети до тях, са показани на Фиг. ~\ref{Fig:Board_b}.
Свържете операционните усилватели към тях, така че \textbf{ зеления етикет на всеки усилвател да е към зеления етикет на платката. Бъдете внимателни, неправилно свързване може да повреди усилвателите.}
\item Свържете рейката, на която е запоен фотодиода, чрез пиновете ``BPW34'' на Фиг.~\ref{Fig:Board_b}. Позицията на рейката е очертана с маркер върху платката.
Внимателно сложете оптичното влакно в черната сламка на чашата. Белият цилиндър трябва да влиза в сламката от по-дългия край, стърчащ от чашата. Проверете, че влакното може да се движи в сламката без голямо усилие.
\item Поставете свободния край на оптичното влакно в отвора на белия цилиндър, намиращ се върху крушката с нажежаема жичка, запоена на платката. 
\item Поставете чашата в близост до платката, така че фотодиода, намиращ се върху рейката, свързана към платката, да влиза в сламката от късия край. 
\item Свържете единия от конекторите за батериите от 9~V , към трите захранващи пина, намиращи се от лявата страна на платката. \\
\textbf{Бъдете внимателни!! Оранжевите етикети на конекторите трябва да са към оранжевите етикети на платката.}
Превключете волтметъра на обхват от 2000~mV и местете вълновода в сламката. Измереното напрежение е $U_\pm$ и трябва да може да достигне 1000~mV.
По този начин, чрез местене на оптичното влакно, можете да променяте фото-тока и да измервате падът на напрежението $U_\pm$ върху резистора $R$.
Ако не можете да достигнете максимална стойност от поне 700~mV или няма никакво напрежение изобщо, моля поискайте помощ от квесторите в залата.
Това е важна начална стъпка от експеримента!
\textbf{Не се опитвайте да премествате белия цилиндър, закачен за крушката.}
\item Поискайте лед и вода и напълнете чашата. Поставете оптичното влакно в сламката от дългия край, стърчащ от чашата.
\item Проверете чашата за течове. Ако има такива, внимателно отделете чашата от платката и поискайте нова.
\item Свържете ``COM'' изхода на платката с ``COM'' входа на волтметъра  V$_2$ (това е същият волтметър, които сте използвали в Глава ~\ref{s:M_b}).
\item Свържете ``OUT'' изхода на платката с ``$\mathrm{V\,\Omega\, m\!A}$'' входа на волтметъра     V$_2$. Така, волтметъра V$_2$  измерва напрежението $U_\mathrm{S}$, което е пропорционално на шума на фото-диода.
\item Свържете другият конектор за батериите 9~V към десния захранващ 3-пин на платката.  \\
\textbf{Бъдете внимателни!! Оранжевите етикети на конекторите трябва да са към оранжевите етикети на платката.}
\item Помолете квесторите за лепенка и залепете 6-те кабела от постановката към масата по следния начин:
\begin{itemize}
\item Първо разположете четирите батерии от 9~V под платката.
\item Залепете ``+'' и ``-'' кабелите над платката, като се убедите, че не контактуват един с друг.
\item Залепете ``IN'' и ``NC'' кабелите над платката и далеч от ``+'' и ``-'' кабелите.
\item Залепете ``OUT'' и ``COM'' кабелите в дясната страна на платката , като се убедите, че не контактуват един с друг.
\end{itemize}
Сега вече вашата експериментална постановка би трябвало да прилича на паяк. 
\item Погледнете измерваното напрежение V$_2$, то трябва да се стабилизира около няколко десетки mV.
\item Започнете измерването на $U_\pm$ и $U_\mathrm{S}$: менете $U_\pm$ чрез внимателно местейки оптичното влакно в сламката малко по-близко до фотодиода.
Промяната на $U_\pm$ между всеки последователни измервания трябва да бъде поне 100~mV.
Търпеливо изчакайте $U_\mathrm{S}$ да се стабилизира, чакайте поне 2 минути.
\emph{Ако някакво смущение промени значителни показанието на волтмеъра, отново изчакайте търпеливо стабилизация.}
Запишете резултатите в таблица с колони:
номер на измерване $i$, $U_\pm$ и $U_\mathrm{S}$. 
\item Представете резултатите графично, където  $U_\pm$ е по абциса (хоризонтална координата) и $U_\mathrm{S}$ е по ордината (вертикална координата) на милиметрова хартия (ако ви е необходима допълнителна милиметрова хартия, помолете квесторите да ви дадат).
Прекарайте права линия, която най-добре показва линейната зависимост
$U_\mathrm{S}=k U_\pm +\mathrm{const}$.
Изберете 2 точки на линията и определете наклона
$k=\Delta U_\mathrm{S}/ \Delta U_\pm$,
където $\Delta$ означава разлика.
Тази математична процедура се нарича линейна регресия.
За нашата постановка $k\simeq 10^{-3}.$
\item Накрая определете заряда на електрона $q_e$
използвайки формулата
\begin{equation}
q_e=2 k \frac{y_1}{Y^2} \frac{R_\mathrm{_L}}{R} C_\mathrm{_L}U_0,
\label{electron_charge_b}
\end{equation}
където $Y=1.01\times 10^{6}$ е пълното усилване на нашия усилвател и $y_1=101$ е усилването на първото стъпало. Можете да намерите стойността на $C_\mathrm{_L}$ върху етикета в десния долен край на платката ($C \equiv C_\mathrm{_L}$), $R_\mathrm{_L}=510~\Omega$ и $R=200~\Omega$. \\
Поздравления! Току що измерихте фундаментална константа и сте сред добра компания! :)
\item Вашето измерване може да е малко по прецизно ако отчетете не-идеалното поведения на операционните усилватели, чрез формулата
\begin{equation}
q_e=2 k \frac{y_1}{(1-\varepsilon)Y^2} \frac{R_\mathrm{_L}}{R} 
C_\mathrm{_L}U_0.
\label{electron_charge_mod_b}
\end{equation}
Малката поправка $\varepsilon(C_\mathrm{_L})$ като функция на капацитета $C_\mathrm{_L}$ е графично представена на Фиг.~\ref{Fig:Eps_b}.
\begin{figure}[h]
\centering
\includegraphics[scale=0.8]{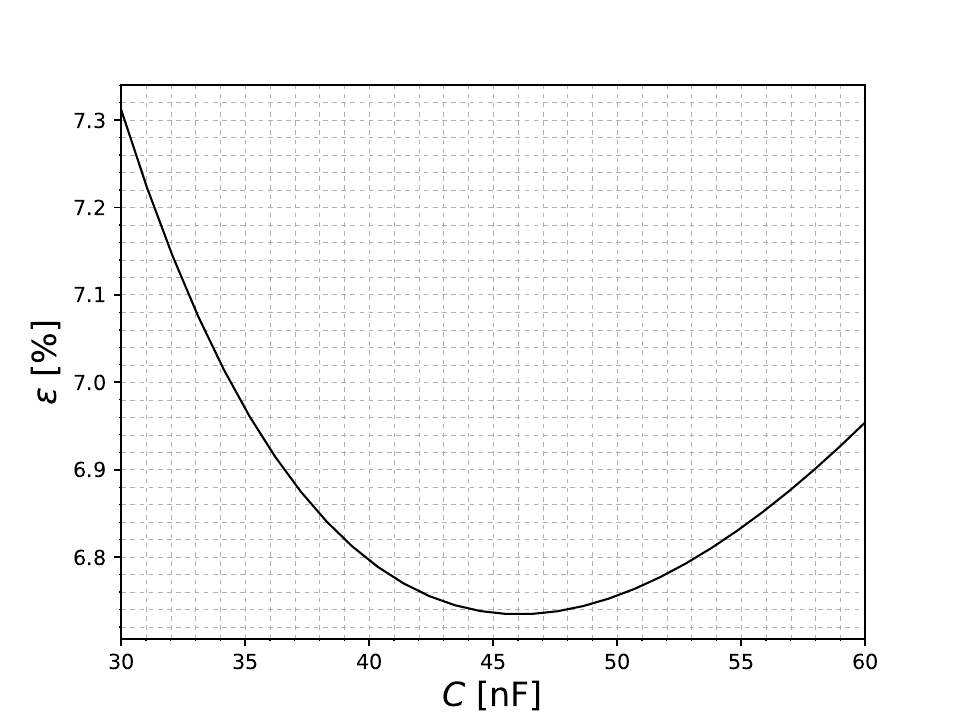}
\caption{Грешката $\varepsilon$ в проценти използвана за определянето на $q_e$ като функция на капацитета $C_\mathrm{_L} \equiv C$.}
\label{Fig:Eps_b}
\end{figure}
Намерете грешката $\varepsilon$ от Фиг. ~\ref{Fig:Eps_b} според стойността $C_\mathrm{_L}$ написана на вашата платка. Не забравяйте да разделите  $\varepsilon$ на 100 преди да изчислите $q_e$ от уравнение (\ref{electron_charge_mod_b}).
Преди един век този метод за измерване на заряда на електрона $q_e$, чрез измерване на шума породен от дискретният заряд на токовите носители е предложен от Уолтър Шотки \cite{Schottky:1918}. По това време той е работил с Макс Планк и е бил вдъхновен от лекция на Алберт Айнщайн на тема електрически флуктоации.

\section{Задача за домашно. XL}
Изведете формули Eq.~(\ref{electron_charge_b}) и Eq.~(\ref{electron_charge_mod_b}) анализирайки веригата и табулираните стойности на параметрите от Таблица~\ref{tbl:values_b}. 
Сметнете усилванията $y_1$ на буфера и пълното усилване на всички стъпала на усилвателя $Y$. 
Различните детайли от веригата са показани на Фигури~\ref{Non-inverting amplifier_b},\ref{Buffer_b}, 
\ref{Differential amplifier_b},
\ref{Inverting amplifier_b},
\ref{Analog multiplier_b}.

\begin{center}
\begin{table}[h]
\begin{tabular}{| c | r |}
		\hline
		&  \\ [-1em]
		Circuit element  & Value  \\ \tableline
			&  \\ [-1em]
			$R$ &200~$\Omega$ \\
			$r_\mathrm{_G}$ & 20~$\Omega$ \\
			$R_\mathrm{F}$ &  1~k$\Omega$  \\
			$C_\mathrm{F}$ &  10~pF  \\ 
			$C_\mathrm{G}$ & 10~$\mu$F \\
			$R_\mathrm{G}$ &  100~$\Omega$  \\ 
			$R_\mathrm{F}^\prime$ & 10~k$\Omega$ \\
			$C_\mathrm{F}^\prime$ & 10~pF \\
			$R_\mathrm{_L}$ & 510~$\Omega$ \\
			$R_1$ &  2~k$\Omega$  \\ 
			$R_2$ & 18~k$\Omega$  \\
			$R_\mathrm{av}$ & 1.5~M$\Omega$ \\
			$C_\mathrm{av}$ & 10~$\mu$F \\
			$R_\mathrm{_V}$ & $\approx 1~\mathrm{M} \Omega$ \\
			$V_\mathrm{CC}$ & +9~V \\
			$V_\mathrm{EE}$ & -9~V \\
\tableline
\end{tabular}
	\caption{Таблица с числените стойности на елементите от веригата.}
	\label{tbl:values_b}
\end{table}
\end{center}

\begin{figure}[t]
\begin{minipage}[t]{0.31\linewidth}
\includegraphics[scale=0.28]{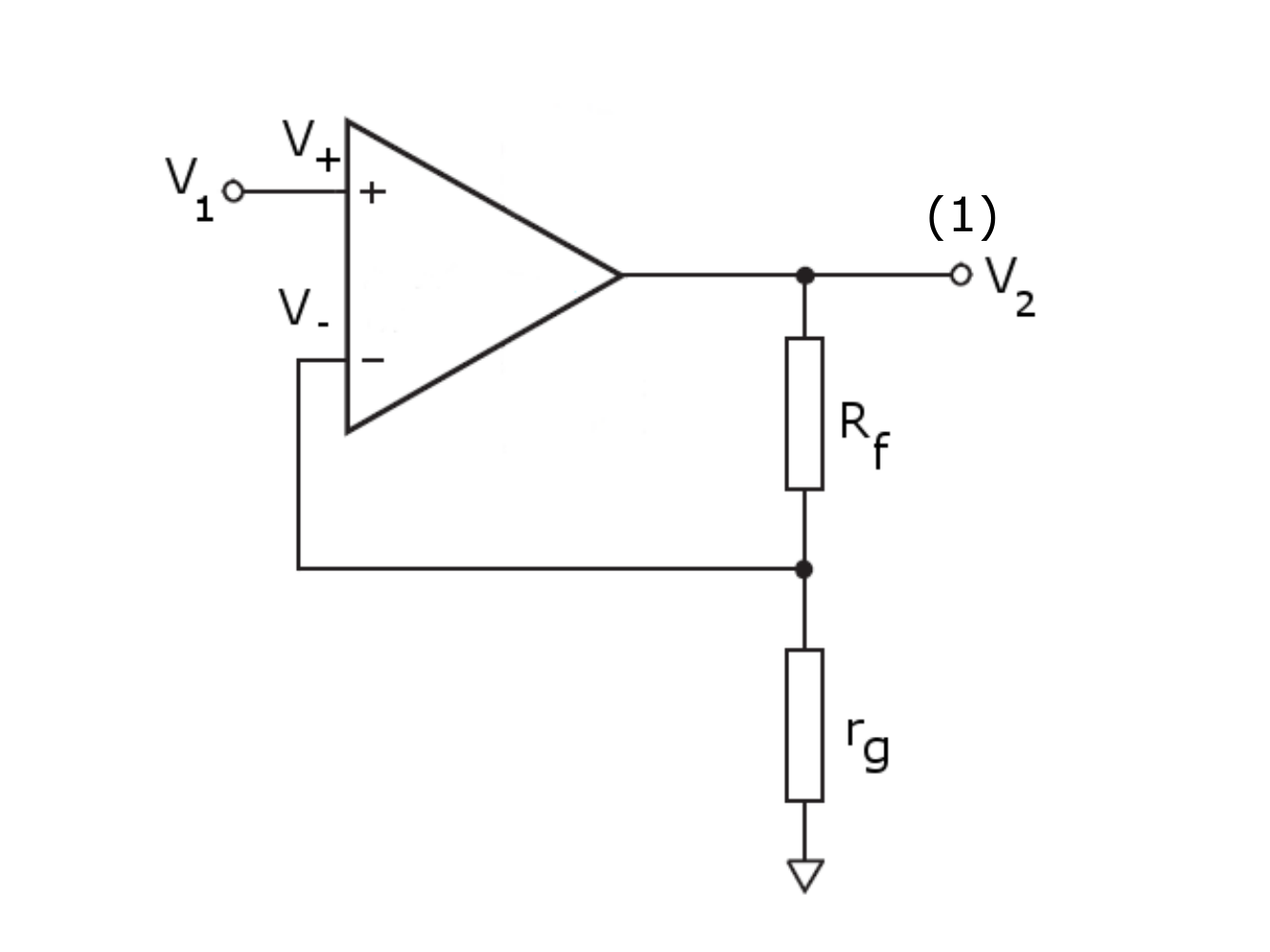}
\caption{Non-inverting amplifier}
\label{Non-inverting amplifier_b}
\end{minipage}
\begin{minipage}[t]{0.31\linewidth}
\includegraphics[scale=0.28]{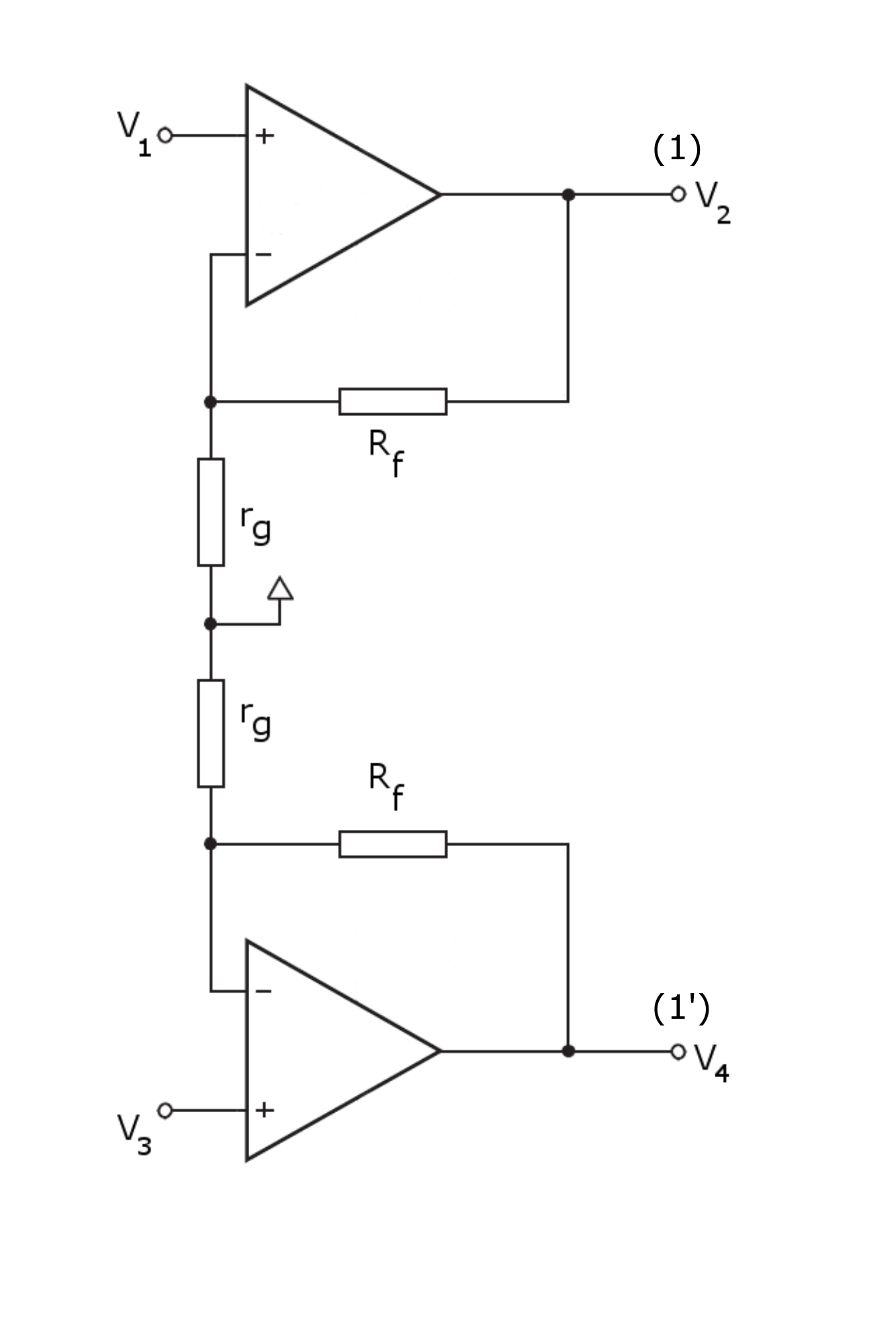}
\caption{Buffer}
\label{Buffer_b}
\end{minipage}
\begin{minipage}[t]{0.36\linewidth}
\includegraphics[scale=0.28]{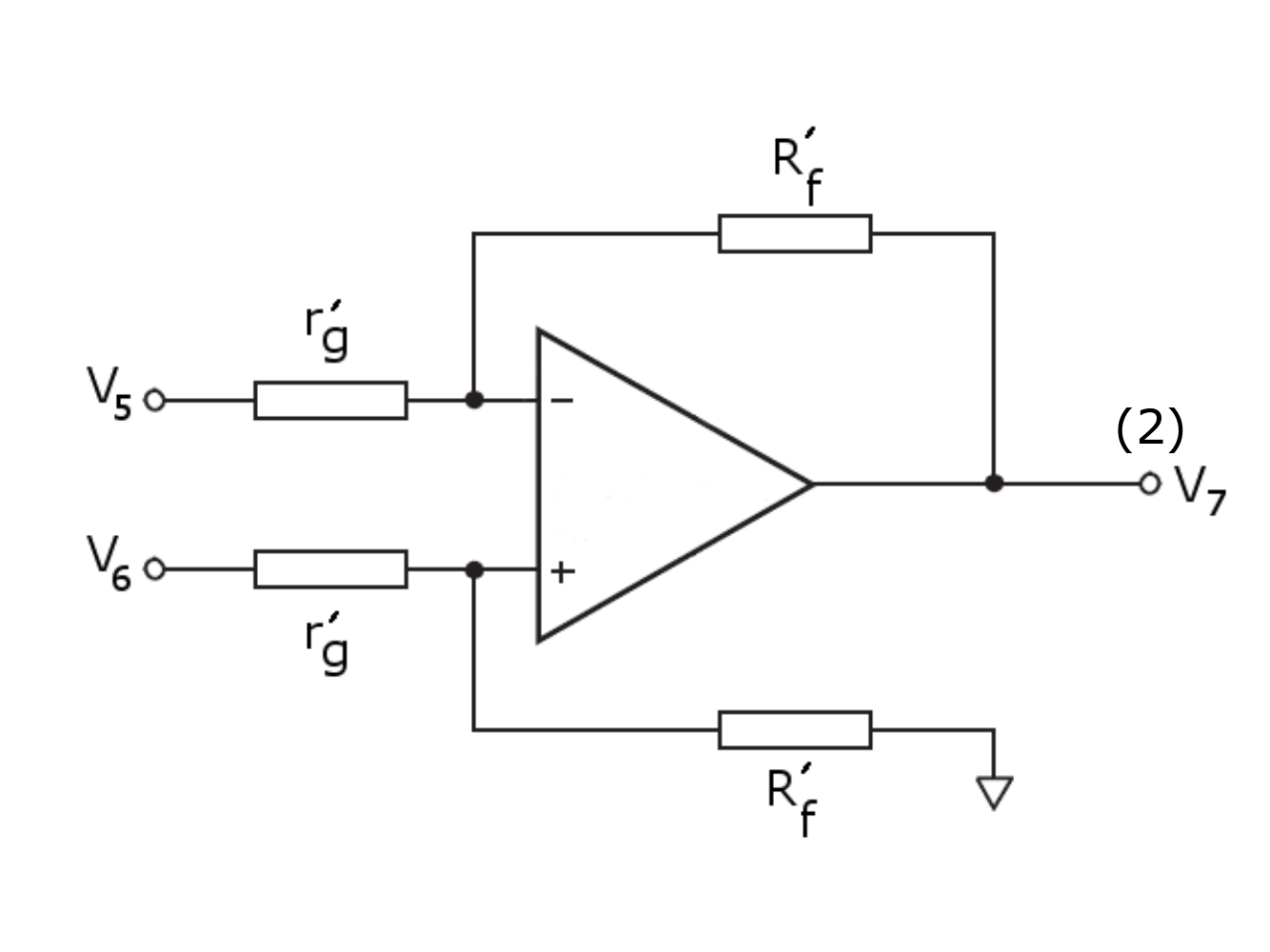}
\caption{Differential amplifier}
\label{Differential amplifier_b}
\end{minipage}
\begin{minipage}[c]{0.4\linewidth}
\includegraphics[scale=0.28]{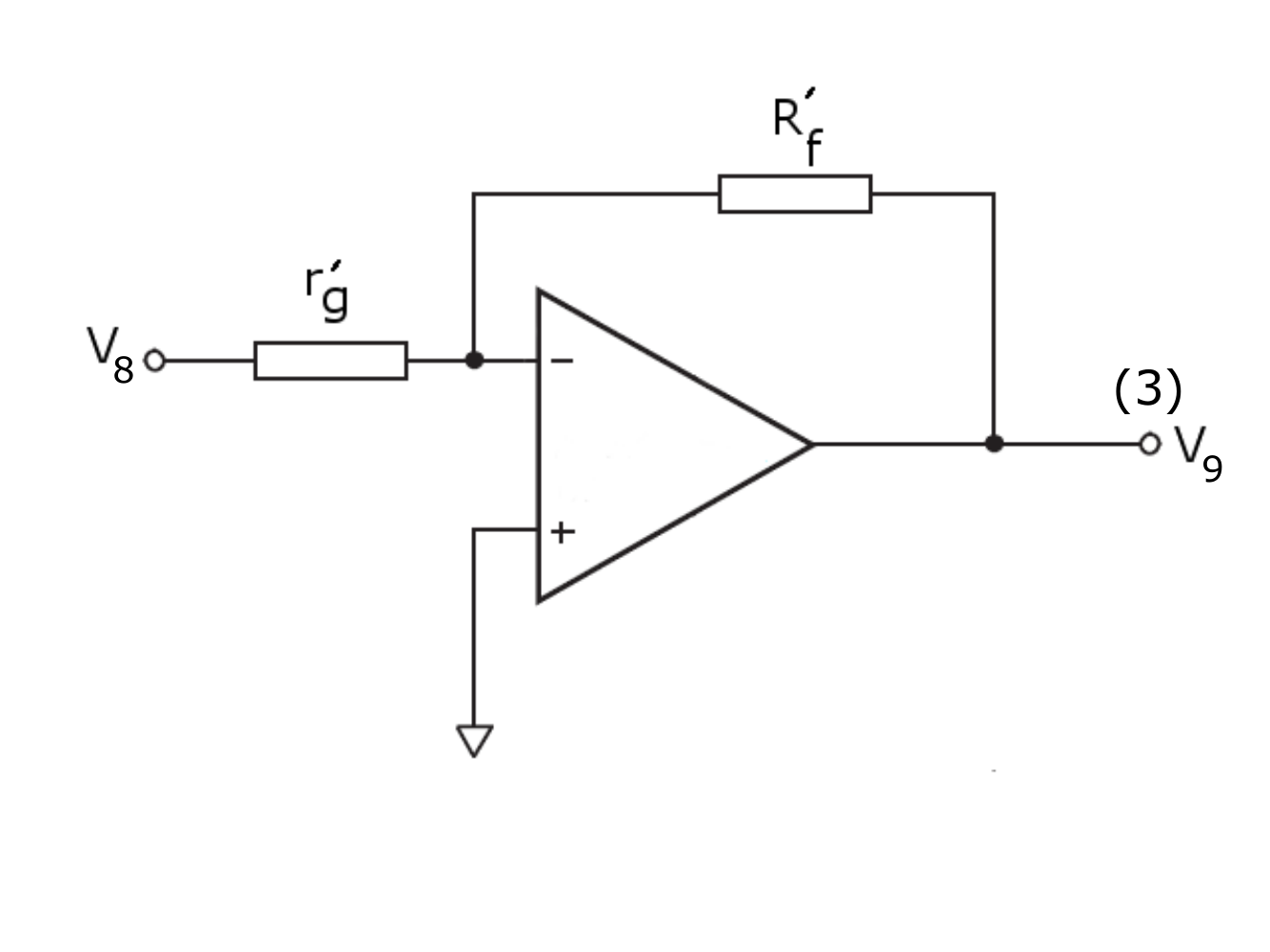}
\caption{Inverting amplifier}
\label{Inverting amplifier_b}
\end{minipage}
\begin{minipage}[c]{0.57\linewidth}
\includegraphics[scale=0.28]{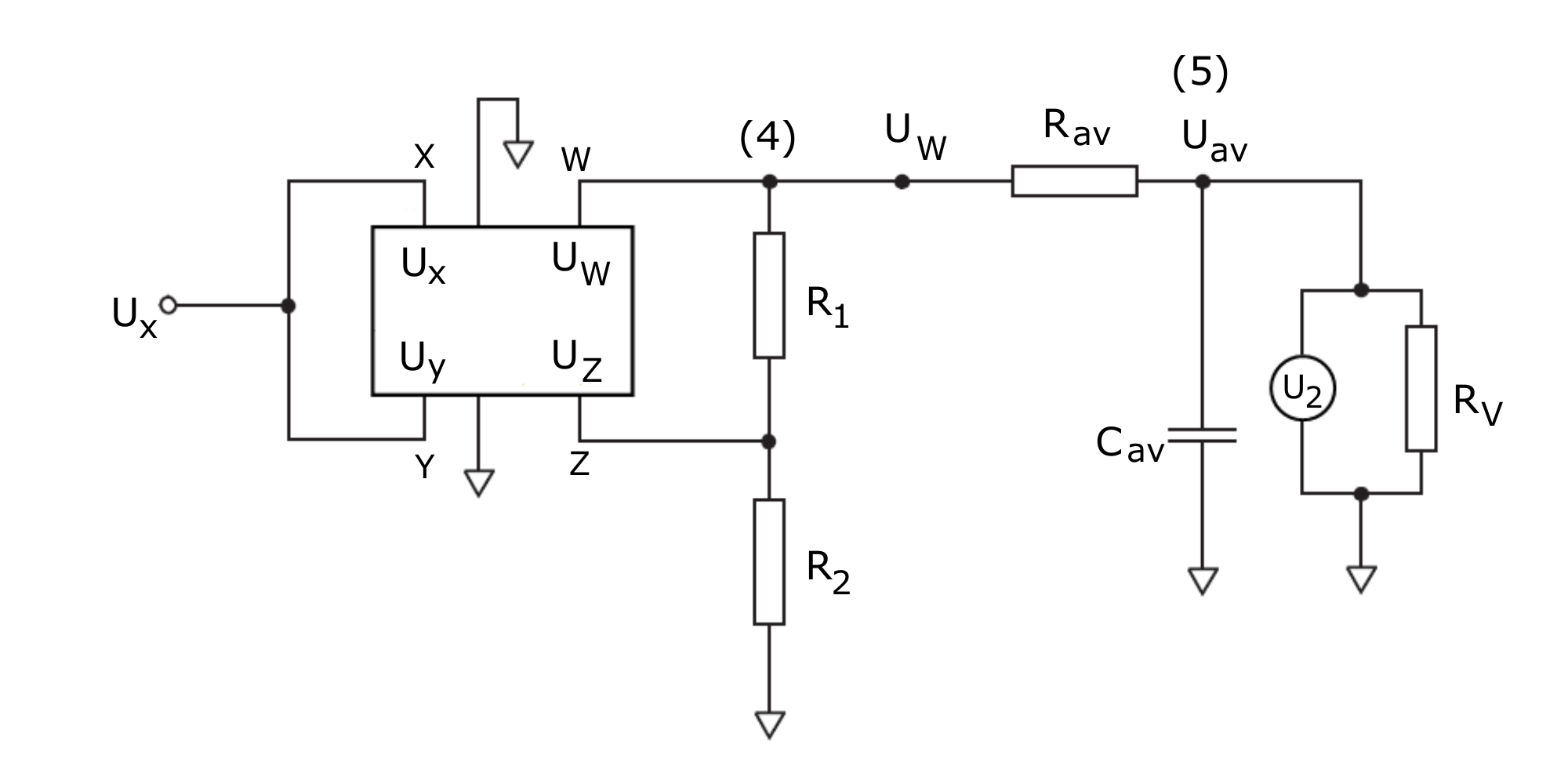}
\caption{Analog multiplier}
\label{Analog multiplier_b}
\end{minipage}
\end{figure}

Решението трябва да бъде изпратено през нощта след Олимпиадата до изгрев Слънце на 
e-mail адреса на Олимпиадата epo@bgphysics.eu.
Най-доброто решение ще спечели наградата на Зомерфелд с паричен еквивалент от DM137. 

\section{Задачи за по-нататъшна работа}

Експерименталната постановка е всъщност lock-in волтметър и с нея може да измервате променлив сигнал по-малък от  $\mathrm{\mu V}$.
Пишете ни как да използвате ``NC'' изхода в този случай.

\end{enumerate}

%% file: EPO6_mk_a.tex
\clearpage
\renewcommand{\figurename}{Слика}
\renewcommand{\tablename}{Табела}
\begin{center}
\textbf{Мерење на електричниот полнеж на електронот $q_e$ користејќи шум на Шотки. 
Задача за 6-та олимпијада по експериментална физика. Софија, 8 Декември 2018}
\end{center}
\normalsize
\indent
\setlength{\leftskip}{1.2cm}
\setlength{\rightskip}{1.2cm}

За оваа олимпијада се поставени неколку последователни експерименти со специјално дизајнирана PCB плоча. 
Нивно последователно изведување дава можност да се измери полнежoт на електронот, 
$q_e.$
Ова се постигнува преку мерење на флуктуациите на $U(t)$  при различно осветлување на фотодиодата.
Мерениот напон е засилен еден милион пати $Y=10^6$. 
Овој засилен напон $YU(t)$ се предава на инструментот, кој како резултат дава, временски усреднета, средно квадратна вредност на напонот,
$U_\mathrm{S}=\left<(Y U(t))^2\right>/U_0$.
Напонот $U_\mathrm{S}$ се мери со миливолтметар.
Доколку се направи серија мерења може да се определи 
$q_e$  користејќи ја добро познатата формула на Шотки за спектрална густина на шум на струја,
$(I^2)_f=2q_e\left<I\right>.$
За учесниците од основните училишта целта е да се анализира аналогното квадрирање на сигналот.
Работата на учесниците е поделена и се оценува по категории S, M, L, XL според годините на учесникот.
За учесници од XL категоријата, задачата содржи проблеми кои се ориентирани кон универзитетско образование по физика и вклучува теориско истражување на PCB плочата  како апликативен уред.
Со ова е претставувена задачата на EPO 6 - “ден на полнежот на електронот“.
ЕПО6 е организирано од Софискиот огранок на унијата на физичари на Бугарија со помош од факултетот по физика од Софискиот универзитет и Друштвото на физичарите на Република Македонија.

\setlength{\leftskip}{0cm}
\setlength{\rightskip}{0cm}
\section{Вовед}

Од самиот почеток,  Олимпиjaдата по Експериментална Физика (ЕPO) e позната широко низ светот.
Сите проблеми на минатите олимпиjади се обjaвени на интернет \cite{EPO1,EPO2,EPO3,EPO4,EPO5} и од самиот почеток учествуваат по 120 учесници на секоjа олимпиjада.
Минатите години учествуваа ученици од средни училишта од 7 држави и разликата помеѓу наjоддалечените градови е повеќе од 4~Mm.

Секоj учесник на EPO од организаторите ja добива како подарок поставката со коjа работи, што ја прави ЕPO различна од другите натпревари.
\begin{itemize}
\item
По завршувањето на олимпиjaдата, дури и учесниците кои имале послаби изработки се во можност да го повторат експериментот и да го достигнат нивото на шампион.
На овоj начин, олимпиjадата директно влиjае на зголемување на квалитетот на наставата по физика во целиот свет.
По завршувањето на школската година, поставката останува во училиштето, каде ученикот учи.
\item
Секоj од проблемите е оригинален и е поврзан со фундаменталната физика или анализирање на технички патент.
\item
Олимпискиот дух е реализиран во EPO во  неговата основна форма и секоj ученик од било кој дел на светот може да учествува на Олимпијадата ако има желба за тоа.
\item
Од друга страна, сличноста со други олимпиjади е што проблемите се директно илустрирани во наставниот материjал и заедно со другите слични натпревари се спротивставува на деградациjата на средношколското образование, што е тенденциjа во целиот свет.
Нема ограничување на броjот на учесници.
\end{itemize}

Накратко ќе ги споменеме проблемите на минатите 5 EPO:
1) Поставката на EPO1 беше всушност ученичка верзија на Американски патент за авто-нула стабилизирани уреди.~\cite{EPO1}
2) Проблемот на втората олимпијада EPO2~\cite{EPO2} беше да се измери Планковата константа преку дифракција на ЛЕД светлина со компакт диск.
3) Современа реализација на патент на НАСА за употреба на NIC (negative impendans converter) за создавање на осцилации на напонот беше задачата на EPO3~\cite{EPO3}
4) EPO4~\cite{EPO4} беше посветен на фундаментална физика - да ја определи брзината на светлината со мерење на електрични и магнетни сили.
Нов елемент беше примена на теорија на катастрофи при анализа на стабилноста на нишало.
5) Темата на EPO5 беше да се измери Болцмановата константа $\kb$ следејќи ја идејата на Ајнштајн за следење на топлинските флуктуации на напонот на еден кондензатор.

EPO6 ја следи традицијата за тема на олимпијадата да биде мерење на некоја фундаментална константа. Инспириран од лекција за флуктуации дадена од Ајнштајн, Волтер Шотки заклучил дека електричниот полнеж на електронот може исто така да биде измерен преку мерење на струен шум во едно електрично коло. 

Денес, после изминат еден век, со појавата на операциони засилувачи со мал шум, идејата на Шотки може да биде применета на задача од средно училиште.
 
На кратко, утврдените традиции на ЕPO се рамнотежа меѓу современите технички пронајдоци и фундаменталната физика. 

\section{Експериментална поставка}
Вашата поставка треба да ги содржи следните елементи:
\begin{enumerate}
\item Печатена електронска плочка PCB со двоен приклучок за фотодиода и држач за литиумска батерија.
\item Четири 9V батерии, два двојни спојници за нивно поврзување.
\item Три 1.5V батерии и држач во кој тие се поставуваат поврзан со потенциометар.
Пластична кеса во која ги има следните елементи:
\begin{itemize}
\item Два операциони засилувачи означени со зелена лепенка и соодветно обележени со бројка
\item Множител означен со портокалова лепенка
\item Кружна литиумска батерија за часовник
\end{itemize}
\item Оптичко влакно со бел цилиндер на едниот крај.
\item Пластична чаша со црна цевка залепена низ нејзе.
Со вас треба да имате два сопствени мултиметри со нивните сонди и калкулатор.

\section{Почетни лесни задачи. Категорија S}

\item Приклучи ги мултиметрите како волтметри.
\item Измери го напонот на 4-те батерии од 9~V со максимaлна точност, користејќи го 20~V подрачје на волтметарот и запишете ги вредностите.
Користејќи го 2000~mV подрачје на волтметарот, измери го напонот на трите 1.5~V батерии.
\item Стави ги трите батери од 1.5~V во држачот за батерии.
Поврзи го потенциометарот, потенциометриски, на еден од вашите мултиметри.
Вртете ја оската на потенциометарот и измерете го интервалот на напоните што ги добивате. Ова ќе биде извор на напон за следните задачи. Промена на поларитетот го менува знакот на напонот.
\item Поврзи ги четирите 9~V батерии на двете дупли спојници со притискање на капачињата.
\end{enumerate}

\section{Аналогно квадрирање на напон. Категорија M}
\begin{enumerate}
\item \textbf{Внимание! Од овој момент па натаму, можно е да се изгорат интегралните кола ако ги поврзете батериите погрешно.}
Постави го PCB-сетот со кој работиш така да, двете жици ``OUT'' и ``COM'' кои се на работ од плочката да бидат на вашата десна страна.
\item Внимателно поврзи ја едната 9~V спојница во десната машка 3 пин спојница во долната зона на PCB-плочата, портокалова со портокалова налепница по налепница свртени една спрема друга.
\textbf{Работи внимателно-ако направиш грешка со поларитетот ќе ги изгориш множителот и  операциониот засилувач.}
\item Поврзи го средниот контакт на потенциометарот со инпут жицата на горниот дел на плочата означена со ``IN''.
Употреби ја жицата со ``крокодил'' спојници. 
\item Поврзи го крајниот контакт на потенциометарот со жицата од заземјувањето на плочката со знак ``COM''.
\item Поврзи го првиот волтметар V$_1$ кој го покажува напонот $U_1$ паралелно со потенциометарот и ``IN''-``COM'' влезовите на плочата.
\item Кога ќе ja заротираш оската на потенциометарот, напонот $U_1$ треба да се промене приближно меѓу 0 и вкупниот напон на батеријата 4.5~V.
\item Вториот волтметар V$_2$, кој го покажува напон $U_2$, поврзи го меѓу излезната жица на плочата ``OUT'' и точката ``COM''. На овој начин $U_2$ е напонот помеѓу ``OUT'' и ``COM''
\item Провери дали ``COM'' електродите на двата мултиметри и плочата се правилно поврзани.
\item Четирите дупли пинови со портокалова налепница во средина на десната страна од PCB-то се означени со ``AD633'' на слика~\ref{fig:Board_m}. Постави го множителот, со портокалова лепенка на него, на плочката така што \textbf{портокаловата налепница од множителот е свртена кон портокаловата налепница на плочата.}
\item Врти ја оската на потенциометарот, почекај 1 мин. и запиши ги напоните $U_1$ and $U_2$.
Смени го поларитетот на изворот и повтори ги мерењата на влезниот напон $U_1$ и излезниот напон $U_2$.
Подреди ги резултатите во табела со колони:
број на мерења $i$, $U_1$ и $U_2$. 
\item Претстави ги резултатите графички на милиметарска хартија каде $U_1$ е претставен на апсциса (хоризонталната координата) и $U_2$ е претставен на ординатата (вертикалната координата).
\item Во табелата додај колона во која ќе пресметаш $(U_1)^2$ и на милиметарска хартија претстави ги резултатите графички, $(U_1)^2$- апсциса и $U_2$-ордината.
Нацртај права линија што поминува најблиску до графички претставените измерени вредности од експериментот.
Оваа приближна права е опишана со равенката
$U_2=(U_1)^2/U_0 + \mathrm{const}.$
Избери две точки од правата, измери ги разликите од проекциите на овие точки на апсцисата  
$\Delta(U_1^2)$, и на ординатата $\Delta(U_2)$ 
и најди го наклонот
$U_0=\Delta(U_1^2)/\Delta(U_2)$.
Овој параметар $U_0$ со димензии на напон е основен параметар за пресметување на полнежот на елекронот $q_e$ како што е опишано во следното поглавје.

\section{Напонска флуктуациона спектроскопија.
 Определување на полнежот на електронот $q_e$. Категорија L} 

\begin{figure}[h]
\centering
\includegraphics[scale=0.3]{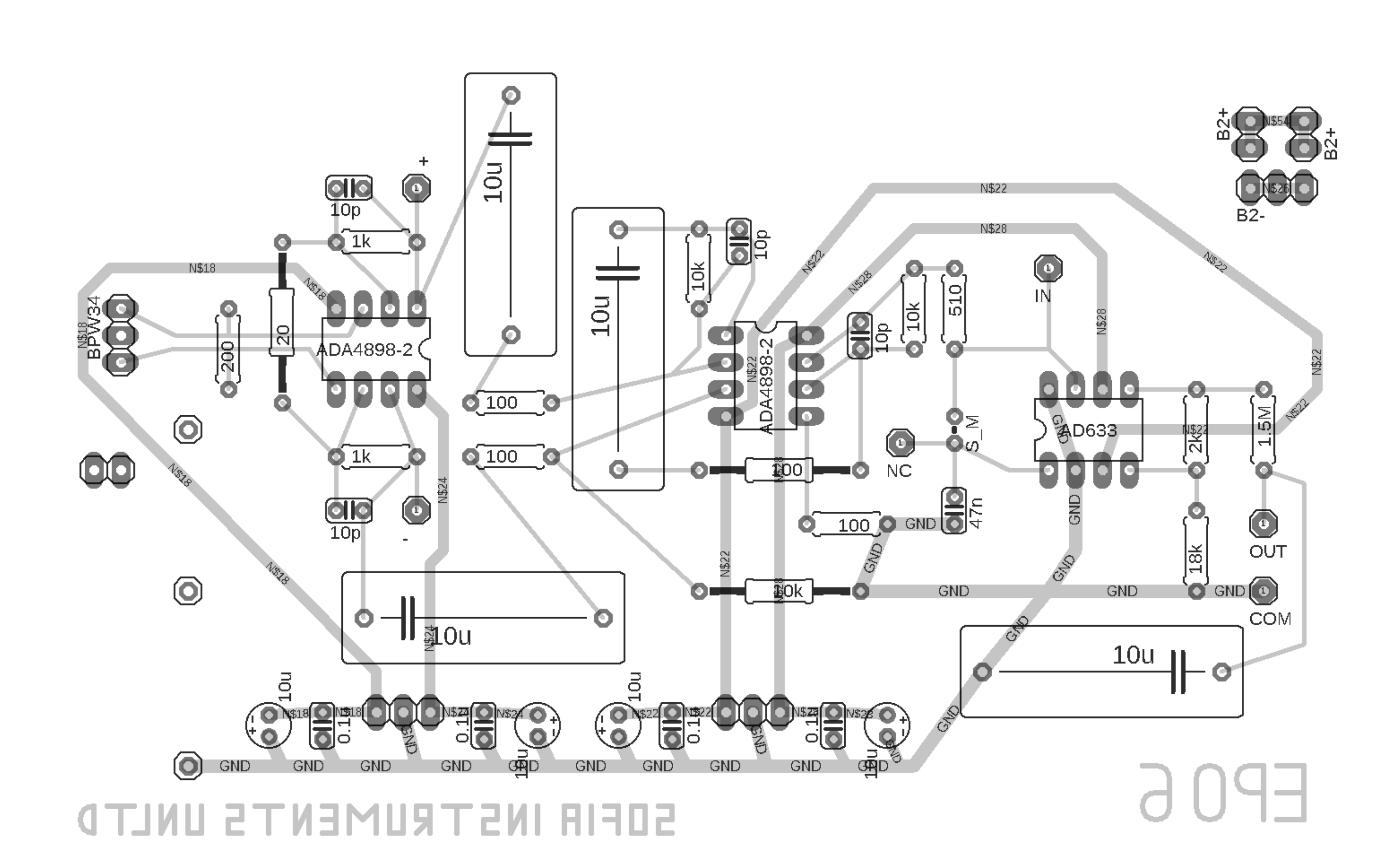}
\caption{Дијаграмот на печатената плочка од експерименталната поставка. Волфрамова сијаличка со бел цилиндар прикачена на неа која е залемена помеѓу  ``B2-'' и ``B2+'' пиновите кои не се означени на плочката. Двата ненумерирани тројни пин конектори се за напонско напојување на плочката.}
\label{fig:Board_m}
\end{figure}

\item Со крокодилки поврзете го држачот на батериите од 4.5~V со двете електроди на сијаличката $B2-$ и $B2+$ пиновите во горниот десен агол од плочката прикажан на сл. 1. Овде поларитетот не е важен. По точно поврзување, сијаличката треба да засвети.
\item Поврзи ја со ``крокодил кабел'', ``($-$)'' точката од плочката со ``COM'' точката од волтметарот V$_1$.
\item Аналогно на ова, поврзи го со друг  ``крокодил-кабел'' ``($+$)'' електродата од плочката со ``$\mathrm{V\,\Omega\, m\!A}$'' влезот на волтметарот V$_1$. На овој начин волтметарот V$_1$ ја покажува потенцијалната разлика $U_{\pm}$ меѓу ``($+$)''  и ``($-$)'' точките и овој напон е пропорционален на приближната вредност на фото-струјата од фото-диодата.
\item Постави ја 3~V саатната литиумска батерија во држачот кај фотодиодата така што \textbf{+ страната на батеријата да е поврзана со + страната на држачот за батеријата.} 
\item Двата дупли пин конектори на левата страна и центарот на плочката означени со зелени лепенки се означени со ``ADA4898-2'' на слика~\ref{fig:Board_m}. Постави ги засилувачите на овие позиции така што \textbf{зелената налепница на секој засилувач е свртена кон зелената налепница на плочата. Бидете внимателни, доколку поврзете погрешно засилувачот ќе прегори.}
\item Внимателно наместете ја чашката така што, фотодиодата влегува во црната сламка на чашката од кратката страна. Едниот крај на влакното со бел цилиндер вметнете го во сламката низ подолгиот крај. Другиот крај на влакното вметнете го во малото отворче на белиот цилиндер кој е прицврстен на сијаличката. Проверете дали влакното се движи лесно низ сламката. \textbf{Внимавајте да не го поместите белиот цилиндер на сијаличкаа бидејќи лесно може да ја скршите.}
\item Поврзете спојница со батерии од 9~V на левиот три-пински конектор за снабдување со напон на PCB-то. \textbf{Внимателно, портокаловите налепници од спојницата треба да одговара на портокаловата налепница на плочката. Обратно поврзување ќе ги изгори засилувачите.} Поставете го мултиметарот во опсег од 2000~mV и полека движете го оптичкото влакно низ сламката. Овој напон $U$ би требало да се менува до 1000~mV. На овој начин, движејќи го оптичкото влакно  се менува фото-струјата преку која го мерите фото-напонот $U_{\pm}$ создаден од фото-струјата која минува низ отпорникот $R$. Ако не можете да достигнете вредност на напонот од најмалку 700~mV или воопшто немате напон, побарајте помош од тестаторите во училницата. Ова е важен основен дел од експерименталното мерење.
\item За да продолжите понатаму побарајте вода и мраз од тестаторите кои треба да се стават во чашата со сламката и фото диодата во нејзе.
\item По ставањето на вода и мраз внимателно погледнете дали доаѓа до истекувања од чашата. Доколку има, внимателно одстранете ја чашката и побарајте промена.
\item Поврзете ја ``COM'' точката на плочката со ``COM'' точката од волтметарот V$_2$ (Волтметарот V$_2$ треба да биде истиот кој го користевте во поглавје IV).
\item Поврзете ја ``OUT'' точката од плочката со ``$\mathrm{V\,\Omega\, m\!A}$'' влезот на волтметарот V$_2$.
Сега, волтметарот V$_2$ го мери напонот $U_S$ кој е пропорционален на струјниот шум од фото-диодата.
\item Поврзи ја другата спојница со 9~V ни батерии на десниот трипински конектор на PCB-то. \textbf{Внимателно, портокаловите налепници од спојницата треба да одговара на портокаловата налепница на плочката. Обратно поврзување ќе ги изгори засилувачите.}
\item Од тестаторите побарајте салотејп и залепете ги на маса сите 6 кабли од плочката на следниот начин
- Прво сите 9~V батерии долу под PCB-то.
- Жичките од + и -  каблите над PCB-то внимавајќи да не се во контакт.
- Залепи ги ``IN'' и ``NC'' каблите над PCB-то подалеку од  + и - каблите.
- Залепи ги ``OUT'' и ``COM''  каблите на десната страна од PCB-то внимавајќи да не се во контакт.
Вака поставените кабли и плочка треба да наликуваат на пајак.
\item Погледнете го напонот на волтметарот $V_2$, би требало да се стабилизра околу триесетина миливолти.
\item Започнете го мерењето на $U_{\pm}$ и $U_S$: менувајте го $U_{\pm}$ со внимателно придвижување на оптичкото влакно во сламката малку по малку кон фотодиодата. Промената на $U_{\pm}$ помеѓу секое посебно мерење треба да биде најмалку 100~mV.  Почекајте барем 1 минута $U_S$ да се стабилизира. Резултатите внесете ги во табела со колони: $i$, $U_{\pm}$ и $U_S$.
\item Графички претставете ги резултатие каде $U_{\pm}$ е поставено на апсцисата и $U_S$ на ординатата на милиметарската хартија. (Ако ви треба уште милиметарска хартија побарајте од тестаторите.) Нацртајте права линија која најдобро соодветствува на линеарна зависност $U_S=kU_{\pm}+const$. Изберете две точки од правата и определете го наклонот на правата со $k=\Delta U_\mathrm{S}/ \Delta U_\pm$, каде $\Delta$ ја означува разликата. Оваа математичка процедура е наречена линеарна регресија. За нашата експериментална поставка $k\simeq 10^{-3}$.
\item Конечно, определете ја вредноста на полнежот на електронот $q_e$ преку следната равенка \begin{equation}
q_e=2 k \frac{y_1}{Y^2} \frac{R_\mathrm{_L}}{R} C_\mathrm{_L}U_0,
\label{electron_charge_m}
\end{equation}
каде $Y=1.01\times 10^{6}$ е вкупното засилување на нашиот засилувач а $y_1=101$ е засилувањето на првиот чекор на множителот. Вредноста на $C_L$ е запишана на налепница на долната десна страна на PCB-то. $C=C_L$, $R_L=510~\Omega$ и $R=200~\Omega$.
Честитки! Само што ја измеривте фундаменталната константа на полнежот на електронот! ;)
\item Вашето мерење може да биде малку по прецизно ако го вклучите неидеалниот ефект на операционите засилувачи преку формулата
\begin{equation}
q_e=2 k \frac{y_1}{(1-\varepsilon)Y^2} \frac{R_\mathrm{_L}}{R} 
C_\mathrm{_L}U_0.
\label{electron_charge_mod_m}
\end{equation}
Малата корекција  $\varepsilon(C_\mathrm{_L})$  како функција од капацитетот $C_L$ е графички претставена на слика~\ref{fig:Eps_m}. Најди ја грешката $\varepsilon$ од слика~\ref{fig:Eps_m} според вашата вредност на $C_L$. Не заборавајте да го поделите $\varepsilon$ со 100 пред да го пресметате $q_e$.
\begin{figure}[h]
\centering
\includegraphics[scale=0.8]{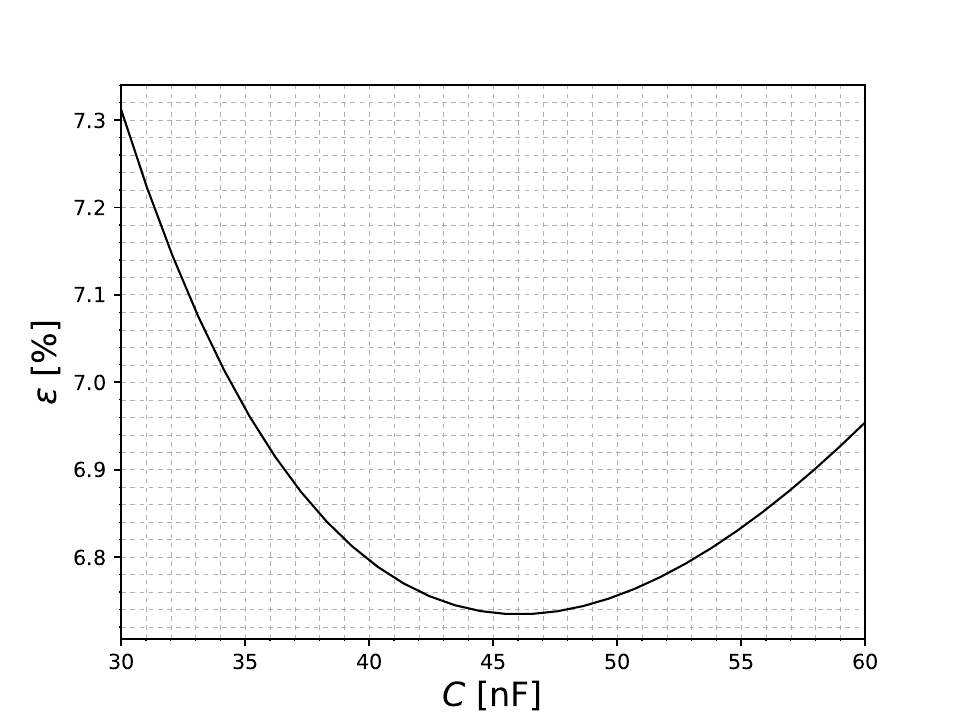}
\caption{Корекција  $\varepsilon(C_\mathrm{_L})$  како функција од капацитетот $C_L$.}
\label{fig:Eps_m}
\end{figure} 

\end{enumerate}

Пред еден век овој метод за определување на полнежот на електронот со мерење на шум бил предложен од Волтер Шотки. Во тој период Волтер Шотки работел со Макс Планк и бил инспириран од предавањата на Алберт Ајнштајн за електричните флуктуации.
 
Честитки! Ти само што ја пресмета фундаменталната константа и си во одлично друштво! :)

\section{Домашна задача. Категорија XL}
\begin{figure}[h]
\begin{minipage}[t]{0.31\linewidth}
\includegraphics[scale=0.28]{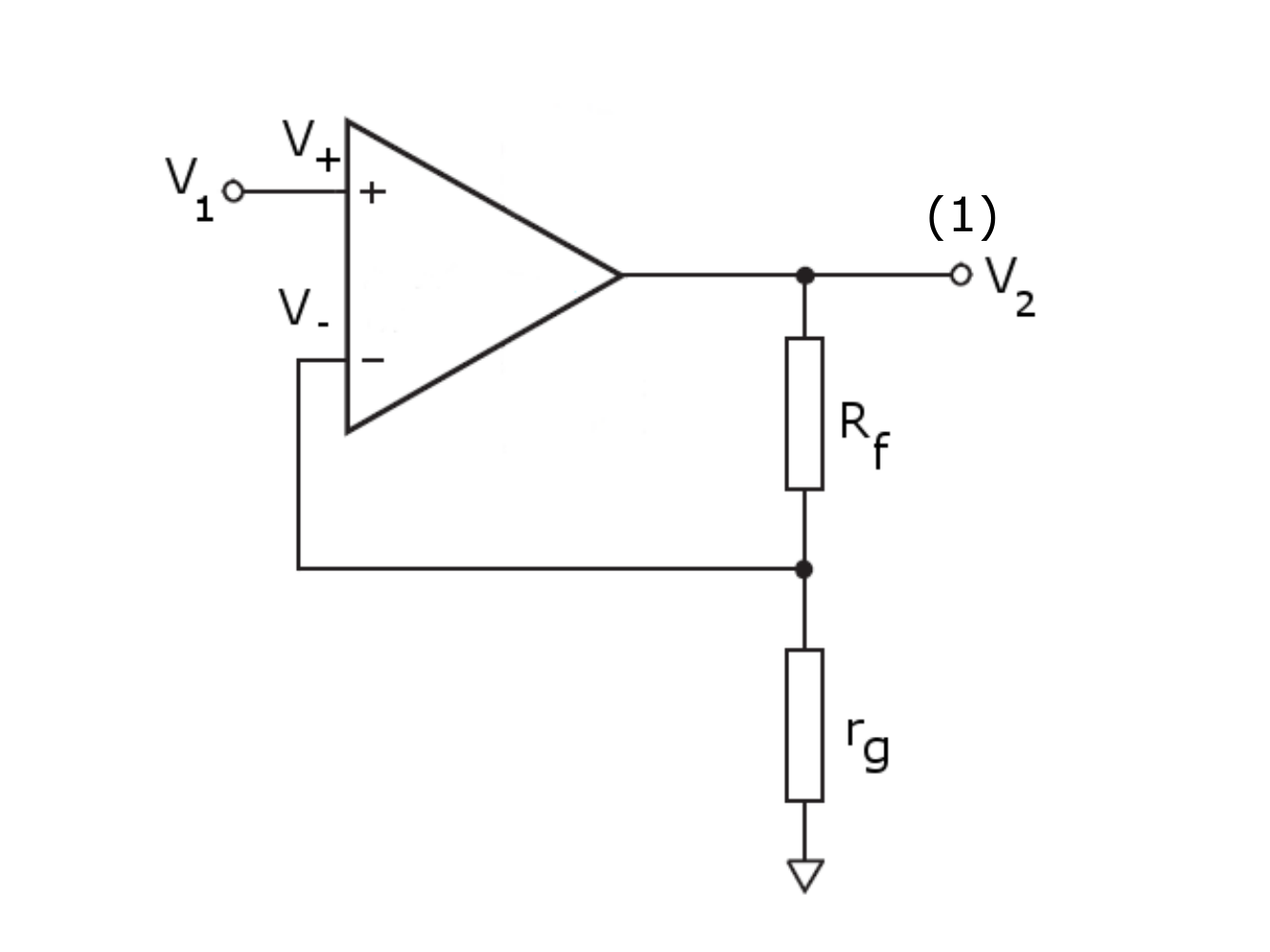}
\caption{Non-inverting amplifier}
\label{Non-inverting amplifier_m}
\end{minipage}
\begin{minipage}[t]{0.31\linewidth}
\includegraphics[scale=0.28]{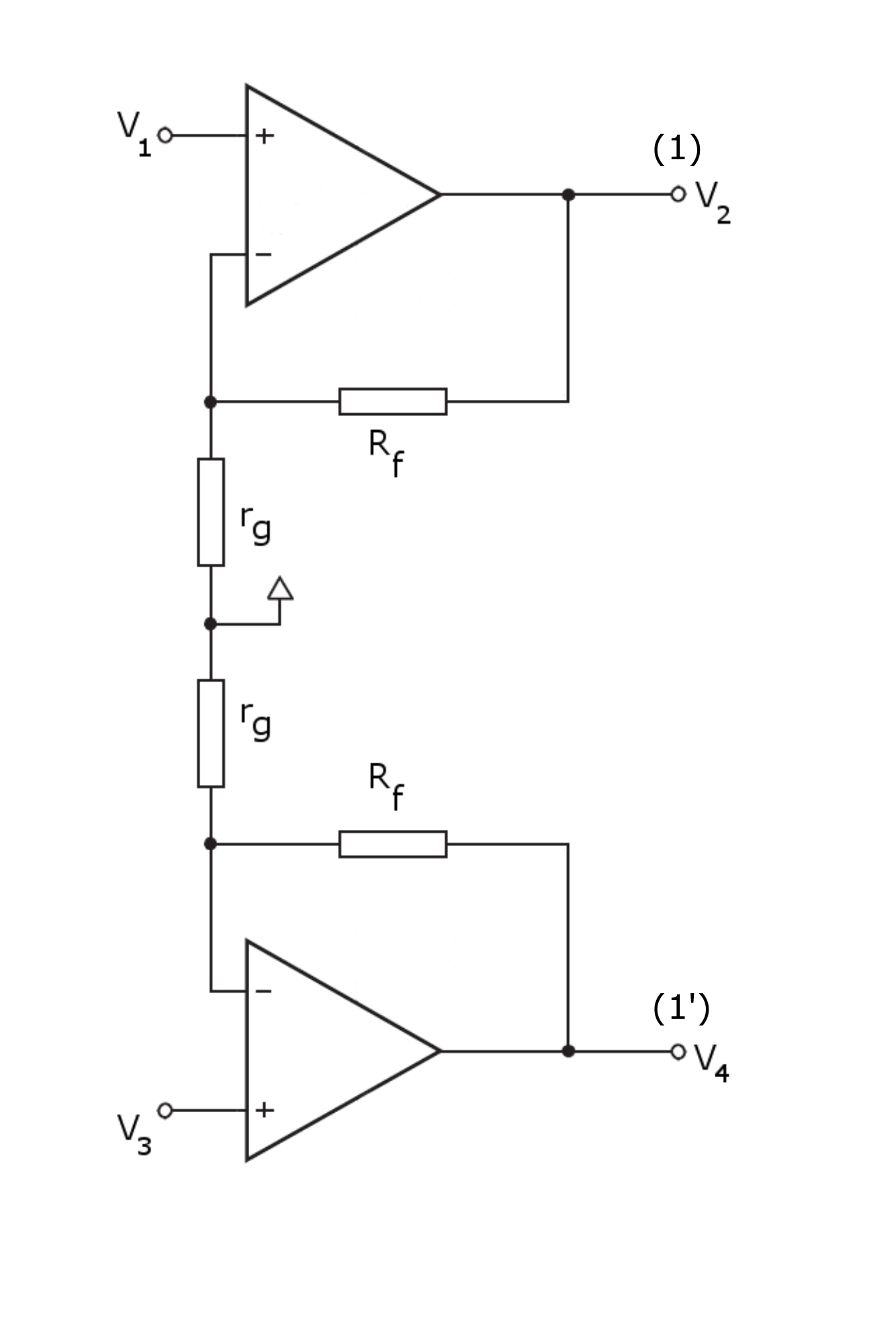}
\caption{Buffer}
\label{Buffer_m}
\end{minipage}
\begin{minipage}[t]{0.36\linewidth}
\includegraphics[scale=0.28]{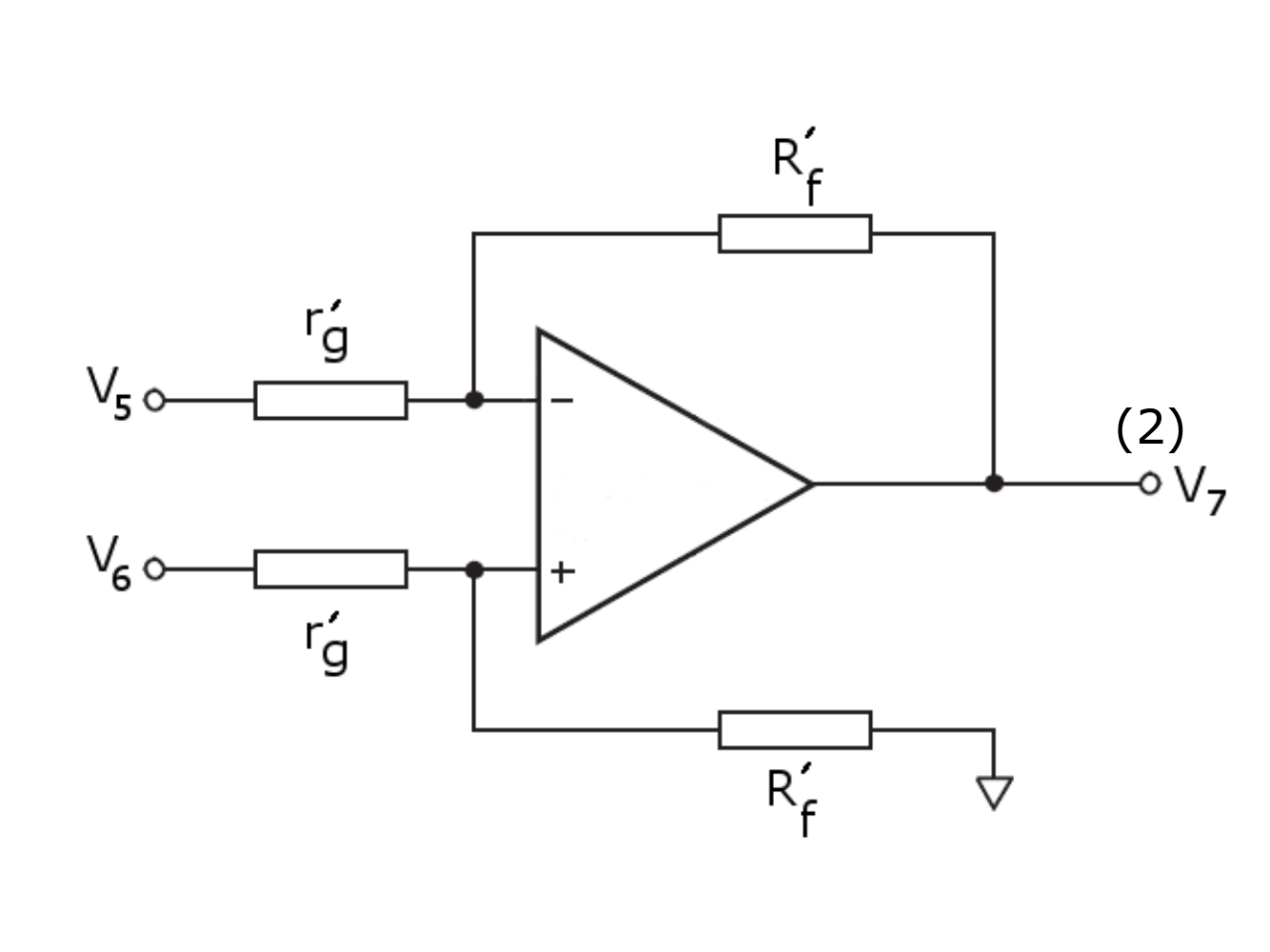}
\caption{Differential amplifier}
\label{Differential amplifier_m}
\end{minipage}
\begin{minipage}[c]{0.4\linewidth}
\includegraphics[scale=0.28]{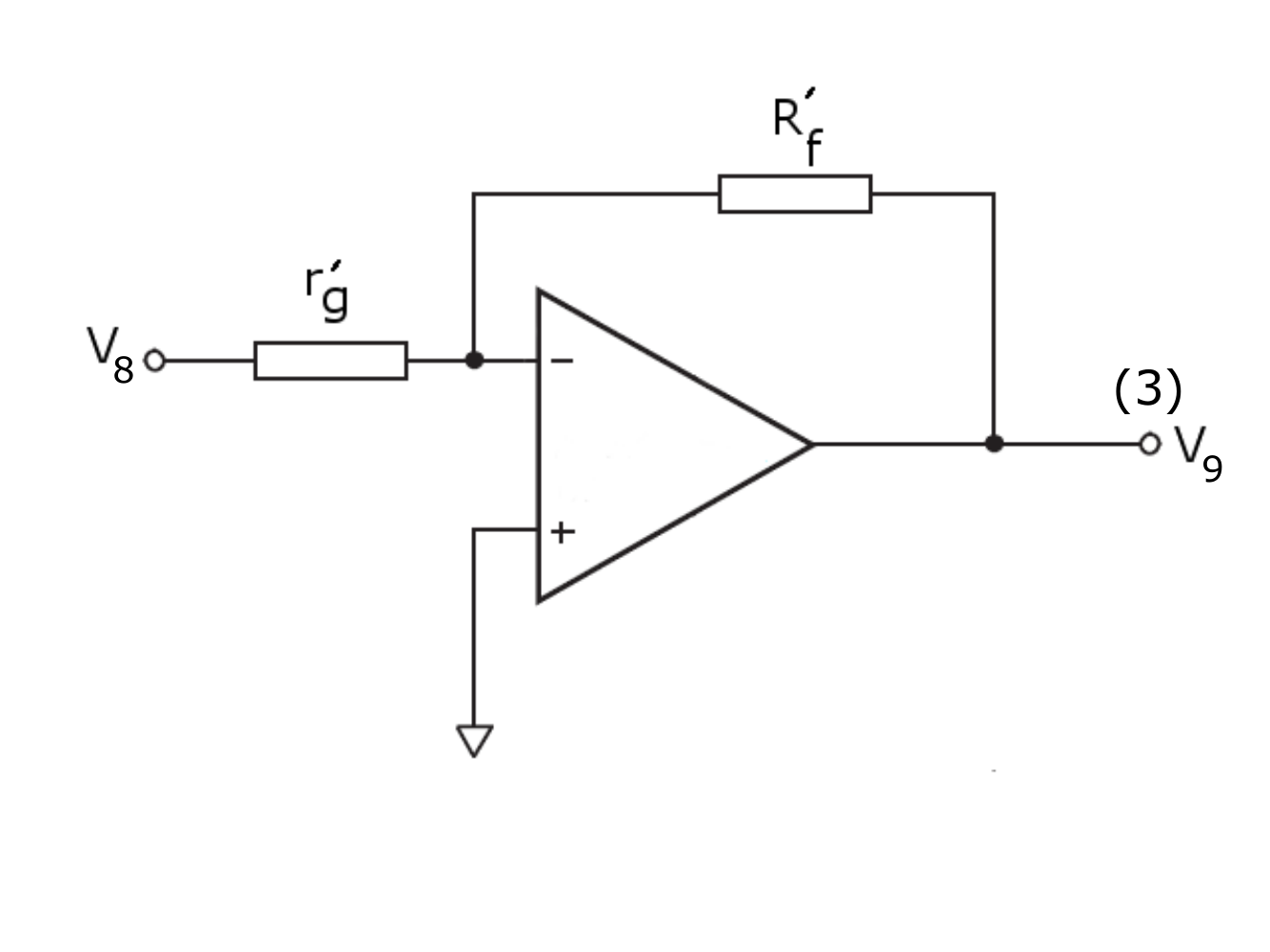}
\caption{Inverting amplifier}
\label{Inverting amplifier_m}
\end{minipage}
\begin{minipage}[c]{0.57\linewidth}
\includegraphics[scale=0.28]{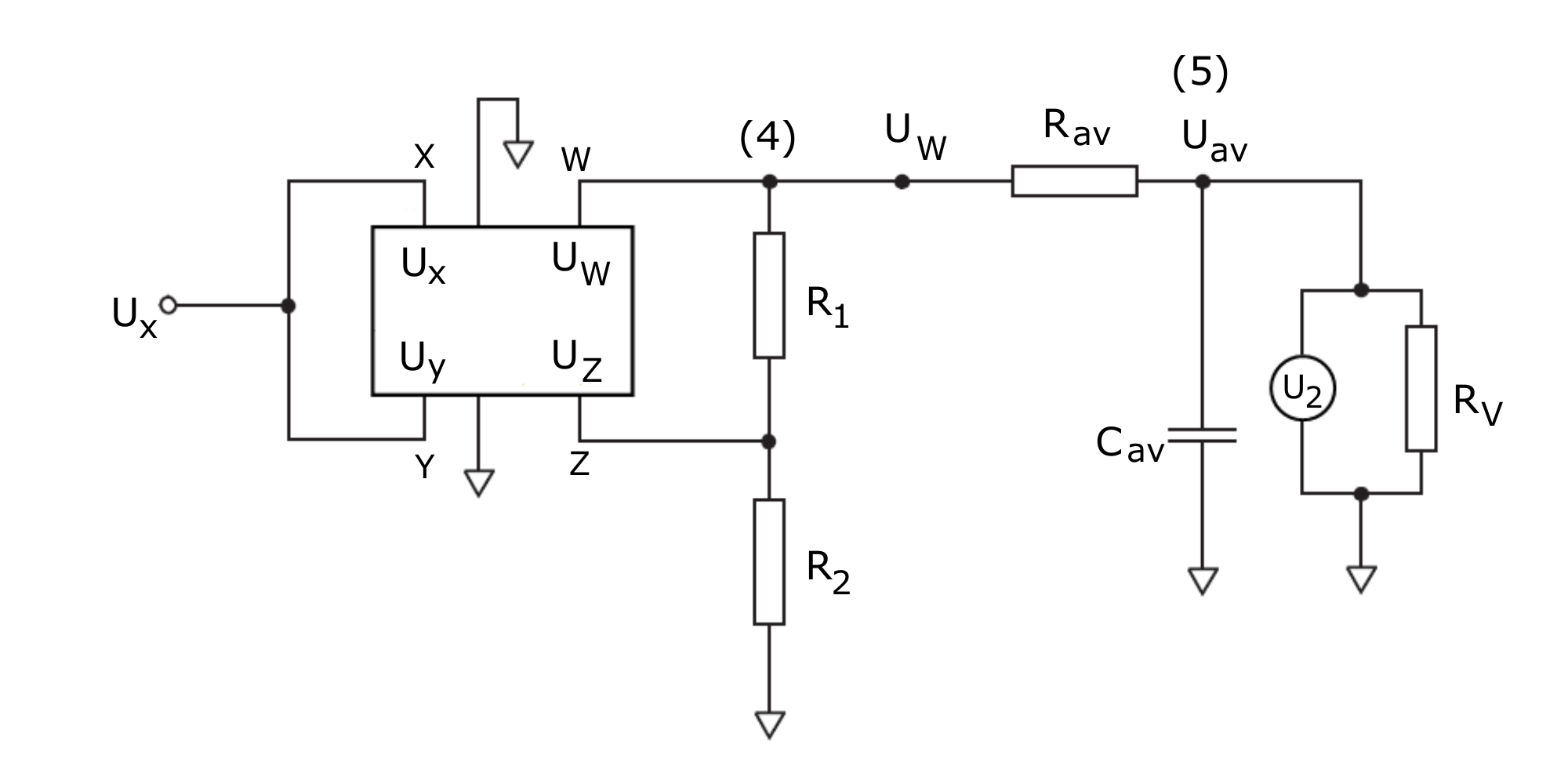}
\caption{Analog multiplier}
\label{Analog multiplier_m}
\end{minipage}
\end{figure}

Теориски добијте ја формулата Eq.~(\ref{electron_charge_m}) 
и Eq.~(\ref{electron_charge_mod_m}) анализирајќи ги колата и табеларните вредности од Табела~\ref{tbl:values_m}. 
Пресметајте го засилувањето $y_1$ на буферот и вкупното засилување на сите степени на засилувачот $Y$. 
Деталите на колата се опишани на сликите
Figs.~\ref{Non-inverting amplifier_m},\ref{Buffer_m}, 
\ref{Differential amplifier_m},
\ref{Inverting amplifier_m},
\ref{Analog multiplier_m}.
\begin{center}
\begin{table}[h]
\begin{tabular}{| c | r |}
		\hline
		&  \\ [-1em]
		Circuit element  & Value  \\ \tableline
			&  \\ [-1em]
			$R$ &200~$\Omega$ \\
			$r_\mathrm{_G}$ & 20~$\Omega$ \\
			$R_\mathrm{F}$ &  1~k$\Omega$  \\
			$C_\mathrm{F}$ &  10~pF  \\ 
			$C_\mathrm{G}$ & 10~$\mu$F \\
			$R_\mathrm{G}$ &  100~$\Omega$  \\ 
			$R_\mathrm{F}^\prime$ & 10~k$\Omega$ \\
			$C_\mathrm{F}^\prime$ & 10~pF \\
			$R_\mathrm{_L}$ & 510~$\Omega$ \\
			$R_1$ &  2~k$\Omega$  \\ 
			$R_2$ & 18~k$\Omega$  \\
			$R_\mathrm{av}$ & 1.5~M$\Omega$ \\
			$C_\mathrm{av}$ & 10~$\mu$F \\
			$R_\mathrm{_V}$ & $\approx 1~\mathrm{M} \Omega$ \\
			$V_\mathrm{CC}$ & +9~V \\
			$V_\mathrm{EE}$ & -9~V \\
\tableline
\end{tabular}
	\caption{Табела со нумерички вредности од елементите од колата.}
	\label{tbl:values_m}
\end{table}
\end{center}

Решението на оваа домашна работа треба да се испрати најдоцна до изгрејсонце, по денот на олимпијадата, по емаил на epo@bgphysics.eu.
Најдоброто решение ќе биде наградено со наградата на Зомерфилд која има парична вредност од DM137. 

\section{Задачи за понатамошна работа}
Експерименталната поставка е всушност lock-in волтметар со кој можете да мерете AC сигнали помали од $\mathrm{\mu V}$.
Можете да ни напишете како за ова би го користеле "NC" додатокот на плочата.

%% file: EPO6_sr_a.tex
\clearpage
\begin{center}
\textbf{Одређивање наелектрисања електрона коришћењем Шоткијевог шума.  Задатак на 6. експерименталној физичкој олимпијади.  Софија 8. децембар 2018.}
\end{center}

\indent
\setlength{\leftskip}{1.2cm}
\setlength{\rightskip}{1.2cm}
Овде је описано неколико узастопних експеримената са постављеним  PCB штампаним плочама, посебно дизајнираним за ове експерименте.
Извођење узастопних експерименталних задатака омогућава одређивања вредности наелектрисања електрона
$q_e.$
Промене напона $U(t)$ треба мерити за различите осветљености фотодиоде.
Напон је појачан милион пута $Y=10^6$. 
Појачан напон $YU(t)$ се примењује на уређају, the који даје резултат од вредности
време у просеку квадрат напона
$U_\mathrm{S}=\left<(Y U(t))^2\right>/U_0$.
Овај напон $U_\mathrm{S}$ се мери унимером.
Серија мерења даје могућност да се одреди 
$q_e$ користећи добро познату Шоткијеву формулу за спектралну густину тренутног шума
$(I^2)_f=2q_e\left<I\right>.$
За ученике средње школе, основни проблем је анализа аналогног квадрирања.
Радови ученика су одвојени и оцењују се у четири категорије S, M, L, XL према узрасту (старости) ученика.
За последњу XL категорију, задаци садрже проблеме оријентисане на програм високог образовања физике и укључују теоријско истраживање постављања PCB комплета као инжењерског уређаја.
Овај проблем је разматран на EPO6, децембра 2018. године ``Day of the Charge''. 
EPO6 организује Подружница града Софије, Удружења физичара Бугарске, у сарадњи са Физичким факултетом Софијског универзитета и Друштвом физичара Републике Македоније.  

\setlength{\leftskip}{0cm}
\setlength{\rightskip}{0cm}

\section{Увод}

Од самог почетка, Олипијада експерименталне физике (the Experimental Physics Olympiad - EPO) је позната широм света; сви проблеми ранијих олимпијада објављени су на интернету~\cite{EPO1,EPO2,EPO3,EPO4,EPO5} а од самог почетка учествује по 120 ученика на свакој олимпијади.
Претходних година учествовали су ученици средњих школа из 7 држава а удаљеност између најудаљенијих градова је већа од  4~Mm.

Описаћемо разлике између EPO и других сличних такмичења.
\begin{itemize}
\item Сваки учесник EPO од организатора добија на поклон комплет опреме са којом је радио. Значи, након завршетка олимпијаде, и ученици који су имали слабије резултате могу да понове експеримент и да достигне ниво шампиона.
На овај начин, олимпијада директно утиче на побиљшање квалитета наставе физике у целом свету. На крају школске године, поставка остаје школи из које долазе ученици.
\item Сваки проблем је оригиналан и повезан са фундаменталном физиком или разумевањем рада техничког патента.
\item Олимпијска идеја је реализована у EPO у њеном основном облику и сви који желе могу да се укључе. Број учесника није ограничен. Са друге стране, сличност са другим олимпијадама је што су директна илустрација наставних материјала и заједно са другим такмичењима се супротстављају деградацији средњошколског образовања, што је светска тенденција. 
\item Једна иста експериментална поставка се даје свим учесницима, али задаци су различити за различите старосне групе, исто као што је и вода у базену подједнако је влажна за све старосне групе на пливачком такмичењу.
\end{itemize}

Укратко ћемо поменути проблеме проблеме претходних 5 EPOs: 
1) Поставка EPO1 је била ученичка верзија Америчког патента ауто-нула стабилизоване једносмерне струје.~\cite{EPO1}
2) Проблем друге олимпијаде  EPO2~\cite{EPO2} био је мерење Планкове константе помоћу дифракције светлости ЛЕД диоде помоћу компакт диска.
3) Савремена реализација патента НАСА за употребу негативног претварача импедансе за генерисање осцилација напона била је задатак EPO3.~\cite{EPO3}
4) EPO4~\cite{EPO4} била је посвећена фундаменталној физици - одређивање брзине светлости мерењем електричних и магнетних сила. Иновативни елемент била је примена теорије катастрофе у анализи стабилности клатна.
5) Тема EPO5 била је мерење Болцманове константе $\kb$ праћењем Ајнштајнове идеје пручавања термичких промена електричног напона кондензатора..

EPO6 прати традицију да тема олимпијада буде мерење неке фундаменталне константе.инспирисана је предавањем о флуктуацијама Ајнштајна, Валтер Шотки је закључио да наелектрисање електрона може да буде измерено помоћу проучавања шума у електричном колу.
Сада, после једног века, појављивањем операционих појачивача слабпг шума, идеја Шоткија може се имплементирати као задатак за средњошколце.

Укратко, устаљене традиције ЕРО су равнотежа савремених техничких достигнућа и фундаменталне физике.

 (\textit{Allah akbar}).

\section{Експериментални комплет}
Експериментални комплет чине следећи делови:
\begin{enumerate}
\item Штампана плоча (ПЦБ) са конектором за фотодиоду и држачем литијумске батерије.
\item Четири батерије од по 9~V и са двоструким спојницама.
\item Три батерије од по 1.5~V и држач за њих са потенциометром.
\item Пластична кеса садржи следеће делове:
\begin{itemize}
\item Два оперативна појачивача на плочама са конекторима са зеленом налепницом, на којој је написан број.
\item Мултипликатор на плочи са конектором са наранџастом налепницом.
\item Литијумска батерија.
\end{itemize}
\item Оптички кабл са белим цилиндром на једном свом крају.
\item Пластична чаша са цевчицом која је провучена кроз њу.
\end{enumerate}
Ви треба да донесете два унимера са кабловима за повезивање и дигитрон.
\section{Почетни (лаганији) задаци. S}
\begin{enumerate}
\item Подеси унимер да мери напон (волтметар).
\item Измери напоне четири батерије од 9~V са максималном прецизношћу (користећи опсег волтметра 20~V) и запишите их.
Исто уради и са батеријама од 1.5~V  (користећи опсег волтметра 2000~mV).
\item Постави три батерије од 1.5~V у њихов држач.
Повежи потенциометар \emph{potentiometrically} са једним унимером.
Окрећи точак потенциометра.
Запиши интервал напона који добијеш.
Ово ће бити вредности напона које ћеш користити у наредним задацима.
Промена поларитета мења знак напона.
\item Повежи четири батерије од 9~V  на два двострука кабла са прикључцима (спојницама).
\section{Аналогно квадрирање напона. M}
\label{s:M_s}
\item \textbf{Пажња! Од овог тренутка постоји могућност прегоревања
интегралних кола, ако неправилно прикључиш батерије.} 
Постави штампану плочу тако да ти ознаке ``OUT'' и ``COM'' буду са десне стране, погледај at Слику.~\ref{Fig:Board_s}.
\item Пажљиво повежи батерије од 9~V на десни 3-pin конектор на штампаној плочи (налепница према налепници - налепнице су окренуте једна према другој).
\textbf{Ради пажљиво -- ако направиш грешку са поларитетом могу да прегоре множитељ (multiplier) и оперативни појачивач.}
\item 
Повежи средњи контакт потенциометра
са улазном жицом повезаном у тачки ``IN''. 
Користи жицу са ``крокодилкама''. 
\item Прикључи другу електроду потенциометра на жицу која долази из ``уземљења'' означеног са ``COM''. 
\item 
Повежи први волтметар V$_1$ који показује $U_1$ на потенциометру на ``IN''--``COM'' улазе кола паралелно.
\item Када вртиш точкић потенциометра, напон $U_1$ требало би да се мења приближно између 0 и максималног напона батерије 4.5~V.
\item Прикључи други волтметар V$_2$, који показује напон $U_2$, између прикључака ``OUT'' и ``COM''.
На овај начин $U_2$ је напон између ``OUT'' и ``COM'' тачака кола.
\item Провери да ли су ``COM'' прикључци оба унимера исправно повезани.
\item Конектори са 4 двострука пина са наранџастом ознаком налазе се на десној страни штампане плоче - означени са  ``AD633'' in Fig.~\ref{Fig:Board_s}.
Повежи множитељ (multiplier) на ``AD633'' тако да  \textbf{наранџаста ознака на множитељу буде окренуте према наранџастој ознаци на плочи}.
\item Окрећи точкић потенциометра, сачекајте 1 минут и запиши вредности напона $U_1$ и $U_2$.
Промени поларитет извора напона и понови мерења улазног напона $U_1$ и излазног напона $U_2$.
Упиши резултате у табелу са колонама :
број мерења $i$, $U_1$ и $U_2$. 
\item Прикажи резултате графички где је $U_1$ апсциса (хоризонтална оса) и $U_2$ ордината (вертикална оса) на милиметарском папиру.
\item Додај колону $(U_1)^2$ у табелу и прикажи резултате графички , $(U_1)^2$ -- апсциса (хоризонтална оса) и $U_2$ -- ордината (вертикална оса) на милиметарском папиру.
Повуци праву линију која пролази најближе експерименталним тачкама.
Ова линија приближно описана једначином
$U_2= (U_1)^2/U_0 + \mathrm{const}.$
Одабери две тачке на правој линији, одредите разлику на апсциси 
$\Delta(U_1^2)$, и на ординати $\Delta(U_2)$ 
одредите нагиб (коефицијент правца)
$U_0=\Delta(U_1^2)/\Delta(U_2)$.
Овај параметар $U_0$ са димензијом напона је од суштинског значаја за одређивање наелектрисања електрона $q_e$ како је описано у следећем одељку.

\section{Флуктуациона спектроскопија. Одређивање наелектрисања електрона $q_e$. L} 

\begin{figure}[h]
\centering
\includegraphics[scale=0.3]{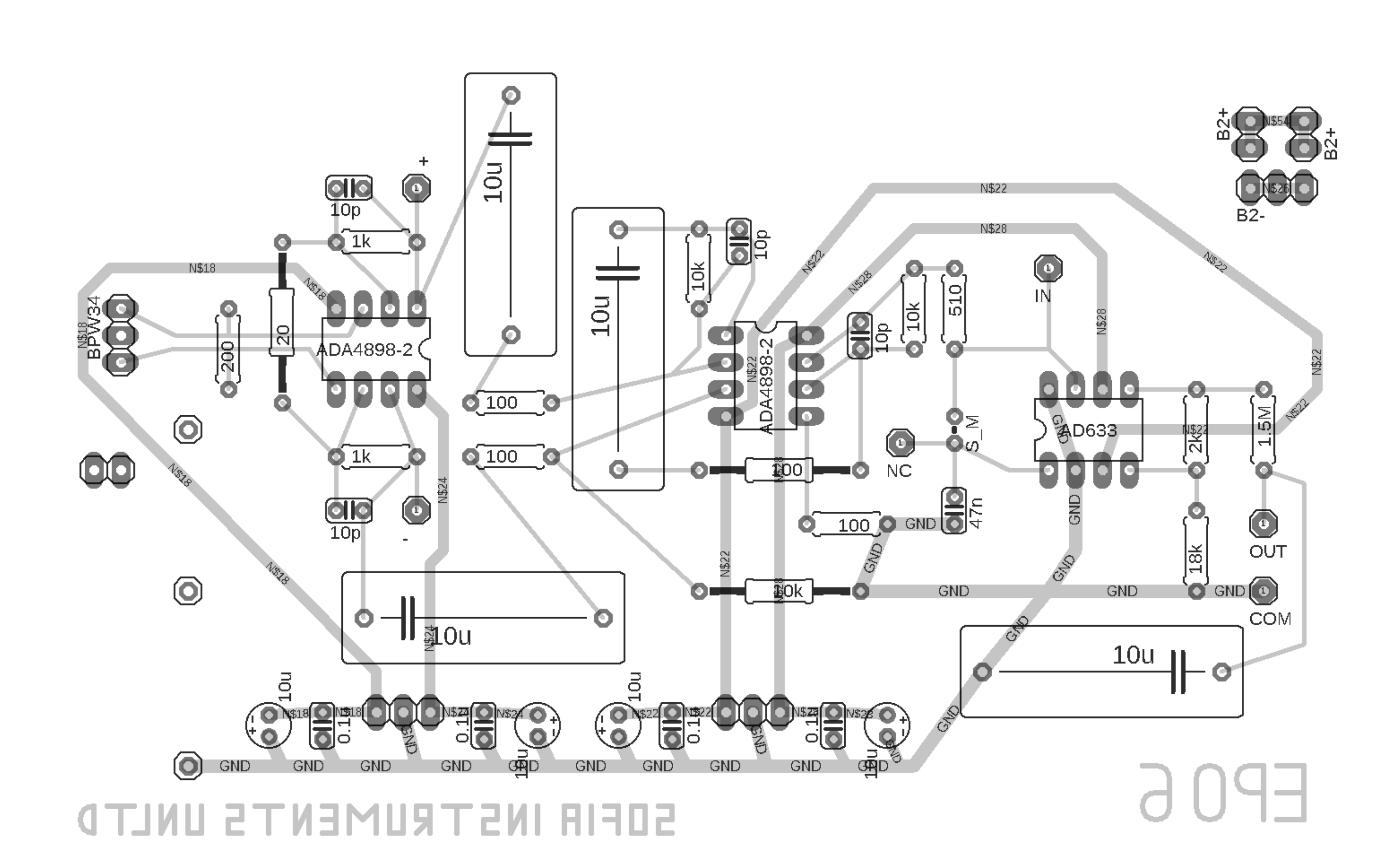}
\caption{Дијаграм штампане плоче експерименталне поставке.
Лед сијалица са белим цилиндром на њој која није приказана овде спаја се између ``B2-'' и ``B2+'' пинова (нису означени на штампаној плочи) на врху плоче са десне стране (горе десно).
Два неозначена 3-пинска прикључка постављена на дну плоче су прикључци за изворе напона}
\label{Fig:Board_s}
\end{figure}
\itemПостави 3 батерије од  1.5~V у њихов држач. 
Повежи две спољне електроде потенциометара са две ''крокодилке''  са лед сијалицом ``B2-'' и ``B2+'' пинове у горњем десном угли плоче приказано на  Fig.~\ref{Fig:Board_s}.
Овде поларитет није важан, ако је правилно прикључена сијалица светли. 
\item Прикључи помоћу ``крокодилке'' ``($-$)'' са``COM'' прикључком волтметра V$_1$.
\item Аналогно помоћу друге ``крокодилке'' повежи ``($+$)'' електроду на плочи са
``$\mathrm{V\,\Omega\, m\!A}$'' прикључком волтметра V$_1$.
На овај начин волтметар V$_1$ пказује потенцијалну разлику $U_{\pm}$ између ``($+$)''  и ``($-$)'' тачака и овај напон је пропорционалан приближној вредности фото струје на фотодиоди.
\item Постави литијумску батерију од  3~V на држач поред фотодиоде тако што је ,
\textbf{ ``($+$)''  страна батерије повезана са  ``($+$)''  страном на држачу батерије.}
\item Два дупла пин конектора на левој и централној страни плоче означена су зеленим налепницама са ``ADA4898-2'' на слици 1 .~\ref{Fig:Board_s}.
Постави појачиваче на ове позиције тако што ће зелена налепница на сваком  ``ADA4898-2'' бити окренута \textbf{ка зеленој налепници на плочи. Буди пажљиви, уколико погрешише појачивач ће прегорети.}
\item Пажљиво прикључи плочасти конектор са причвршћеном фотодиодом и држачем литијумске батерије на своје место ``BPW34'' in Fig.~\ref{Fig:Board_s} на штампаној плочи као што је означено фломастером на плочи.
Пажљиво намести чашу тако да фотодиода уђе у црну цевчицу са краће стране.
Један крај оптичког влакна са белим цилиндром убаци у дужи крај цевчице. 
Провери да ли оптички кабл може да се креће кроз цевчицу без значајног трења.
\itemДруги крај оптичког кабла уметни у мали отвор на белом цилиндру који је причвршћен на сијалицу. Пази да не помериш бели цилиндар на сијалици да се не би поломила.
\item Повежи прикључак батерије од 9~V на лежиште левог 3-пинског конектора за снабдевање напоном на плочи. \\
\textbf{Буди пажљив! Нараџаста налепница спојнице треба да одговара нараџастој налепници на плочи у супротном би могло да дође до прегоревања појачивача.}
Постави унимер на опсег од 2000~mV и полако померај оптичко влакно низ цевчицу. Овај напон $U_\pm$ би требао да се мења до 1000~mV.
На овај начин померањем оптичког влакна мења се фото струја преко које меримо фотонапон $U_\pm$ добијен фотострујом која пролази кроз отпорник $R$.
Ако не може да се достигне напон од миниму 700~mV или уопште нема напона, потражи помоћ од дежурног у учионици. 
Ово је важан део за експериментално мерење.
\item Да би продужили даље потражите воду и лед од дежурног који треба да је сипа у чашу са цевчицом и фотодиодом. 
\item Након сипања воде и леда у чашу обрати пажњу да ли долази до цурења. Ако цури склони чашу и потражи другу.
\item Повежи ``COM'' тачку на плочи са ``COM'' тачком волтметра V$_2$ (волтметар V$_2$ мора бити исти који си користио у Поглављу.~\ref{s:M_s}).
\item Повежи ``OUT'' тачку на плочи са ``$\mathrm{V\,\Omega\, m\!A}$'' на волтметру V$_2$.
Сада, волтметар V$_2$ мери напон $U_\mathrm{S}$ који је пропорционалан струјном шуму фотодиоде.
\item Повежи повежи другу спојницу са батеријама од 9~V на десни 3-пински конкектор на плочи. \\
\textbf{Буди пажљив! Нараџаста налепница на спојници треба да одговара нараџастој налепници на плочи. У супротном прегореће појачивач.}
\item Потражи од дежурног лепљиву траку и залепи на столу свих 6 каблова експерименталне поставке на следећи начин:
\begin{itemize}
\item Прво место сва четири кабла батерије од 9~V испод плоче.
\item Залепи ``+'' и ``-'' каблове изнад плоче, обезбеђујући да нису у контакту.
\item Залепи ``IN'' и ``NC'' каблове изнад плоче и удаљене од ``+'' и ``-'' каблова.
\item  Залепи``OUT'' и ``COM'' каблове на дасну страну плоче водећи рачуна да нису у контакту.
\end{itemize}
Овако постављени каблови и плоча треба да личе на паука.
\item Погледај напон на волтметру V$_2$, требало би да се стабилизује око тридесетак mV.
\item Започни мерење $U_\pm$ и $U_\mathrm{S}$: мењај $U_\pm$ са минималним померањем оптичког кабла у цевчици мало по мало према фотодиоди.
Промена $U_\pm$ између сваке вредности напона треба да буде најмање 100~mV.
Причекај најмање 2 минута да се вредност $U_\mathrm{S}$ стабилизује.
\emph{Уколико се било какво поремећај значајно мења показивање волтметара, поново сачекај стрпљиво да се стабилизује.}
Резултате упиши у табелу са колонама:
број мерења $i$, $U_\pm$ и $U_\mathrm{S}$. 
\item Графички представи резултате токо да је  $U_\pm$ на апсциси (хоризонтална координата а$U_\mathrm{S}$ на одринати (вертикална координата) на милиметарском папиру (ако немаш потражи од дежурног).
Нацртај праву линију која најбоље представља линеарну зависност
$U_\mathrm{S}=k U_\pm +\mathrm{const}$.
Изабери два тачке на правој и одреди нагиб праве
$k=\Delta U_\mathrm{S}/ \Delta U_\pm$,
где $\Delta$ означава разлику.
Ова математичка процедура названа је линеарна регресија.
За нашу експерименталну поставку је $k\simeq 10^{-3}.$
\item Коначно, одреди наелектрисање електрона $q_e$
помоћу формуле
\begin{equation}
q_e=2 k \frac{y_1}{Y^2} \frac{R_\mathrm{_L}}{R} C_\mathrm{_L}U_0,
\label{electron_charge_s}
\end{equation}
где је $Y=1.01\times 10^{6}$ је укупно појачање нашег појачивача,
и $y_1=101$ је амплификација првог корака мултипликатора.
Вредност $C_\mathrm{_L}$ можеш да пронађеш на налепници која се налази у доњем десном углу плоче ($C \equiv C_\mathrm{_L}$), $R_\mathrm{_L}=510~\Omega$ and $R=200~\Omega$. \\
Честитамо! Управо си одредио фундаменталну константу и у добром си друштву
! :)
\item Твоје мерење може да буде мало прецизније уколико укључимо не-идеалне ефекте  оперативних појачивача користећи формулу:
\begin{equation}
q_e=2 k \frac{y_1}{(1-\varepsilon)Y^2} \frac{R_\mathrm{_L}}{R} 
C_\mathrm{_L}U_0.
\label{electron_charge_mod_s}
\end{equation}
Мала корекција $\varepsilon(C_\mathrm{_L})$ као функција капацитета $C_\mathrm{_L}$ је графички приказана на слици.~\ref{Fig:Eps_s}.
\begin{figure}[h]
\centering
\includegraphics[scale=0.8]{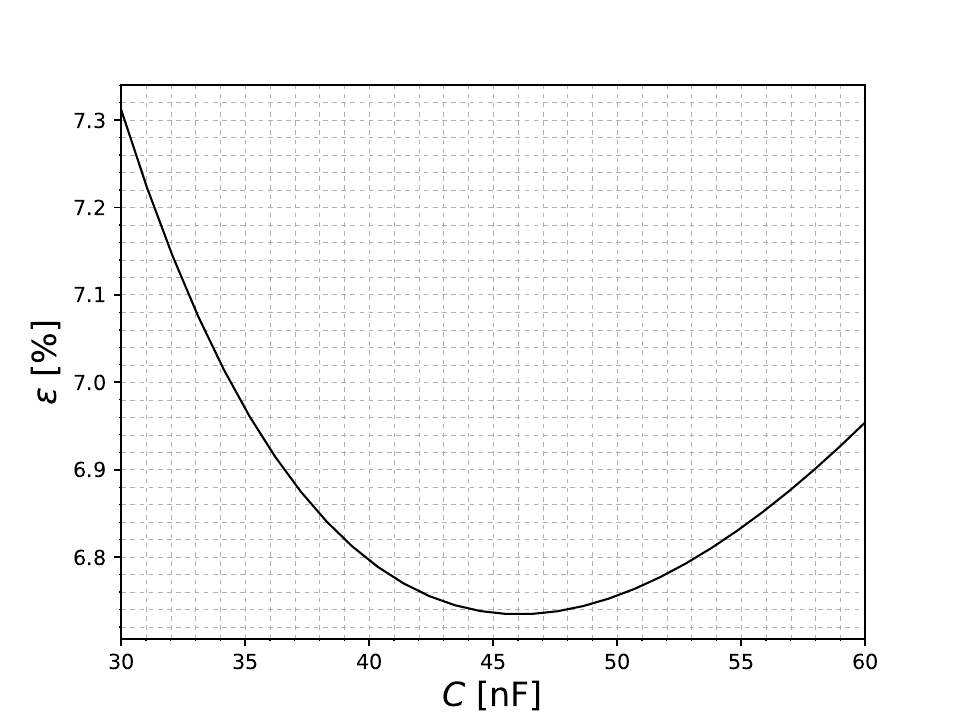}
\caption{Error $\varepsilon$ in percent in determination of $q_e$ as a function of the  
capacitor $C_\mathrm{_L} \equiv C$.}
\label{Fig:Eps_s}
\end{figure}
Пронађи грешку $\varepsilon$ from Fig.~\ref{Fig:Eps_s} у односу на твоју вредност $C_\mathrm{_L}$ записану на плочи и не заборави да поделиш са  $\varepsilon$  100 пре узрачунавања $q_e$ from Eq.~(\ref{electron_charge_mod_s}).

За овај метод одређивања наелектрисања електрона $q_e$
се определио Волтер Шотки.\cite{Schottky:1918}
Волтер Шотки је радио са Максом Планком и био инспирисан предавањима Алберта Ајнштајна у вези са електричним флуктуацијама.

\section{Домаћи задатак. XL}
Изведи формуле Eq.~(\ref{electron_charge_s}) 
and Eq.~(\ref{electron_charge_mod}) анализирајући елементе кола и табличне вредности из табеле Table~\ref{tbl:values_s}. 
Израчунај појачања $y_1$ бафера и укупног појачања свих корака појачивача $Y$. 
Приказани су различити детаљи кола на сликама
Figs.~\ref{Non-inverting_amplifier_s},\ref{Buffer_s}, 
\ref{Differential_amplifier_s},
\ref{Inverting_amplifier_s},
\ref{Analog_multiplier_s}.

\begin{center}
\begin{table}[h]
\begin{tabular}{| c | r |}
		\hline
		&  \\ [-1em]
		Circuit element  & Value  \\ \tableline
			&  \\ [-1em]
			$R$ &200~$\Omega$ \\
			$r_\mathrm{_G}$ & 20~$\Omega$ \\
			$R_\mathrm{F}$ &  1~k$\Omega$  \\
			$C_\mathrm{F}$ &  10~pF  \\ 
			$C_\mathrm{G}$ & 10~$\mu$F \\
			$R_\mathrm{G}$ &  100~$\Omega$  \\ 
			$R_\mathrm{F}^\prime$ & 10~k$\Omega$ \\
			$C_\mathrm{F}^\prime$ & 10~pF \\
			$R_\mathrm{_L}$ & 510~$\Omega$ \\
			$R_1$ &  2~k$\Omega$  \\ 
			$R_2$ & 18~k$\Omega$  \\
			$R_\mathrm{av}$ & 1.5~M$\Omega$ \\
			$C_\mathrm{av}$ & 10~$\mu$F \\
			$R_\mathrm{_V}$ & $\approx 1~\mathrm{M} \Omega$ \\
			$V_\mathrm{CC}$ & +9~V \\
			$V_\mathrm{EE}$ & -9~V \\
\tableline
\end{tabular}
	\caption{Табела нумеричких вредности струјног кола.}
	\label{tbl:values_s}
\end{table}
\end{center}

\begin{figure}[t]
\begin{minipage}[t]{0.31\linewidth}
\includegraphics[scale=0.28]{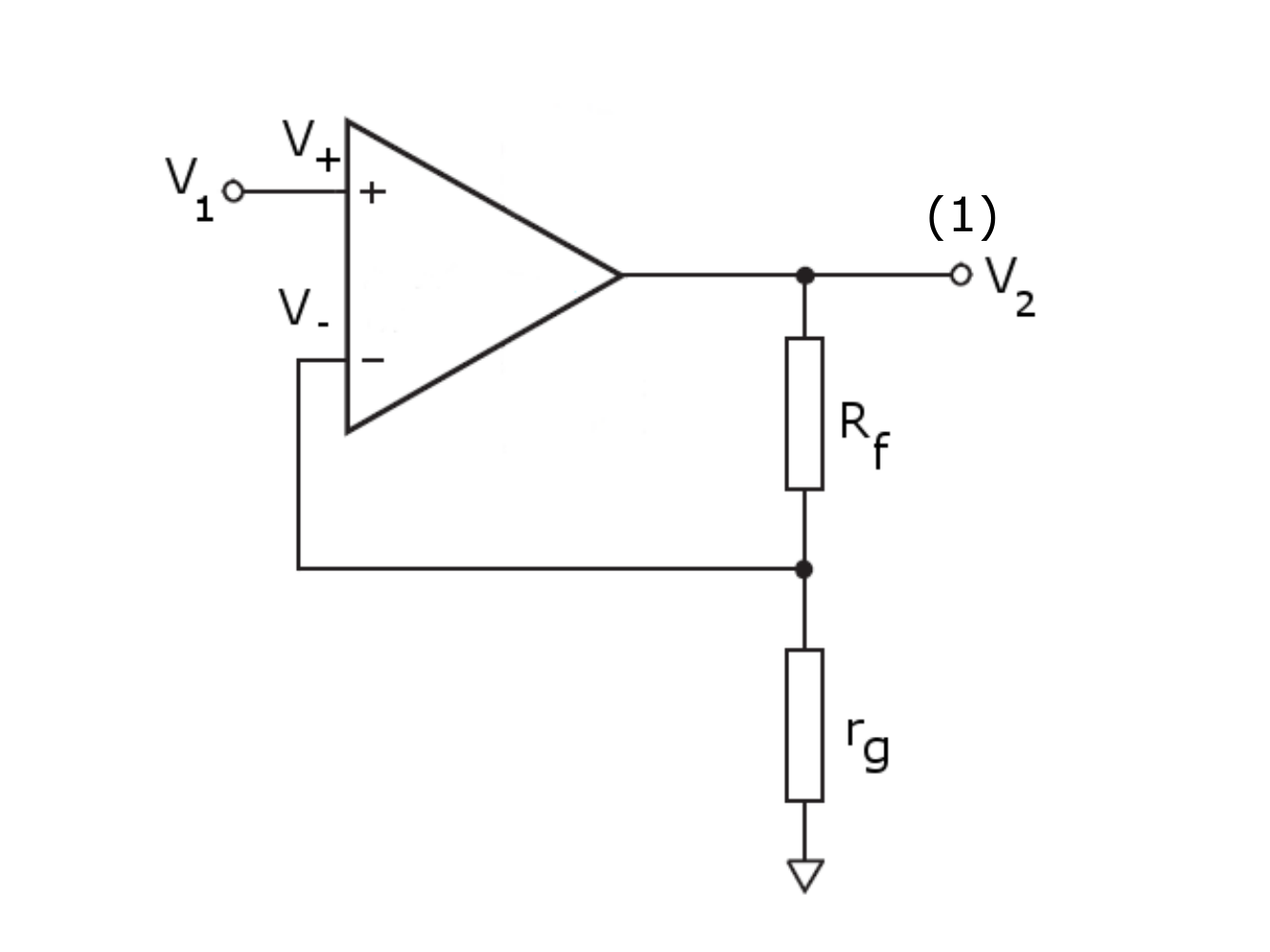}
\caption{Non-inverting amplifier}
\label{Non-inverting_amplifier_s}
\end{minipage}
\begin{minipage}[t]{0.31\linewidth}
\includegraphics[scale=0.28]{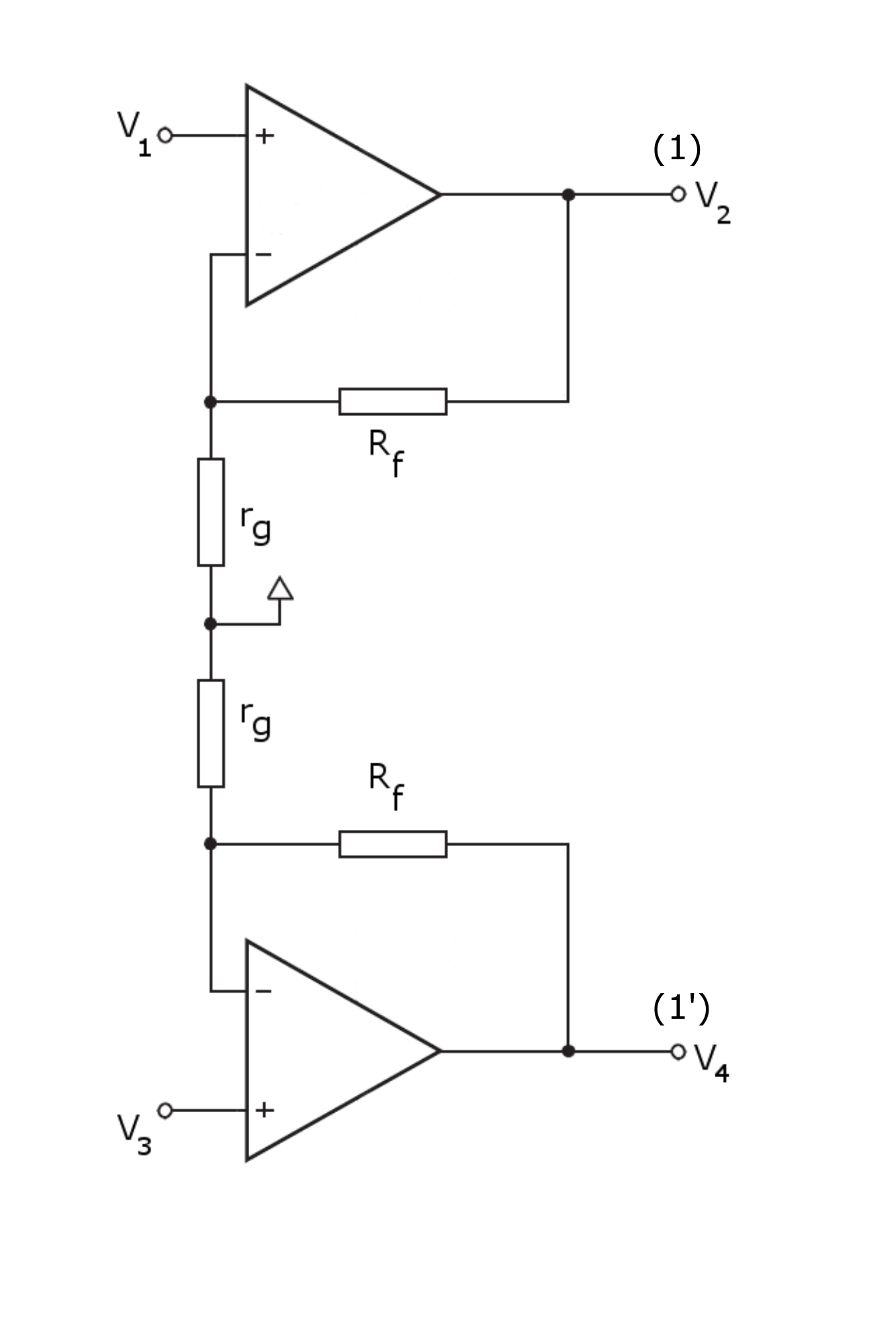}
\caption{Buffer}
\label{Buffer_s}
\end{minipage}
\begin{minipage}[t]{0.36\linewidth}
\includegraphics[scale=0.28]{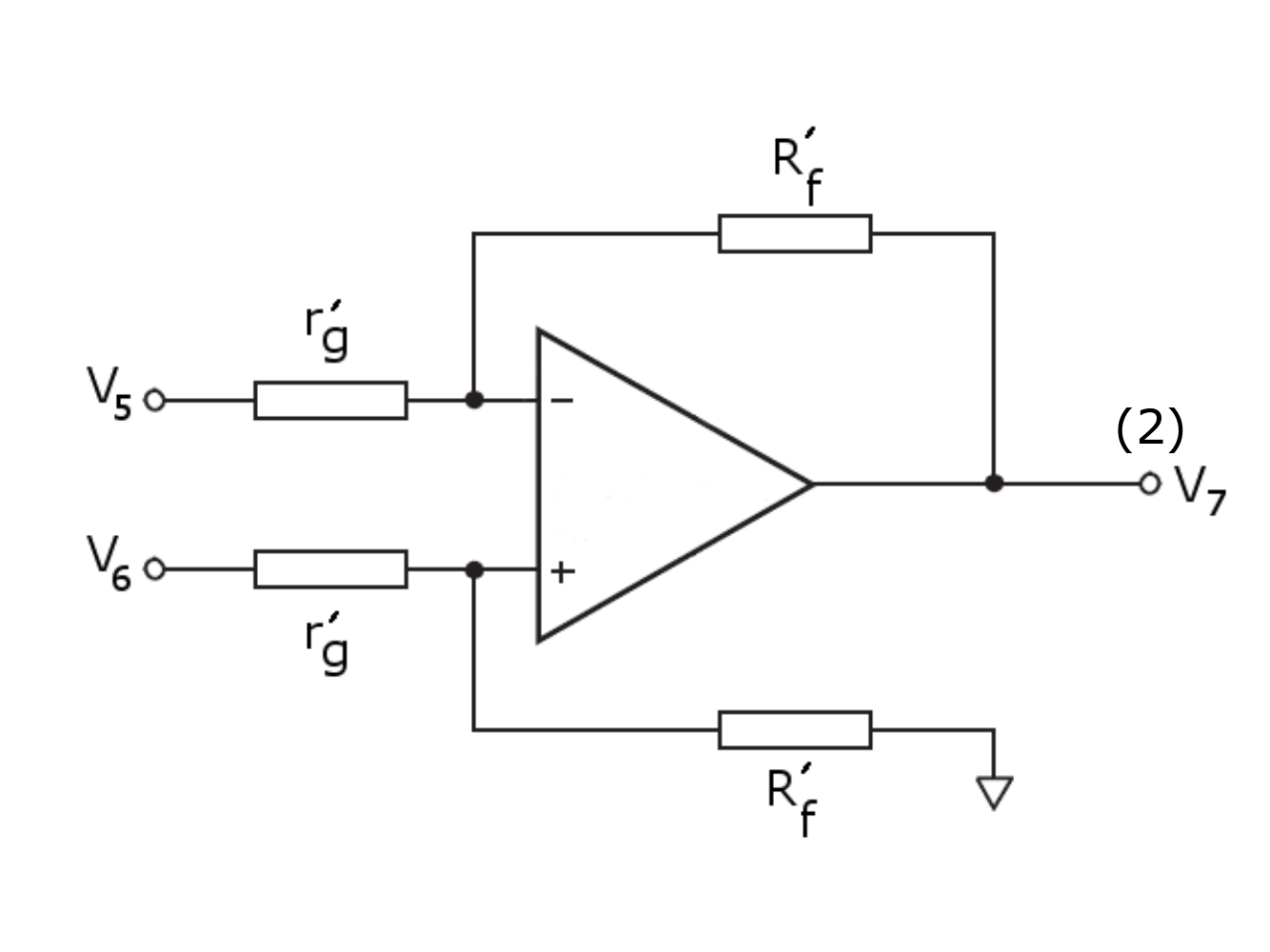}
\caption{Differential amplifier}
\label{Differential_amplifier_s}
\end{minipage}
\begin{minipage}[c]{0.4\linewidth}
\includegraphics[scale=0.28]{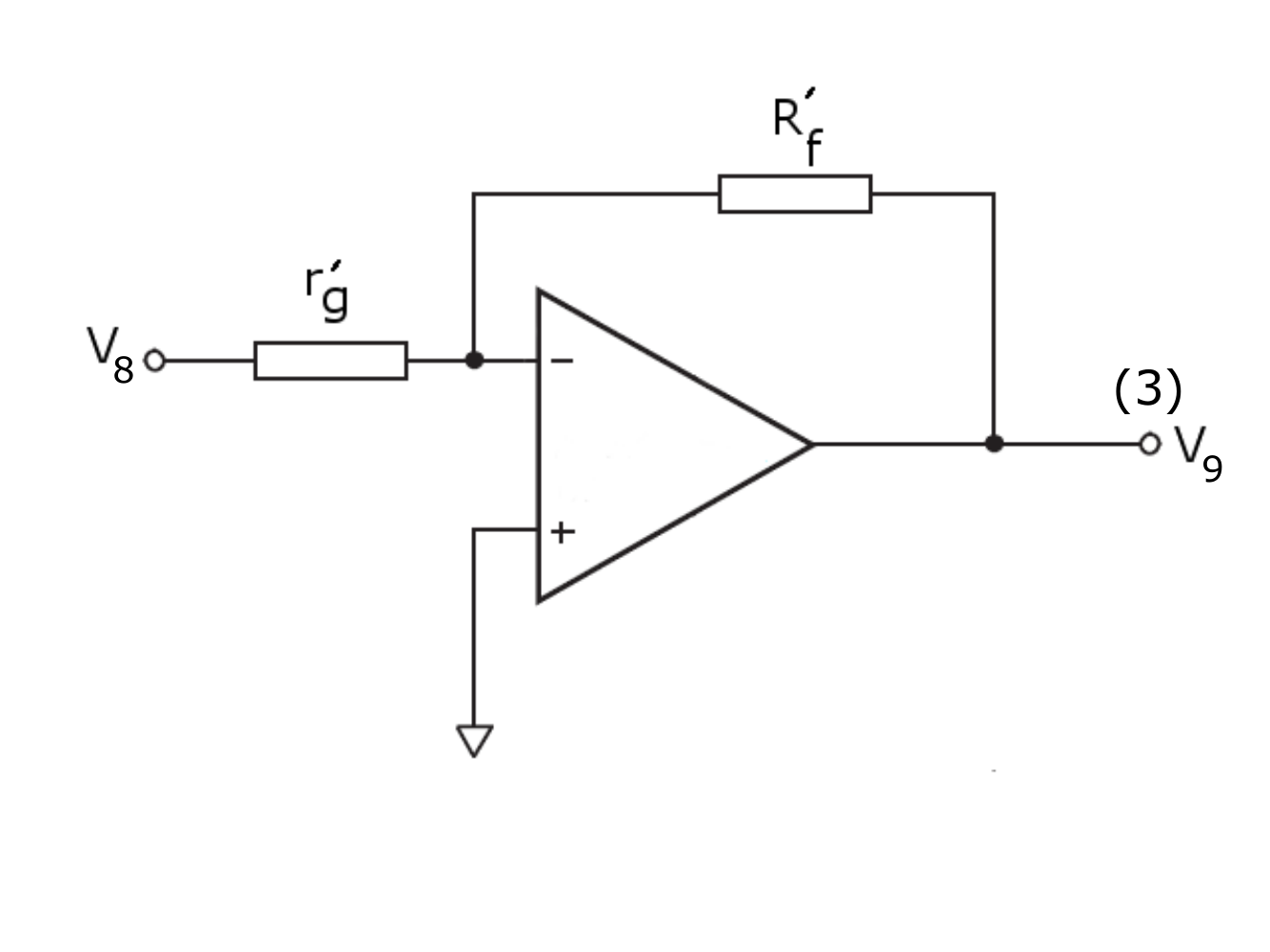}
\caption{Inverting amplifier}
\label{Inverting_amplifier_s}
\end{minipage}
\begin{minipage}[c]{0.57\linewidth}
\includegraphics[scale=0.28]{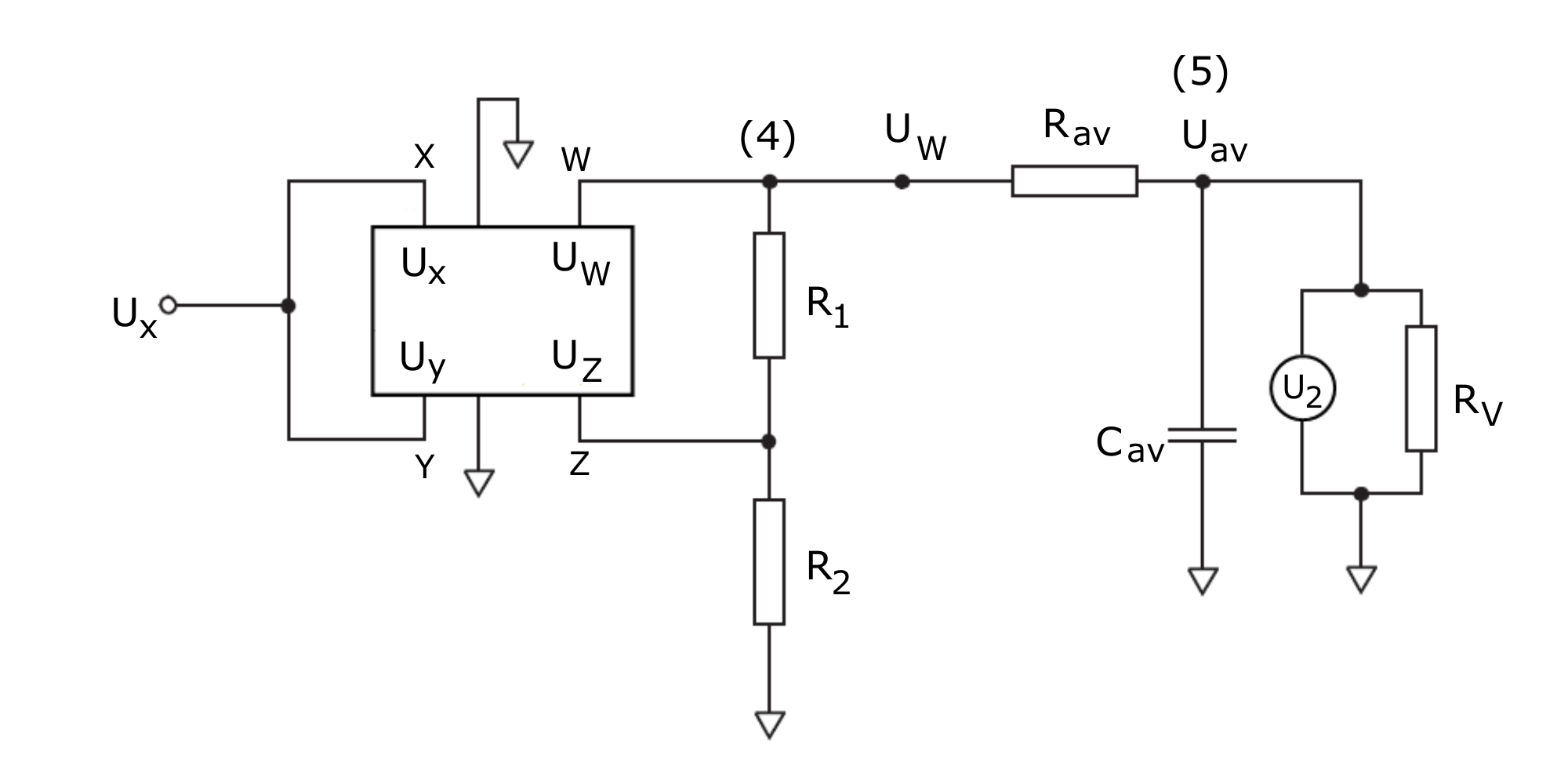}
\caption{Analog multiplier}
\label{Analog_multiplier_s}
\end{minipage}
\end{figure}

Решење мораш послати током ноћи после олипијаде, до јутра, на 
 e-mail epo@bgphysics.eu.
Најбоље решење ће добити  Sommerfeld price with a monetary equivalent of DM137. 

\section{Проблем за будући рад - будућност}
Експериментална поставка је уставри lock-in voltmeter и њиме може да се мери AC сигнал мањи од $\mathrm{\mu V}$.
Можете нам написати како да користимо ``NC'' излаз у овом случају.

\end{enumerate}

%% file: EPO6_ru_a.tex
\clearpage
\renewcommand{\figurename}{Фигура}
\renewcommand{\tablename}{Таблица}
\begin{center}
\textbf{Измерение заряда электрона $q_e$ по методу Шоттки (дробовой шум). 
  Задание Шестой олимпиады по экспериментальной физике EPO6.
  София. 8~декабря 2018~г.}
\end{center}
\normalsize
\indent
\setlength{\leftskip}{1.2cm}
\setlength{\rightskip}{1.2cm}
Измеряются флуктуации напряжения $U(t)$ при различной освещенности фотодиода.
Коэффициент усиления напряжения $Y=10^6$. 
Усиленное напряжение $YU(t)$ подается на устройство,
что приводит к усредненному по времени значению квадрата напряжения
$U_\mathrm{S}=\left<(Y U(t))^2\right>/U_0$.
Это напряжение $U_\mathrm{S}$ измеряется с помощью мультиметра.
Последовательность измерений позволяет определить значение~$q_e$,
используя хорошо известную формулу Шоттки для спектральной плотности
флуктуаций тока
$(I^2)_f=2q_e\left<I\right>$.
Для младшей группы школьников основная задача заключается в анализе квадратичной зависимости.
По сложности задания разделены на 4~категории: S, M, L и XL в соответствии с возрастом участников.
Для самых старших участников задание содержит вопросы, относящиеся к университетской программе по физике, включая теоретическое исследование устройства представленной печатной платы с точки зрения ее функционального назначения.
Это задание олимпиады EPO6 ``День заряда'' (декабрь 2018).
EPO6 организовано Софийским отделением Болгарского союза физиков
в сотрудничестве с Физическим факультетом Софийского университета
и Союзом физиков Республики Македонии.

\setlength{\leftskip}{0cm}
\setlength{\rightskip}{0cm}


\section{Введение}

С первых лет своего существования Олимпиада по экспериментальной физике (EPO)
получила мировую известность;
все задания Олимпиад были опубликованы
в Интернете~\cite{EPO1,EPO2,EPO3,EPO4,EPO5},
а число участников составляло 120~человек.
В недавних Олимпиадах приняли участие школьники старших классов
из 7~стран, любопытно, что расстояние между наиболее удаленными
городами более чем 4~Мм.

В чем основные отличия EPO от подобных соревнований?
\begin{itemize}
\item Экспериментальную установку, с которой работают участники EPO,
  все они получают в качестве подарка от организаторов.
  Таким образом, после окончания Олимпиады
  даже неудачно выступившие участники могут самостоятельно повторить эксперимент
  и достичь уровня победителя.
  Так Олимпиада повышает уровень преподавания в мировом масштабе.
  По окончанию учебного года экспериментальная установка остается в той школе,
  где обучался участник Олимпиады.

\item Каждое из заданий оригинально и связано
  с фундаментальной физикой или физическими принципами технического патента.

\item Олимпийская идея воплощена в EPO в своей изначальной форме:
  каждый желающий может принять участие в соревновании вне зависимости
  от места своего проживания.
  У нас нет ограничения на число участников.
  С другой стороны, сходство с другими Олимпиадами в том,
  что задания являются прямой иллюстрацией изучаемого предмета,
  и олимпиада EPO наряду с подобными соревнованиями
  смягчает повсеместную тенденцию к деградации школьного образования.

\item Всем участникам выдается одинаковое экспериментальное оборудование,
  но сами задания различаются в зависимости от возраста участников,
  так же как и в плавательном бассейне:
  вода одинаково мокрая для всех возрастных категорий спортсменов.
\end{itemize}

Мы кратко упомянем задания пяти предыдущих Олимпиад:
1)~На EPO1 установка представляла собой студенческую версию
американского патента стабилизированного усилителя постоянного тока
с технологией "auto-zero" \cite{EPO1}.
2)~Задание EPO2~\cite{EPO2} заключалось в измерении постоянной Планка
путем наблюдения дифракции света, испущенного светодиодом, на лазерном компакт-диске.
3)~Современная реализация патента NASA на использование преобразователя с отрицательным импедансом для генерации колебаний напряжения являлась заданием на~EPO3~\cite{EPO3}.
4)~EPO4~\cite{EPO4} была посвящена фундаментальной физике~--- определению скорости света путем измерения электрических и магнитных сил.
Инновация заключалась в приложении теории катастроф
к анализу устойчивости маятника.
5)~Темой EPO5 было измерение постоянной Больцмана~$\kb$, восходящее к идее Эйнштейна об исследовании тепловых флуктуаций величины электрического напряжения в конденсаторе.

Олимпиада EPO6 продолжает традицию измерения фундаментальных постоянных.
Воодушевленный лекцией Эйнштейна о флуктуациях Вальтер Шоттки понял,
что заряд электрона также может быть определен при исследовании шумов напряжения.
Сейчас, век спустя, благодаря появлению малошумящих операционных усилителей
идея Шоттки может быть реализована в качестве задания для учеников выпускных классов.

Если кратко, то установившиеся традиции~--- есть равновесное состояние на границе современного технического прогресса и фундаментальной физики.

\section{Экспериментальная установка}

Экспериментальная установка состоит из следующих частей:
\begin{enumerate}

\item Печатная плата (PCB) с двойным разъемом с фотодиодом  и держателем для литиевой батарейки.

\item Четыре батарейки 9~V и два двухконтактных разъема для них.

\item Три батарейки 1.5~V и держатель для них с впаянным потенциометром.

\item Пластиковый пакет со следующими деталями:
\begin{itemize}

\item Два операционных усилителя на платах с зелеными этикетками с подписанными номерами.

\item Усилитель на плате с оранжевой этикеткой.
\item Литиевая батарейка.
\end{itemize}

\item Оптоволоконный световод с белым цилиндром на одном из концов.

\item Пластиковый стаканчик с поперечно приклеенной трубочкой.
\end{enumerate}

С собой Вы должны принести два дополнительных мультиметра с щупами и калькулятор.

\section{Первые простые задания. S}
\begin{enumerate}

\item Установите мультиметры в режим измерения напряжения (как вольтметры).

\item Измерьте напряжение каждой из 4~батареек с максимальной точностью (используйте диапазон 20~В вольтметра).
  Тоже самое выполните для трех батареек 1.5~В (используя диапазон 2000~мВ вольтметра).
\item Поместите три батарейки 1.5~В в держатель батареек.
Подключите потенциометр \emph{потенциометрически} к одному из ваших мультиметров.
Покрутите ось потенциометра.
Измерьте полученный интервал напряжений.
Это будет вашим источником напряжения для следующих заданий.
Изменение полярности потенциометра меняет знак напряжения на противоположный.

\item Присоедините четыре батарейки 9~В к двум сдвоенным коннекторам, защелкнув электроды.

\section{Исследование квадратичной зависимости. M}
\label{s:M_r}

\item \textbf{Внимание! С этого момента есть риск сжечь интегральные схемы, если батереи подсоединены неверно!}
Расположите установку таким образом, что оба провода ``OUT'' и ``COM'' на краю платы были с правой стороны, как на рис.~\ref{Fig:Board_r}.
\item Аккуратно и соблюдая полярность подключите батарейку 9~В к источнику напряжения через трехконтактный разъем (папа) на печатной плате, наклейка к наклейке (обе наклейки~--- на разъеме и на плате~--- должны оказаться рядом).
\textbf{Действуйте аккуратно~--- если Вы ошибетесь с полярностью, могут сгореть мультиптикатор и операционный усилитель.}
\item Подсоедините средний контакт потенциометра к входному проводу,
расположенному наверху установки с меткой ``IN''.
Используйте провод с клеммами типа ``крокодил''.
\item Другой контакт потенциометра подсоедините к проводу, подходящему к ``земле'' цепи (метка ``COM''). 
\item Подсоедините первый вольтметр V$_1$, измеряющий напряжение $U_1$, параллельно ко входам потенциометра ``IN''--``COM''.
\item При вращении оси потенциометра напряжение $U_1$ должно изменяться приблизительно от 0 до полного напряжения батареи 4.5~В.
\item Подсоедините второй потенциометр V$_2$, показывающий напряжение $U_2$, между проводом--выходом цепи `OUT'' и общим контактом ``COM''.
Таким образом, $U_2$~--- это напряжение между точками ``OUT'' и ``COM''.
\item Убедитесь, что контакты ``COM'' обоих мультиметров и установки подсоединены правильно.
\item Восьмиконтактные разъемы с оранжевыми наклейками под ними (на правой стороне печатной платы) обозначены ``AD633'' на Рис.~\ref{Fig:Board_r}.
Подсоедините мульпликатор к ``AD633'' таким образом, чтобы \textbf{оранжевые наклейки на мультипликаторе оказались рядом с оранжевыми наклейками на плате}.
\item Поверните ось потенциометра, подождите 1 минуту и запишите значения напряжений $U_1$ и $U_2$.
Измените полярность источника напряжения и повторите измерения с входным напряжением $U_1$ и выходным напряжением $U_2$.
Запишите результаты измерений в таблицу со следующими столбцами:
порядковый номер измерения $i$, $U_1$ и $U_2$. 
\item Представьте результаты на графике, где $U_1$~--- на оси абсцисс (горизонтальная координата) и $U_2$~--- на оси ординат (вертикальная координата) на миллиметровой бумаге.
\item К таблице добавьте столбец $(U_1)^2$ и представьте результаты на графике: $(U_1)^2$~--- абсцисса (горизонтальная ось) и $U_2$~--- ордината (вертикальная ось) на миллиметровой бумаге.
Проведите прямую, проходящую как можно лучше через экспериментальные точки.
Эта прямолинейная (аппроксимирующая) зависимость описывается уравнением
$U_2= (U_1)^2/U_0 + \mathrm{const}$.
Выберите две точки на прямой, измерьте разность абсцисс~$\Delta(U_1^2)$ и ординат~$\Delta(U_2)$,
определите наклон прямой
$U_0=\Delta(U_1^2)/\Delta(U_2)$.
Этот параметр $U_0$, имеющий размерность напряжения (В), необходим для определения заряда электрона~$q_e$, как описано в следующем задании.

\begin{figure}[h]
\centering
\includegraphics[scale=0.3]{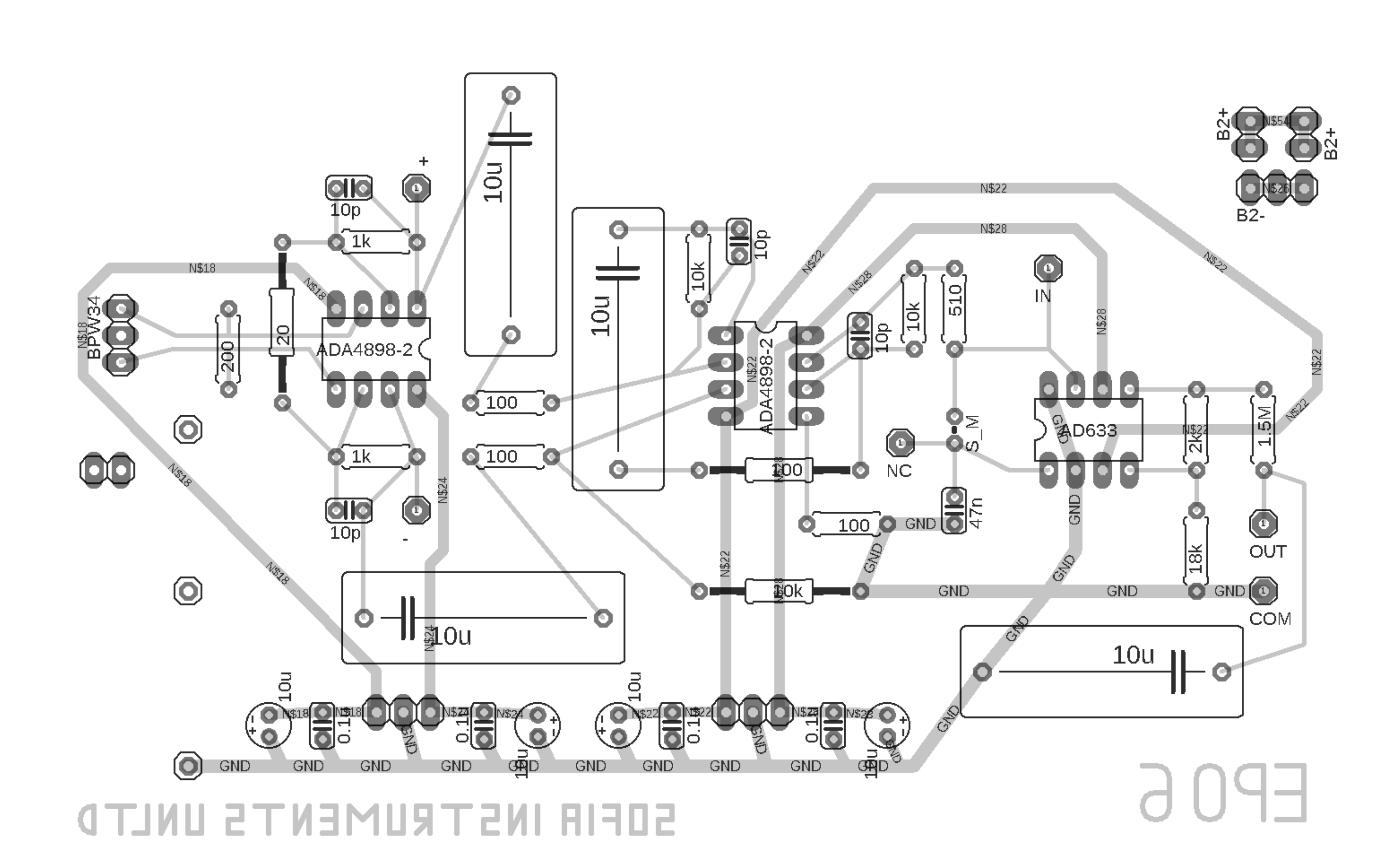}
\caption{Схема печатной платы для экспериментальной установки.
  Лампа накаливания с присоединенным белым цилиндром не показана,
  она впаяна между контактами ``B2-'' и ``B2+'' (не подписаны на печатной плате) в правом верхнем углу платы.
  Два неподписанных разъема на три контакта, расположенные внизу платы,~--- это контакты для подачи напряжения.}
\label{Fig:Board_r}
\end{figure}

\section{Измерение флуктуаций (флуктоскопия) напряжения. Определение заряда электрона~$q_e$. L} 
\item Поместите 3 батареи 1.5~В в держатель батарей. 
  При помощи проводов с двумя крокодилами подсоедините два внешних контакта потенциометра к контактам лампы ``B2-'' и ``B2+'' в правом верхнем углу установки, показанном на рис.~\ref{Fig:Board_r}.
В этом случае полярность неважна, при правильном соединении лампа должна загореться.
\item С помощью провода с ``крокодилами'' соедините провод~``($-$)'' установки с точкой ``COM'' вольтметра~V$_1$.
\item Аналогично другим проводом с ``крокодилами'' соедините контакт ``($+$)'' установки с  
``$\mathrm{V\,\Omega\, m\!A}$'' входом вольтметра V$_1$.
В этом случае V$_1$ показывает разность потенциалов $U_{\pm}$ между точками ``($+$)''  и ``($-$)''. Это напряжение пропорционально среднему фототоку фотодиода.
\item Поместите литиевую батарею 3~В в соответствующий держатель на двойном разъеме платы с припаянным фотодиодом (слева платы),
\textbf{``($+$)'' батареи должен быть соединен с ``($+$)'' на держателе батареи.}
\item Два 8-контактных разъема с левой стороны и в центре печатной платы
  с зелеными наклейками помечены как ``ADA4898-2'' на Рис.~\ref{Fig:Board_r}.
Подсоедините два операционных усилителя к двум ``ADA4898-2'' таким образом, чтобы \textbf{зеленая наклейка каждого операционного усилителя располагалась рядом с зеленой наклейкой на плате. Будьте осторожны: перевернутое подсоединение может привести к повреждению усилителя.}
\item Аккуратно подсоедините разъем с припаянными фотодиодом и литиевой батарейкой к контактам на плате в месте ``BPW34'' на рис.~\ref{Fig:Board_r} печатной платы.
Аккуратно вставьте световод в черную трубочку в пластиковом стакане.
Белый цилиндр должен вставляться в черную трубочку с более длинной стороны.
Убедитесь, что световод может двигаться внутри трубочки без значительного трения. 

\item Вставьте другой конец световода в белый цилиндр, присоединенный к лампочке.
\item Вставьте фотодиод в черную трубочку с короткой стороны, одев пластиковый стакан на фотодиод, припаянный к разъему на плате.
\item Подсоедините один из контактов батареи 9~В к левому 3-контактному разъему на печатной плате.
\textbf{Будьте внимательны! Оранжевые наклейки на контактах батареи должны совпасть с оранжевыми наклейками на плате.}
Переключите вольтметр в диапазон 2000~мВ и подвигайте конец световода с белым цилиндром внутри трубочки.
Соответствующее напряжение $U_\pm$ должно изменяться вплоть до 1000~мВ.
Таким способом, двигая световод, Вы изменяете фототок и измеряете напряжение~$U_\pm$, создаваемое фототоком, протекающим через резистор~$R$.
Если Вам не удается достичь значений по-крайней мере в 700~мВ или совсем нет напряжения, попросите помощи у помощников в аудитории.
Это важный предварительный этап измерений в эксперименте.
\textbf{Не пытайтесь двигать белый цилиндр, присоединенный к лампе, так как его легко повредить.}
\item Попросите лёд и воду, для того чтобы поместить их в пластиковый стакан с черной трубочкой. Вставьте светодиод со стороны трубочки, противоположной фотодиоду.
\item После того, как вода и лёд помещены в стакан, убедитесь, что нет никаких протечек. Если протечки все же есть, аккуратно снимите и попросите заменить стакан.

\item Присоедините ``COM'' выход карты ко входу ``COM'' вольтметра V$_2$ (вольтметр V$_2$ должен быть в точности в том состоянии, как в части~\ref{s:M_r}).
\item Соедините выход ``OUT'' печатной платы с входом ``$\mathrm{V\,\Omega\, m\!A}$'' вольтметра V$_2$.
Теперь вольтметр V$_2$ измеряет напряжение $U_\mathrm{S}$, которое пропорционально дробовому шуму фотодиода.
\item Подсоедините другой контакт батареи 9~В к 3-контактному разъему источника напряжения на правой стороне печатной платы.\\
\textbf{Будьте внимательны! Оранжевые наклейки на подсоединяемом разъеме должны совпасть с оранжевыми наклейками на плате.}
\item Попросите у помощников в аудитории клейкую ленту (scotch tape) и приклейте все 6~проводов к столу в следующем порядке:
\begin{itemize}
\item Во-первых, поместите все 4 батареи 9~В ниже печатной платы.
\item Приклейте провода ``+'' и ``-'' выше платы, убедившись, что между ними нет контакта.
\item Приклейте провода ``IN'' и ``NC'' выше и дальше от платы, чем провода ``+'' и ``-''.
\item Приклейте провода ``OUT'' и ``COM'' с правой стороны печатной платы, убедившись, что между ними нет контакта.
\end{itemize}
Теперь Ваша экспериментальная установка похожа на паука.
\item Посмотрите на измеренное напряжение V$_2$, оно должно стабилизироваться около значения в несколько десятков мВ.
\item Начните этап измерения $U_\pm$ и $U_\mathrm{S}$: изменяйте $U_\pm$, аккуратно передвигая световод внутри трубочки ближе к фотодиоду.
  Шаг изменения $U_\pm$ между последовательными измерениями должен быть не менее 100~мВ.
  Ожидайте по крайней мере 2~минуты, чтобы значение $U_\mathrm{S}$ стабилизировалось.
  \emph{Если из-за какого-либо возмущения, показания вольтметра существенно изменились, снова подождите 2~минуты для стабилизации.}
Поместите результаты в таблицу со следующими столбцами:
порядковый номер измерения $i$, $U_\pm$ и $U_\mathrm{S}$. 
\item Представьте результаты на графике, где $U_\pm$~--- абсцисса (горизонтальная ось) и $U_\mathrm{S}$~--- ордината (вертикальная ось) на миллиметровой бумаге (если Вам нужна дополнительная миллиметровая бумага, попросите у помощников).
Проведите прямую, наилучшим образом описывающую линейную зависимость
$U_\mathrm{S}=k U_\pm +\mathrm{const}$.
Выберите 2~точки на прямой и определите ее наклон
$k=\Delta U_\mathrm{S}/ \Delta U_\pm$,
где $\Delta$ означает разность.
Эта математическая процедура называется линейная регрессия.
Для нашей установки $k\simeq 10^{-3}.$
\item Наконец, определите заряд электрона~$q_e$,
воспользовавшись формулой
\begin{equation}
q_e=2 k \frac{y_1}{Y^2} \frac{R_\mathrm{_L}}{R} C_\mathrm{_L}U_0,
\label{electron_charge_r}
\end{equation}
где $Y=1.01\times 10^{6}$~--- полное усиление нашего усилителя,
и $y_1=101$~--- усиление первого каскада мультипликатора.
Вы можете найти значение $C_\mathrm{_L}$, написанное на наклейке в правом нижнем углу печатной платы ($C \equiv C_\mathrm{_L}$), $R_\mathrm{_L}=510~\Omega$ и $R=200~\Omega$. \\
Поздравляем! Вы только что измерили фундаментальную константу и Вы в отличной компании! :)
\item Ваше измерение могло быть чуть точнее, если учесть эффект неидеальности  операционного усилителя, использовав формулу
\begin{equation}
q_e=2 k \frac{y_1}{(1-\varepsilon)Y^2} \frac{R_\mathrm{_L}}{R} 
C_\mathrm{_L}U_0.
\label{electron_charge_mod_r}
\end{equation}
Небольшая поправка $\varepsilon(C_\mathrm{_L})$ является функцией емкости $C_\mathrm{_L}$, как показано на Рис.~\ref{Fig:Eps_r}.
\begin{figure}[h]
\centering
\includegraphics[scale=0.8]{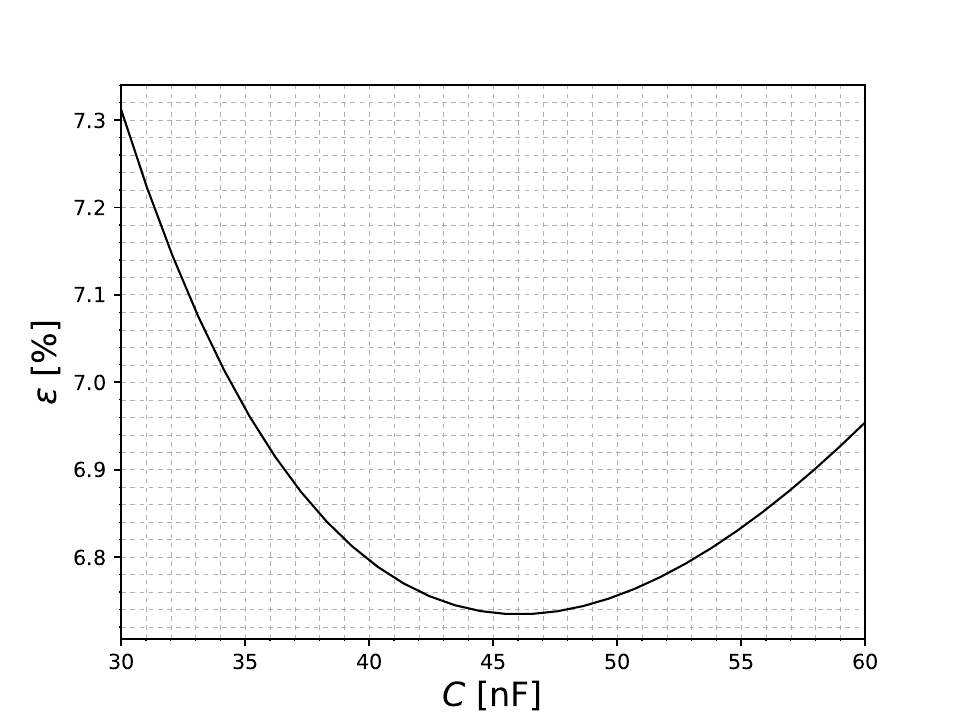}
\caption{Ошибка $\varepsilon$ (в процентах) в измерении $q_e$ как функции емкости конденсатора $C_\mathrm{_L} \equiv C$.}
\label{Fig:Eps_r}
\end{figure}
Определите ошибку $\varepsilon$ из Рис.~\ref{Fig:Eps_r}, воспользовавшись значением $C_\mathrm{_L}$, написанным на Вашей печатной плате и не забудьте поделить $\varepsilon$ на 100, прежде чем посчитать $q_e$ из уравнения~(\ref{electron_charge_mod_r}).

Сто лет назад этот метод определения заряда электрона~$q_e$
путем измерения дробового шума (шума Шоттки)
был предложен Вальтером Шоттки.\cite{Schottky:1918}
В то время Вальтер Шоттки работал с Максом Планком
и был воодушевлен лекцией Альберта Эйнштейна об электрических флуктуациях.

\section{Домашнее задание. XL}
Выведите уравнения~(\ref{electron_charge_r}) 
и (\ref{electron_charge_mod_r}), используя схему и значения параметров из таблицы~\ref{tbl:values_r}. 
Рассчитайте усиление~$y_1$ повторителя напряжения и полное усиление~$Y$ на всех ступенях усилителя. 
Различные части схемы представлены на Рис.~\ref{Non-inverting amplifier_r}, \ref{Buffer_r}, 
\ref{Differential amplifier_r},
\ref{Inverting amplifier_r},
\ref{Analog multiplier_r}.

\begin{center}
\begin{table}[h]
\begin{tabular}{| c | r |}
		\hline
		&  \\ [-1em]
		Элемент цепи  & Значение  \\ \tableline
			&  \\ [-1em]
			$R$ &200~$\Omega$ \\
			$r_\mathrm{_G}$ & 20~$\Omega$ \\
			$R_\mathrm{F}$ &  1~k$\Omega$  \\
			$C_\mathrm{F}$ &  10~pF  \\ 
			$C_\mathrm{G}$ & 10~$\mu$F \\
			$R_\mathrm{G}$ &  100~$\Omega$  \\ 
			$R_\mathrm{F}^\prime$ & 10~k$\Omega$ \\
			$C_\mathrm{F}^\prime$ & 10~pF \\
			$R_\mathrm{_L}$ & 510~$\Omega$ \\
			$R_1$ &  2~k$\Omega$  \\ 
			$R_2$ & 18~k$\Omega$  \\
			$R_\mathrm{av}$ & 1.5~M$\Omega$ \\
			$C_\mathrm{av}$ & 10~$\mu$F \\
			$R_\mathrm{_V}$ & $\approx 1~\mathrm{M} \Omega$ \\
			$V_\mathrm{CC}$ & +9~V \\
			$V_\mathrm{EE}$ & -9~V \\
\tableline
\end{tabular}
	\caption{Значения элементов цепей.}
	\label{tbl:values_r}
\end{table}
\end{center}

\begin{figure}[t]
\begin{minipage}[t]{0.31\linewidth}
\includegraphics[scale=0.28]{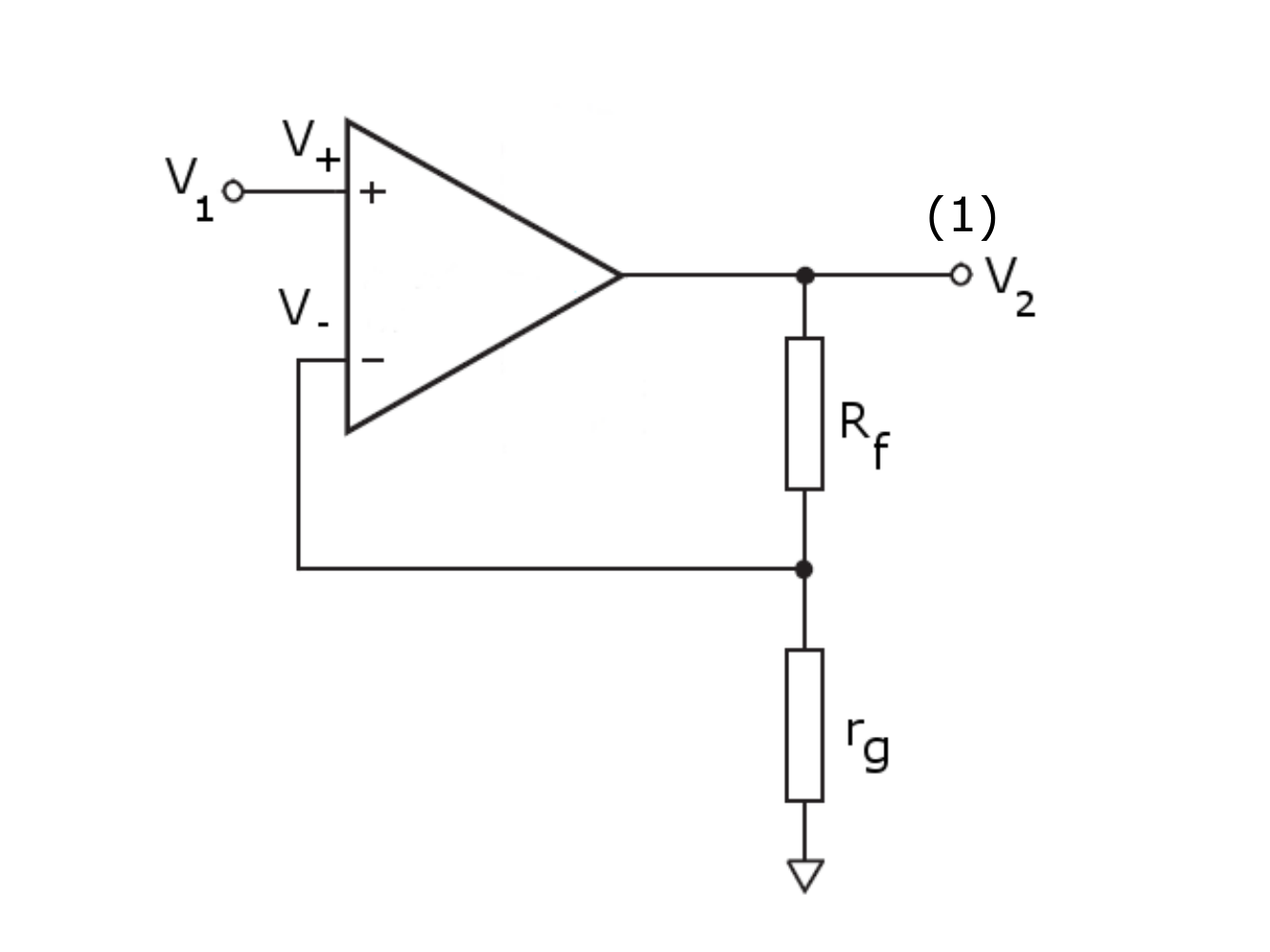}
\caption{Неинвертирующий усилитель}
\label{Non-inverting amplifier_r}
\end{minipage}
\begin{minipage}[t]{0.31\linewidth}
\includegraphics[scale=0.28]{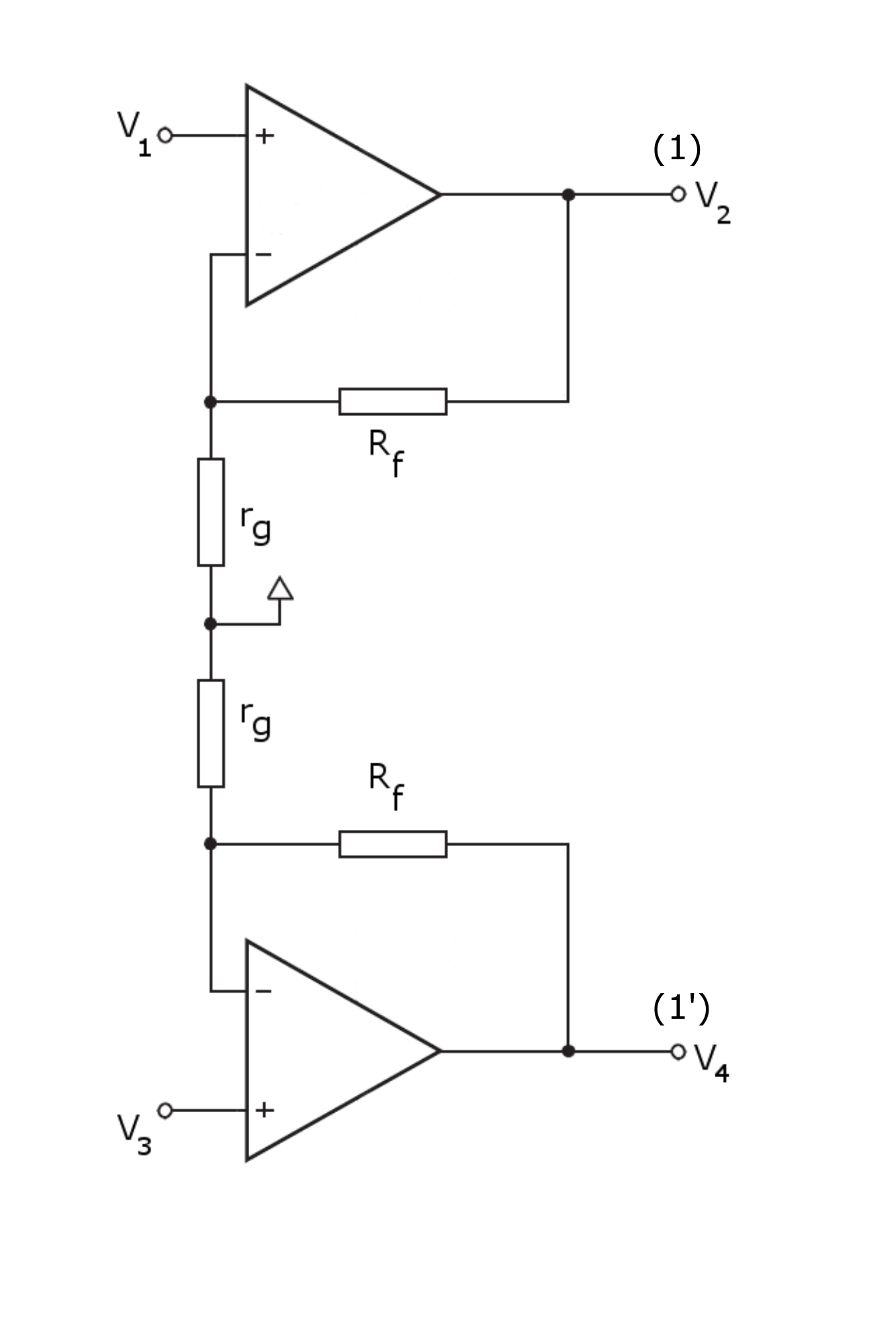}
\caption{Буфер}
\label{Buffer_r}
\end{minipage}
\begin{minipage}[t]{0.36\linewidth}
\includegraphics[scale=0.28]{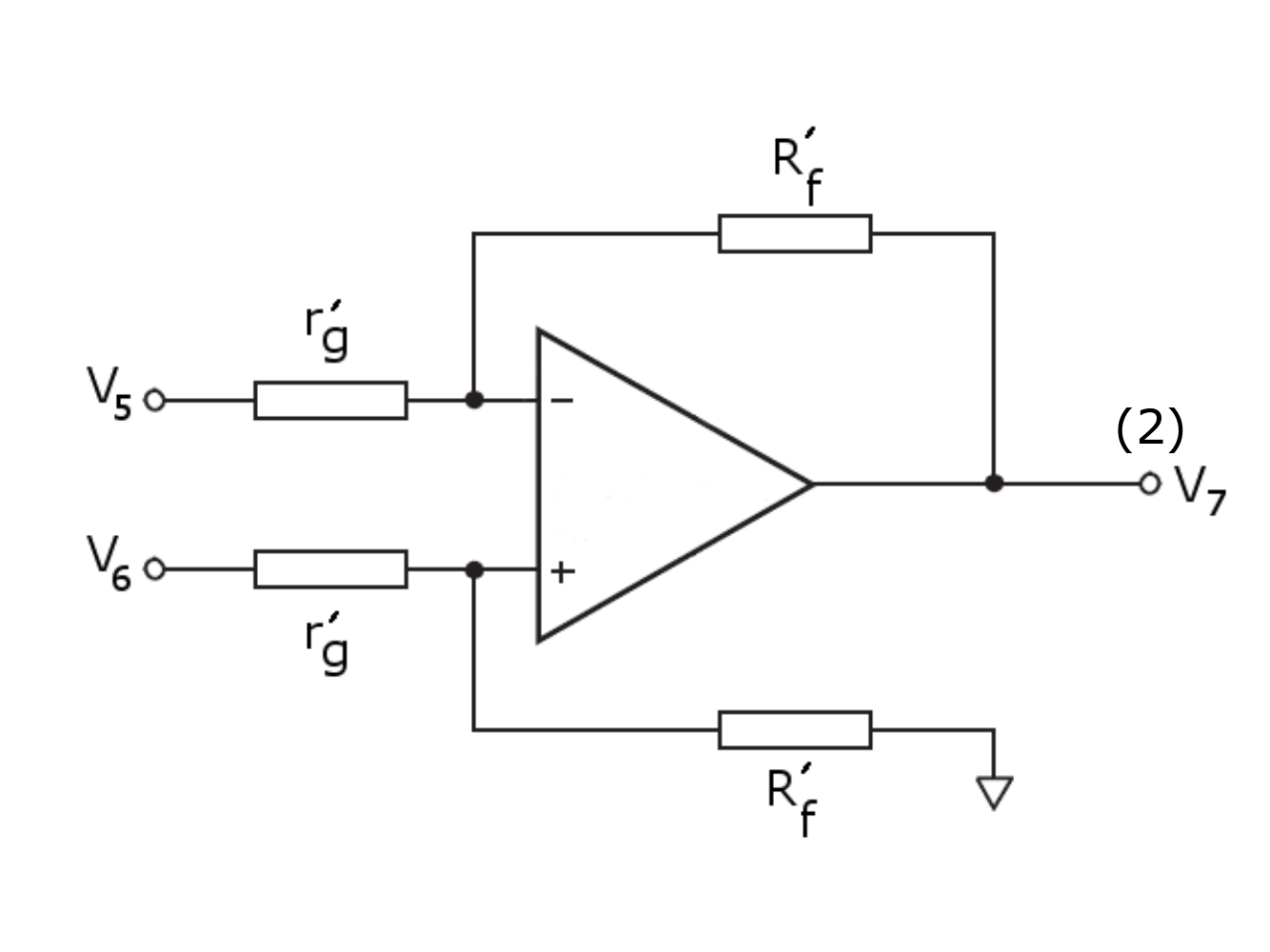}
\caption{Дифференциальный усилитель}
\label{Differential amplifier_r}
\end{minipage}
\begin{minipage}[c]{0.4\linewidth}
\includegraphics[scale=0.28]{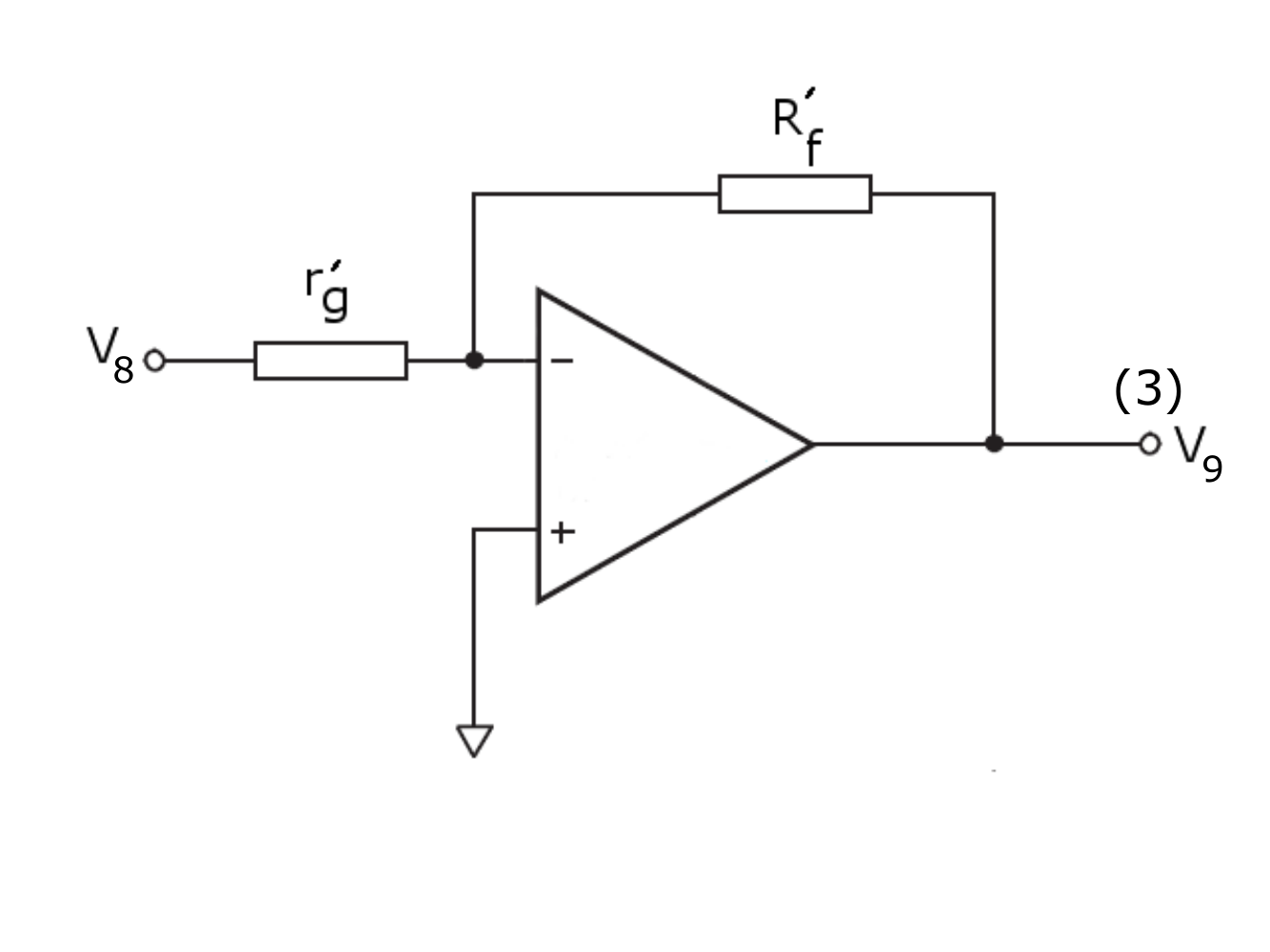}
\caption{Инвертирующий усилитель}
\label{Inverting amplifier_r}
\end{minipage}
\begin{minipage}[c]{0.57\linewidth}
\includegraphics[scale=0.28]{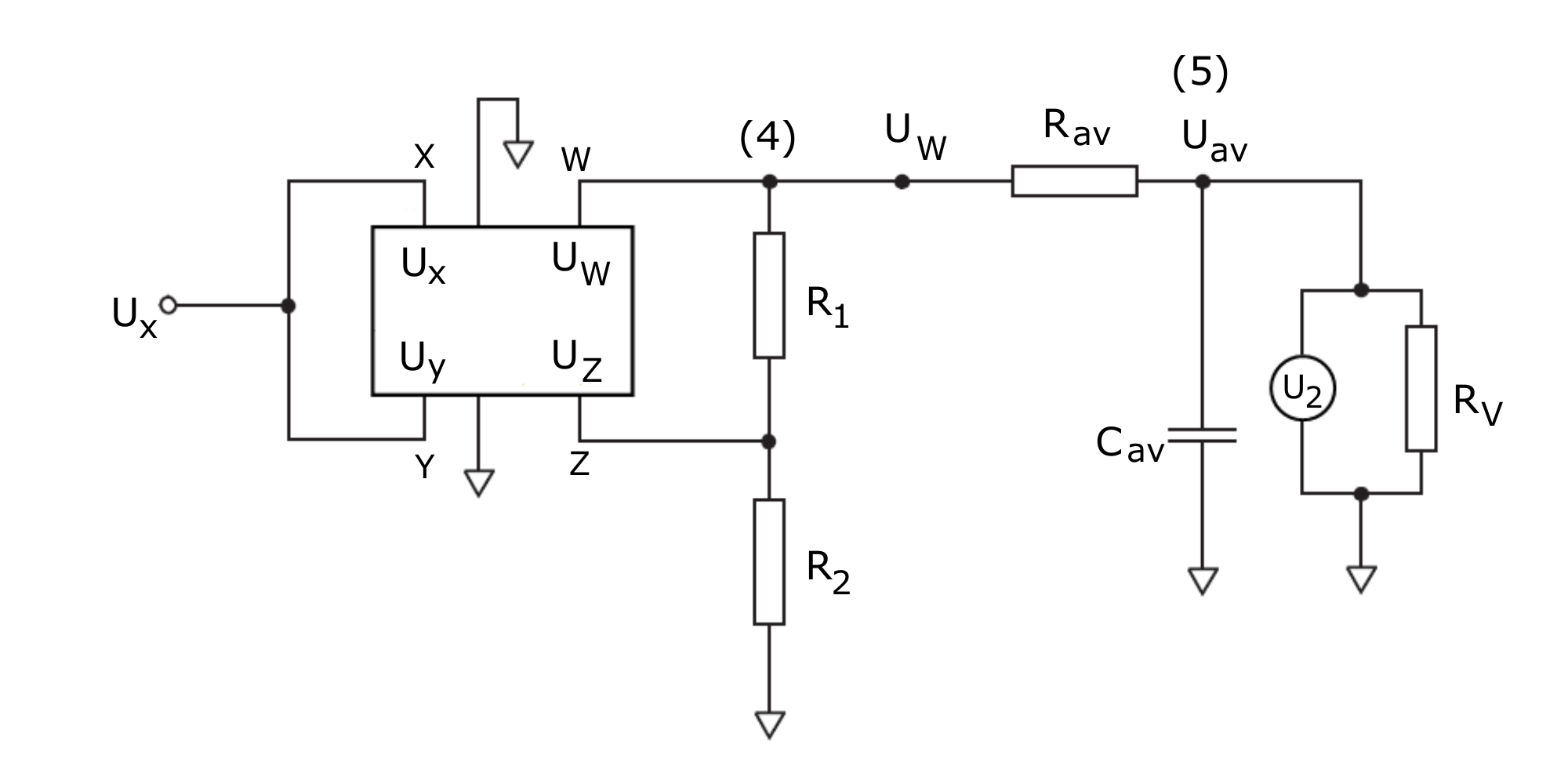}
\caption{Аналоговый усилитель}
\label{Analog multiplier_r}
\end{minipage}
\end{figure}

Решение должно быть отправлено в течение ночи после олимпиады, не позднее восхода, на адрес Олимпиады epo@bgphysics.eu.
Лучшее решение будет награждено призом Зоммерфельда с денежным эквивалентом
DM137.


\section{Вопросы для дальнейшей работы}
Экспериментальная установка в действительности представляет собой синхронный вольтметр (lock-in voltmeter), при помощи которого Вы сможете измерять переменные сигналы, меньшие чем $\mathrm{\mu V}$.
Вы можете сообщить нам, как использовать выход ``NC'' в этом случае.
\end{enumerate}

%% file: EPO6_de_a.tex
\clearpage
\renewcommand{\figurename}{Abb.}
\renewcommand{\tablename}{Тab.}
\begin{center}
\textbf{Messung der Elementarladung $q_e$ mit Hilfe des Shottky-Rauschens; 
Problem der 6.-Experimentalphysikolympiade, 8.12.2018, Sofia}
\end{center}
\normalsize
\indent
\setlength{\leftskip}{1.2cm}
\setlength{\rightskip}{1.2cm}
Diese Anleitung erkl\"art, wie man mit Hilfe eins PCBs (printed circuit board) die Elementarladung 
$q_e$ 
bestimmt. 
Die Grundidee basiert auf der sogenannten Shottky-Gleichung. 
F\"allt Licht auf eine Photodiode, entsteht dabei ein Photostrom mit einem charakteristischen Rauschen, dem Shottky-Rauschen. 
Dieser Strom wird an einem Widerstand in eine Spannung $U(t)$ umgewandelt und anschlie\ss end millionenfach verst\"arkt 
$y\approx{}10^6$. 
Die verst\"arkte Spannung $yU(t)$ wird zun\"achst mit einem Multimeter gemessen und anschlie\ss end von einem 
Multiplizierer quadriert. Diese Spannung $U_S=\left<(y U(t))^2\right>/U_0$ wird erneut durch ein Multimeter gemessen. 
Indem man die Helligkeit des Lichts an der Photodiode variiert, erh\"alt man eine Messreihe mit verschiedenen 
Werten von $U_S$ und den korrespondierenden Werten von $U_\pm$. 
Auf Grund der Shottky-Gleichung ist die aus $\frac{\Delta U_S}{\Delta U_\pm}$ resultierende Steigung proportional zu $q_e$.
Die Sch\"uler werden in vier Altersgruppen S, M, L, XL aufgeteilt. W\"ahrend  j\"ungere Sch\"uler nur die Funktionsweise des Multiplizieres untersuchen, 
ist die die letzte Kategorie XL am Universit\"atsstoff orientiert und beinhaltet theoretische Untersuchungen am PCB-Setup. 
EPO6 wird von einem Ast der Physikerunion in Bulgarien organisiert in Kooperation mit der physikalischen Fakult\"at der Universit\"at 
in Sophia und der physikalischen Gesellschaft von Makedonien.

\setlength{\leftskip}{0cm}
\setlength{\rightskip}{0cm}
\section{Einleitung}

Seit ihrem Beginn ist die experimentell Physikolympiade (EPO) weltweit bekannt. Sie hatte ungef\"ahr 120 Teilnehmer. 
Alle bisherigen Aufgabenstellungen wurden im Internet ver\"offentlicht \cite{EPO1,EPO2,EPO3,EPO4,EPO5}. 
In den letzten Jahren haben Sch\"uler aus 7 L\"andern teilgenommen und manche mussten mehr als 4000~km zur\"ucklegen.

Was ist der Hauptunterschied zwischen EPO und anderen \"ahnlichen Wettbewerben?
\begin{itemize}
\item{} Jeder Teilnehmer erh\"alt das Setup, mit dem er gearbeitet hat, als ein Geschenk von den Organisatoren. Dadurch k\"onnen alle Teilnehmer, selbst wenn sie nicht ganz erfolgreich waren, nach dem Wettbewerb das Experiment erneut durchf\"uhren. Am Ende des Schuljahres bleibt das Setup in der Schule der Teilnehmers.
\item{} Jede Aufgabe wurde von uns selbst erstellt und hat eine Verbindung zur Fundamentalphysik, oder hilft beim Verstehen eines technischen Patentes.
\item{} Die olympische Idee wird durch EPO in seiner Ursprungsform realisiert. Jeder, der Lust hat mitzumachen, kann das tun. Es gibt keine Obergrenze f\"ur die Teilnehmerzahl. Allerdings gibt es auch eine Gemeinsamkeit mit anderen Olympiaden: die Aufgabenstellungen sind direkte Illustrationen von Unterrichtsmaterial, wodurch die weltweite Tendenz der immer schlechter werdenden weiterf\"uhrenden Schulen ausgeglichen werden soll.
\item{} Jeder Teilnehmer erh\"alt das gleiche Setup, allerdings bekommen verschiedene Altersgruppen verschiedene Aufgabenstellungen.
\end{itemize}

An dieser Stelle m\"ochten wir noch kurz die Aufgabenstellungen der letzen EPOs auflisten:
1)     Das Setup von EPO1 war eine Sch\"ulerversion des amerikanischen Patents f\"ur auto-zero und chopper stabilisierte Verst\"arker.
2)      Das Problem von EPO2 war das Messen  der Plank-Konstante durch die Beugung von Licht einer LED an einer CD.
3)      Das Setup f\"ur EPO3 war das kurzfristige Umsetzen eines NASA-Patents, das als negativer Impedanzwandler zum Erzeugen von Spannungsoszillationen genutzt wird.
4)      EPO4 war der Fundamentalphysik gewidmet: Das Messer der  Lichtgeschwindigkeit, durch das Messen von elektrischen und magnetischen Kr\"aften. Das innovative Element war die Anwendung der Chaostheorie in der Analyse der Stabilit\"at eines Pendels.
5)      Das Thema der EPO5 war das Messen der Boltzmann-Konstante, mit Einsteins Idee das thermische Rauschen der Spannung eines Kondensators zu nutzen.
Die aktuelle EPO6 folgt der Tradition der Olympiaden eine Fundamentalkonstante zu messen. Durch eine Vorlesung von Einstein kam Walter Shotky darauf, dass die Elementarladung durch Spannungsrauschen bestimmt werden kann. Jetzt, ein Jahrhundert sp\"ater, 
kann Shottkys Idee von Sch\"ulern umgesetzt werden mit Hilfe von kaum rauschenden Operationsverst\"arkern. Zusammengefasst: die ausgebildete Tradition ist eine Mixtur zwischen aktuellen technischen Neuerfindungen und Fundamentalphysik.

\section{Experimenteller Aufbau}

Dein Setup sollte aus folgenden Gegenst\"anden bestehen:
\begin{enumerate}
\item{} PCB (printed circuit board) mit zweireihigen Leiterplatinensteckverbinder, einer Photodiode und einem Lithiumbatteriehalter
\item{} Vier 9~V Batterien und zwei Verbindungsclips f\"ur diese
\item{} Drei 1.5~V Batterien und ein Halter, an dem ein Potentiometer befestigt ist.
\item{} Plastikt\"ute, die folgende Gegenst\"ande enth\"alt:
\begin{itemize}
\item{}  Zwei Operationsverst\"arker auf einer Verbindungsplatine (SOIC-to-DIP-Adapter) mit jeweils einem gr\"unen Label mit einer Zahl darauf
\item{} Multiplizierer auf einer Verbindungsplatine mit einem orangenen Label
\item{} Lithiumbatterie
\end{itemize}
\item{} Optisches Glasfaserkabel mit einem wei\ss en Zylinder am Ende
\item{} Eine Plastiktasse mit einem dar\"uber geklebten Strohalm.
\end{enumerate}
Du solltest zus\"atzlich zwei Multimeter haben mit Verbindungskabeln und einen Taschenrechner.

\section{Aufgabe S: Einstieg}
\label{s:S_d}
\begin{enumerate}
\item  Stelle die Multimeter so ein, dass sie die Spannung messen.
\item  Miss die Spannungen der vier 9~V Batterien mit maximaler Genauigkeit und schreibe diese auf. Mache das selbe mit den drei 1.5~V  Batterien.
\item Platziere die drei 1.5~V Batterien in ihrem Halter. Verbinde das Potentiometer mit einem der Multimeter, indem 
du dessen Spannung an einer der \"au\ss eren Eektroden und der Elektrode in der Mitte misst. Rotiere die Axe des
 Potentiometers. Miss das Intervall der Spannung, welche du erh\"altst. Dieses Intervall wird deine 
 Spannungsquelle f\"ur deine n\"achste Aufgabe sein. Wenn du die Polarit\"at umdrehst, \"andert sich 
 das Vorzeichen der Spannung.
\item Verbinde die vier 9~V Batterien mit den zwei Verbindungsclips.

\section{Aufgabe M:   Multiplizierer}
\label{s:S_d}
\item \textbf{Vorsicht! Ab jetzt besteht die Gefahr, den IC (integrated circuit) kaputt zu machen, indem du die Batterien falsch verbindest.} 
Orientiere das Setup mit dem du arbeitest so, dass beide Kabel ``OUT`` und ``COM``  an der Kante auf der rechten Seite sind, schau dir   Abbildung ~\ref{Fig:Board_d} an.
\item Verbinde die 9~V-Batterie-Verbindungsclips vorsichtig mit dem rechten Spannungsversorger. Dieser ist ein Dreipin-male-Stecker auf der PCB-Platine, der mit einem orangenen Kleber gekennzeichnet ist. 
\textbf{Vorsicht: Die beiden orangenen Kleber m\"ussen zueinander zeigen.}
\item 
Verbinde den Mittelkontakt des Potentiometers mit dem Eingangskabel oben am Setup mit dem Label  ``IN``. Nutze die ``Krokodilkabel``.
\item 
Verbinde die andere Elektrode des Potentiometers mit dem Kabel, das vom ``ground`` des Schaltkreises, mit der Beschriftung ``COM`` kommt.
\item 
Verbinde das erste Voltmeter $V_1$ mit Spannung $U_1$ mit dem Potentiometer und den ``IN``--``COM``      Eing\"angen gleichzeitig.
\item 
Wenn du die Potentiometeraxe rotierst, sollte die Spannung U1 sich ungef\"ahr zwischen 0 und der absoluten Batteriespannung 4.5~V bewegen.
\item 
Verbinde das zweite Voltmeter $V_2$, das die Spannung $U_2$ angibt, zwischen dem Ausgangskabel des Schaltkreises, 
das mit ``OUT`` beschriftet ist, und der Erdung ``COM``. $U_2$ ist also die Spannung zwischen ``OUT`` und ``COM``.
\item 
\"Uberpr\"ufe, ob die ``COM``-Elektroden beider Multimeter mit dem Setup richtig verbunden sind.
\item Schau dir den doppelreihigen Stecker, mit jeweils vier Pins in einer Reihe an, der mit einem gr\"unen Kleber versehen ist. Auf dem Schaltkreis in Abbildung ~\ref{Fig:Board_d} wird das Bauteil ``AD633`` genannt. 
                              Verbinde den Multiplizierer ``AD633`` so, \text{dass das orangene Label auf dem Multiplizierer das orangene Label auf dem Board anschaut}.
\item 
Rotiere die Potentiometeraxe, warte eine Minute und schreibe die Spannungen $U_1$ und $U_2$ auf. Ver\"andere die Polarit\"at der Spannungsquelle und wiederhole die Messung. Erstelle eine Tabelle mit Nummer des Experiments $i$, Spannung $U_1$ und  Spannung $U_2$.

\item Stelle das Ergebnis graphisch auf einem Milimeterpapier dar- dabei ist $U_1$ ist die x-Achse und $U_2$ die y-Achse
\item 
 F\"uge der Tabelle eine Spalte $(U_1)^2$ hinzu und stelle die Ergebnisse graphisch dar, dieses Mal mit $(U_1)^2$ auf der x-Achse und $U_2$  auf der y-Achse. Erstelle eine gerade Linie, die m\"oglichst nah an alle Messpunkte herankommt. Diese angen\"aherte Linie wird durch die Gleichung $U_2= (U_1)^2/U_0 + \mathrm{const}$ beschrieben. Suche dir zwei Punkte auf der Linie aus und messe 
 $\Delta y=\Delta(U_1^2)$    und $\Delta   x = \Delta(U_2)$      und berechne dadurch die Steigung $U_0=\Delta(U_1^2)/\Delta(U_2)$. Der so bestimmte Parameter $U_0$ mit der Einheit V ist essentiell f\"ur die Bestimmung der Elementarladung, wie im n\"achsten Teil noch genauer beschrieben wird.

\section{Aufgabe L: Elektronenladung $q_e$ bestimmen } 

\begin{figure}[h]
\centering
\includegraphics[scale=0.3]{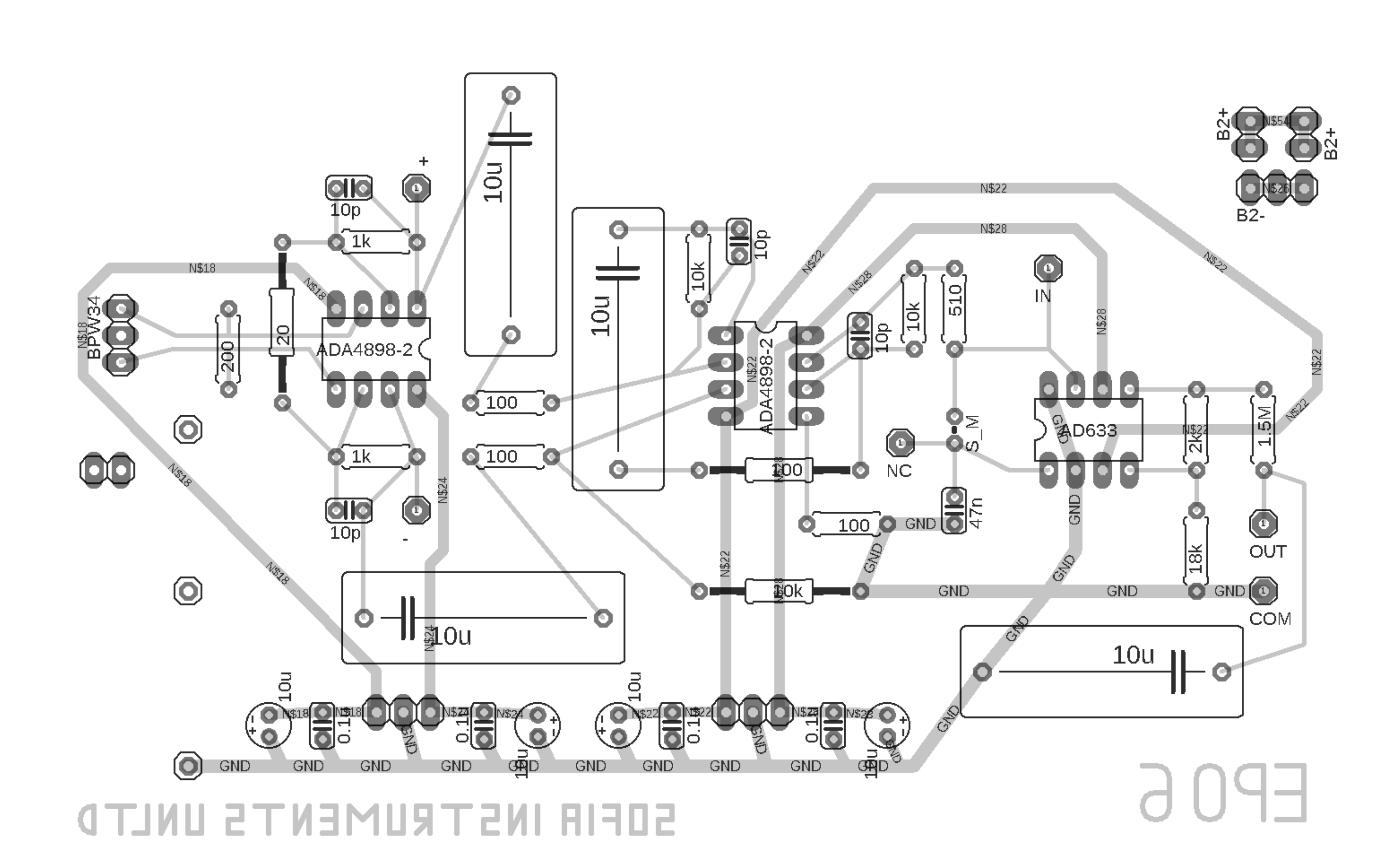}
\caption{Ein Diagramm des PCBs. Die Gl\"uhbirne mit dem wei\ss en Zylinder darauf,
welche zwischen die ``B2-`` und ``B2+`` Pins gel\"otet wurde, ist hier nicht zu sehen. Die zwei nicht benannten Dreierpins, welche am unteren Ende des Schaltkreises sind, sind die Spannungsversorgungspins.}
\label{Fig:Board_d}
\end{figure}
\item Stecke die drei 1.5~V Batterien in ihre Batteriehalter. Verbinde das Potentiometer durch zwei Krokodilkabel mit den beiden \"au\ss eren Elektroden der Lampe (also mit  den  ``B2-``- und ``B2+``-Pins im rechten oberen Eck des Setups Abbildung ~\ref{Fig:Board_d}. Hierbei spielt die Polarit\"at keine Rolle. Bei richtiger Verbindung sollte die Lampe leuchten. 
\item Verbinde mit einem Krokodilkabel das ``($-$)`` Kabel des Setups mit dem ``COM`` Point von $V_1$.
\item 
Verbinde analog die ``($+$)``-Elektrode  des Aufbaus mit dem ``$\mathrm{V\,\Omega\, m\!A}$`-Eingang des Voltmeters $V_1$. Dadurch zeigt $V_1$ die Potentialdifferenz $U_{\pm}$ zwischen der ``($+$)``   Elektrode       und der ``($-$)``  Elektrode an. 
\item 
 Platziere die 3~V Lithiumbatterie in ihre Batteriehalterung auf dem zweireihigen Leiterplatinensteckverbinder, auf welchen die Photodiode gel\"otet ist. Die ``($+$)``-Seite der Batterie sollte mit der ``($+$)``-Seite des Batteriehalters verbunden sein.
\item 
Wende dich nun den beiden doppelreihigen Steckern mit jeweils vier Pins in einer Reihe und  mit gr\"unen Klebern zu. Sie hei\ss en Operationsverst\"arker und sind in Abbildung ~\ref{Fig:Board_d} mit    ``ADA4898-2``     beschriftet.  Verbinde die beiden Operationsverst\"arker (die kleinen, gr\"un gelabelten Bauteile) ``ADA4898-2`` so mit der Platine,  dass die gr\"unen Label jedes Operationsverst\"arkers zu den gr\"unen Labeln auf der Platine zeigen. \textbf{Sei sehr vorsichtig! Eine falsche Verbindung kann zur Besch\"adigung der Verst\"arker f\"uhren.}
\item
Verbinde vorsichtig das Bauelement mit der gel\"oteten Photodiode und dem Lithiumbatteriehalter auf seinen Platz, der in Abbildung 1 zu sehen und zus\"atzlich mit einem Marker gekennzeichnet. F\"uhre das Glasfaserkabel vorsichtig in das l\"angere Ende des schwarzen Strohhalms. \"Uberpr\"ufe, ob sich das Kabel ohne gro\ss e Reibung durch den Strohhalm bewegen l\"asst.

\item F\"uhre das andere Ende des Glasfaserkabels in den wei\ss en Zylinder, der an der Lampe befestigt ist.

\item 
Positioniere den Plastikbecher neben das Verbindungsst\"uck zwischen Photodiode und Platine. F\"uhre nun die Photodiode in das kurze Ende des Strohhalms, so dass sie konstant gek\"uhlt wird.  Das verhindert st\"orendes thermisches Rauschen.
\item 
Verbinde eine der Klemmen der 9~V Batterie mit der linken Dreipinspannungsquelle auf dem PCB. \textbf{Achtung! Die orangenen Aufkleber der Klemmen der Batterien sollten zu den orangenen Aufklebern des Aufbaus zeigen.} 
Schalte das   Voltmeter auf  Millivolt und bewege das Glasfaserkabel durch den Strohalm. Diese Spannung ist nun $U_\pm$ und sollte bis zu 1000~mV variieren. 
Durch das Bewegen des Glasfaserkabels \"anderst du den Photostrom und misst die verst\"arkte Photospannung $U_\pm$, die der durch den Widerstand $R$ flie\ss ende Photostrom, generiert. 
Wenn du es nicht schaffst, eine Spannung von mindestens 700~mV zu messen, z\"ogere nicht, die Aufseher im Pr\"ufungssaal um Hilfe zu bitten. Gerade dieser Teil stellt eine wichtige Voraussetzung f\"ur dieses Experiment dar. \textbf{Versuche nicht, gewaltsam den an der Gl\"uhbirne befestigten Zylinder zu bewegen, da sonst die Gefahr besteht die Gl\"uhbirne zu zerbrechen.} 

\item Frage nach Eis und Wasser, um es in deinen Plastikbecher mit dem Strohhalm zu f\"ullen. Positioniere das Glasfaserkabel am Ende des Strohhalms weit entfernt von der Photodiode.

\item 
Nachdem du Eis und Wasser in deinen Becher gef\"ullt    hast, \"uberpr\"ufe ihn auf Lecks. Falls welche vorhanden sein sollten, entfernte vorsichtig den Becher und frage nach einem neuen. 

\item 
Verbinde den ``COM``-Ausgang des PCBs mit dem COM Eingang deines Voltmeter $V_2$.
\item 
Verbinde den ``OUT``-Ausgang des PCBs mit dem ``$\mathrm{V\,\Omega\, m\!A}$''-Eingang des Voltmeters $V_2$. Nun misst das Voltmeter $V_2$ die Spannung $U_\mathrm{S}$, weche gerade proportional zum Stromrauschen der Photodiode ist. 

\item Verbinde eine der Klemmen der 9~V Batterie mit der rechten Dreipinspannungsquelle des PCBs. \textbf{Achtung! Die orangenen Aufkleber der Klemmen der Batterien sollten erneut sollten zu den orangenen Aufklebern des Aufbaus zeigen.}

\item 
Frag den Aufseher nach Tesafilm und klebe alle sechs Kabel des experimentellen Aufbaus wie folgt auf den Tisch:

\begin{itemize}
\item Platziere zun\"achst alle vier 9~V Batterien unterhalb des PCBs
\item 
klebe die ``$+$``- und ``$-$``-Kabel \"uber den PCB und  stelle sicher, dass sie sich nicht ber\"uhren
\item klebe die ``IN``- und  ``NC``-Kabel \"uber den PCB und weg von den ``$+$`` und ``$-$`` Kabeln 
\item 
Klebe die ``OUT```-  und ``COM``-Kabel auf die rechte Seite des PCBs  und stelle sicher, dass sie sich nicht ber\"uhren

\end{itemize}
Nun sollte dein experimenteller Aufbau wie eine wohlgeformte Spinne aussehen.
\item 
Betrachte nun die gemessene Spannung des Voltmeters $V_2$. Sie sollte sich in der  Gr\"o\ss enordnung von 10-90~mV stabilisieren.

\item 
Beginne $U_\pm$ und $U_\mathrm{S}$ zu messen: ver\"andere $U_\pm$ vorsichtig indem du das Glasfaserkabel im Strohhalm ein St\"uckchen n\"aher zur Photodiode r\"uckst. Die \"Anderung von $U_\pm$ zwischen zwei nachfolgenden Messungen sollte mindestens 100~mV betragen. Warte geduldig darauf, dass  sich $U_\mathrm{S}$ wieder stabilisiert, mindestens jedoch 2 Minuten. \emph{Sollten St\"orungen jeglicher Art den angezeigten Wert des Voltmeters betr\"achtlich ver\"andern, warte nochmals auf die Stabilisation (habe Geduld!).} Schreibe die Ergebnisse in einer Tabelle auf mit Spalten f\"ur die Anzahl der Messungen $i$,  $U_\pm$ und $U_\mathrm{S}$. 

\item 
Stelle die Ergebnisse graphisch dar, wobei $U_\pm$ auf der x-Achse, $U_\mathrm{S}$ auf der y-Achse aufgetragen wird (Zus\"atzliches Millimeterpapier wird von den Betreuern bereitgestellt). Zeichne die Gerade, die am besten die lineare Abh\"angigkeit $U_\mathrm{S}=k U_\pm +\mathrm{const}$ widerspiegelt. W\"ahle zwei Punkte auf der Geraden und bestimme die Steigung k, wobei $\Delta$ f\"ur die Differenz steht. Diese mathematische Prozedur wird lineare Regression genannt. F\"ur unseren Aufbau gilt $k\simeq 10^{-3}$.

\item  Bestimme zuletzt die Elektronenladung $q_e$ mit Hilfe folgender Formel:
\begin{equation}
q_e=2 k \frac{y_1}{y^2} \frac{R_\mathrm{_L}}{R} C_\mathrm{_L}U_0,
\label{electron_charge_d}
\end{equation}
wobei y die gesamte Verst\"arkung unseres Verst\"arkers angibt und $y_1 = 101$ den Wert der ersten Verst\"arkung unseres Multiplizierers darstellt. Den Wert $C_L$ findest du an der unteren rechten Seite des PCBs auf einem Aufkleber ($C \equiv C_\mathrm{_L}$), $R_\mathrm{_L}=510~\Omega$ and $R=200~\Omega$. \\
Herzlichen Gl\"uckw\"unsch, du hast soeben die Elementarladung bestimmt!

\item 
Deine Messungen k\"onnen durch die Miteinberechnung von nicht-idealen Effekten des Operationsverst\"arkers durch folgende Formel pr\"azisiert werden:

\begin{equation}
q_e=2 k \frac{y_1}{(1-\varepsilon)Y^2} \frac{R_\mathrm{_L}}{R} 
C_\mathrm{_L}U_0.
\label{electron_charge_mod_d}
\end{equation}

Die kleine Korrektur $\varepsilon(C_\mathrm{_L})$ als Funktion der Kapazit\"at $C_\mathrm{_L}$ ist graphisch in Abbildung ~\ref{Fig:Eps_d} angegeben. 
\begin{figure}[h]
\centering
\includegraphics[scale=0.8]{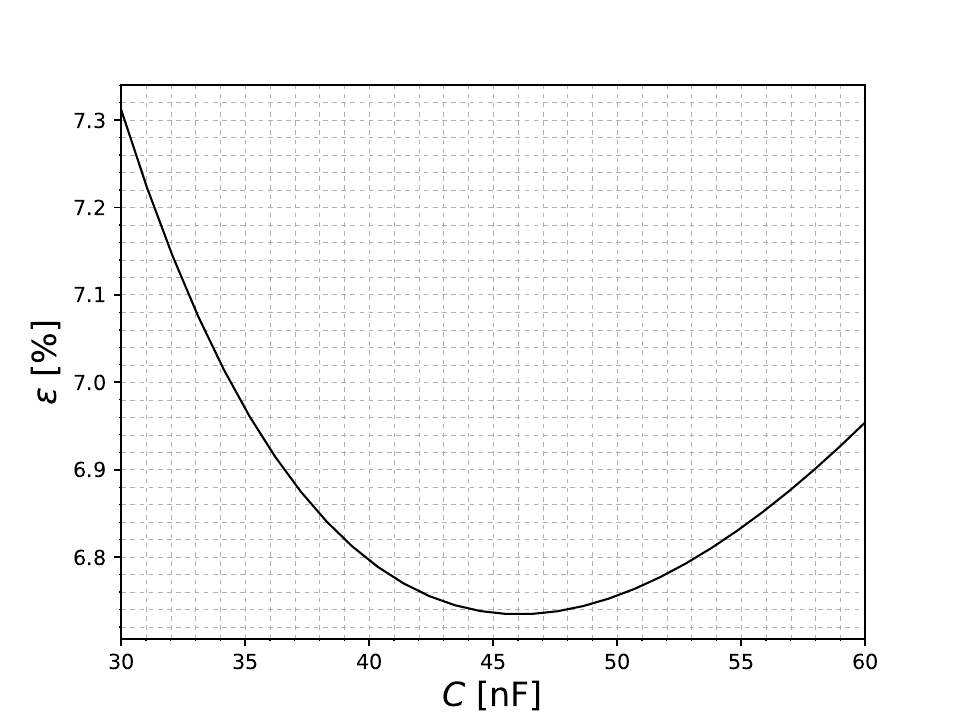}
\caption{Fehler $\varepsilon$ in Prozent zur Bestimmung von $q_e$ als Funktion des Kondensators $C_\mathrm{_L} \equiv C$.}
\label{Fig:Eps_d}
\end{figure}

Finde den durch den auf dem PCB vermerkten Wert $C_L$ verursachten Fehler $\varepsilon$ aus Abbildung~\ref{Fig:Eps_d}   und vergesse nicht $\varepsilon$  durch 100 zu teilen, bevor du $q_e$ mit der Gleichung ~(\ref{electron_charge_mod_d}) berechnest.
Ein Jahrhundert zuvor wurde diese Methode zur Bestimmung der Elektronenladung $q_e$ von Walter Schottky vorgeschlagen. Zu dieser Zeit arbeitete Walter Schottky mit Max Planck zusammen und wurde durch eine Vorlesung Albert Einsteins \"uber elektrische Fluktuationen dazu inspiriert. 

\section{Aufgabe XL: Hausaufgabe} 
Leite die Formeln ~(\ref{electron_charge_d}) und ~(\ref{electron_charge_mod_d}) her, indem du den Schaltkreis analysierst und unter   Zuhilfenahme der Werte der Tabelle ~\ref{tbl:values_d}. Berechne die Verst\"arkungen $y_1$ des Puffers und die gesamte Verst\"arkung Y von allen Schritten des Verst\"arkers. Die verschiedenen Details des Schaltkreises sind in Abbildung ~\ref{Non-inverting amplifier_d} abgebildet. 
\ref{Buffer_d}, 
\ref{Differential amplifier_d},
\ref{Inverting amplifier_d},
\ref{Analog multiplier_d}.

\begin{center}
\begin{table}[h]
\begin{tabular}{| c | r |}
		\hline
		&  \\ [-1em]
		Bauelement  & Wert  \\ \tableline
			&  \\ [-1em]
			$R$ &200~$\Omega$ \\
			$r_\mathrm{_G}$ & 20~$\Omega$ \\
			$R_\mathrm{F}$ &  1~k$\Omega$  \\
			$C_\mathrm{F}$ &  10~pF  \\ 
			$C_\mathrm{G}$ & 10~$\mu$F \\
			$R_\mathrm{G}$ &  100~$\Omega$  \\ 
			$R_\mathrm{F}^\prime$ & 10~k$\Omega$ \\
			$C_\mathrm{F}^\prime$ & 10~pF \\
			$R_\mathrm{_L}$ & 510~$\Omega$ \\
			$R_1$ &  2~k$\Omega$  \\ 
			$R_2$ & 18~k$\Omega$  \\
			$R_\mathrm{av}$ & 1.5~M$\Omega$ \\
			$C_\mathrm{av}$ & 10~$\mu$F \\
			$R_\mathrm{_V}$ & $\approx 1~\mathrm{M} \Omega$ \\
			$V_\mathrm{CC}$ & +9~V \\
			$V_\mathrm{EE}$ & -9~V \\
\tableline
\end{tabular}
	\caption{Tabelle mit den numerischen Werten der einzelnen Komponenten des Schaltkreises.}
	\label{tbl:values_d}
\end{table}
\end{center}

\begin{figure}[t]
\begin{minipage}[t]{0.31\linewidth}
\includegraphics[scale=0.28]{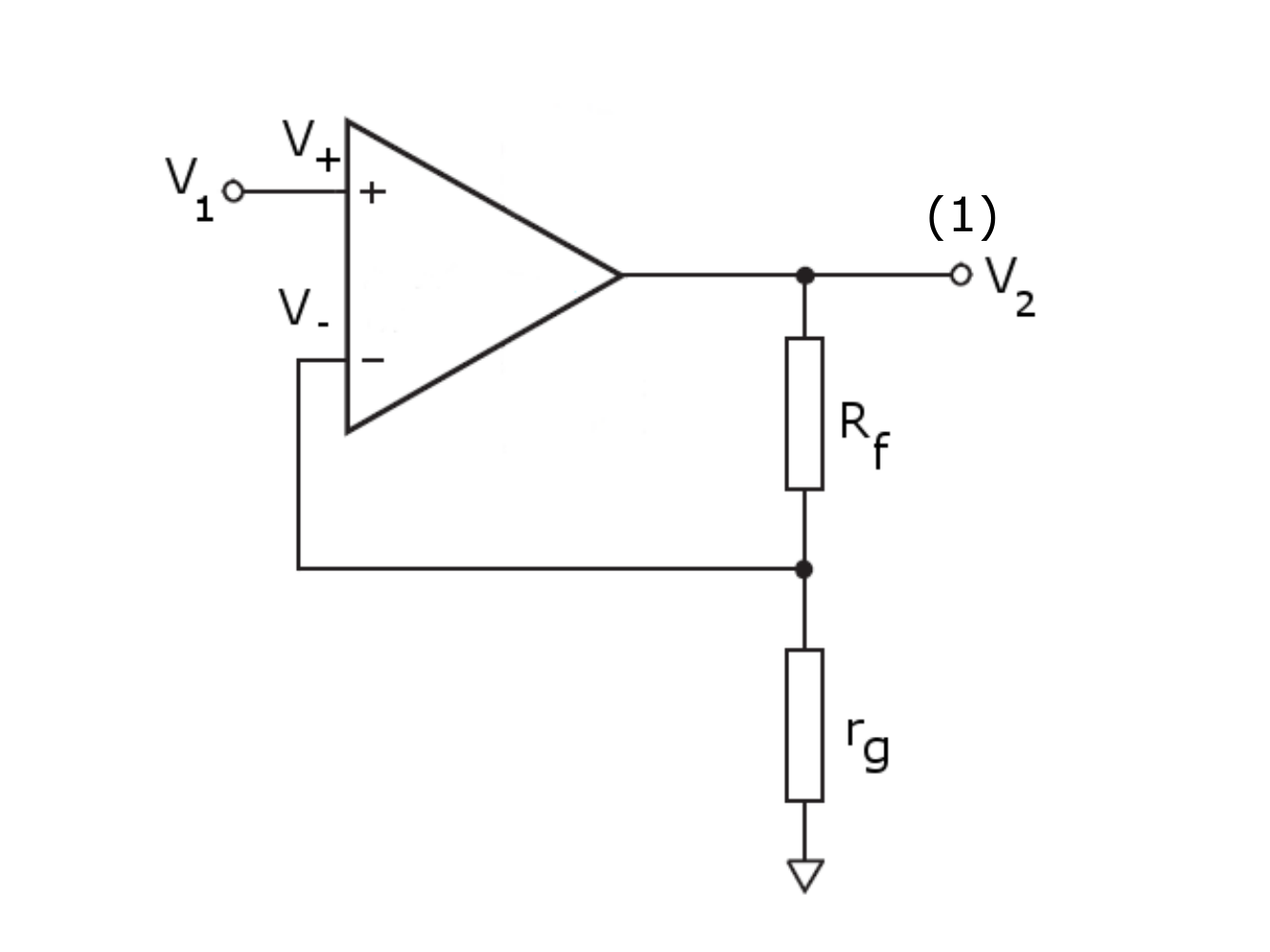}
\caption{Nichtinvertierender Verst\"arker}
\label{Non-inverting amplifier_d}
\end{minipage}
\begin{minipage}[t]{0.31\linewidth}
\includegraphics[scale=0.28]{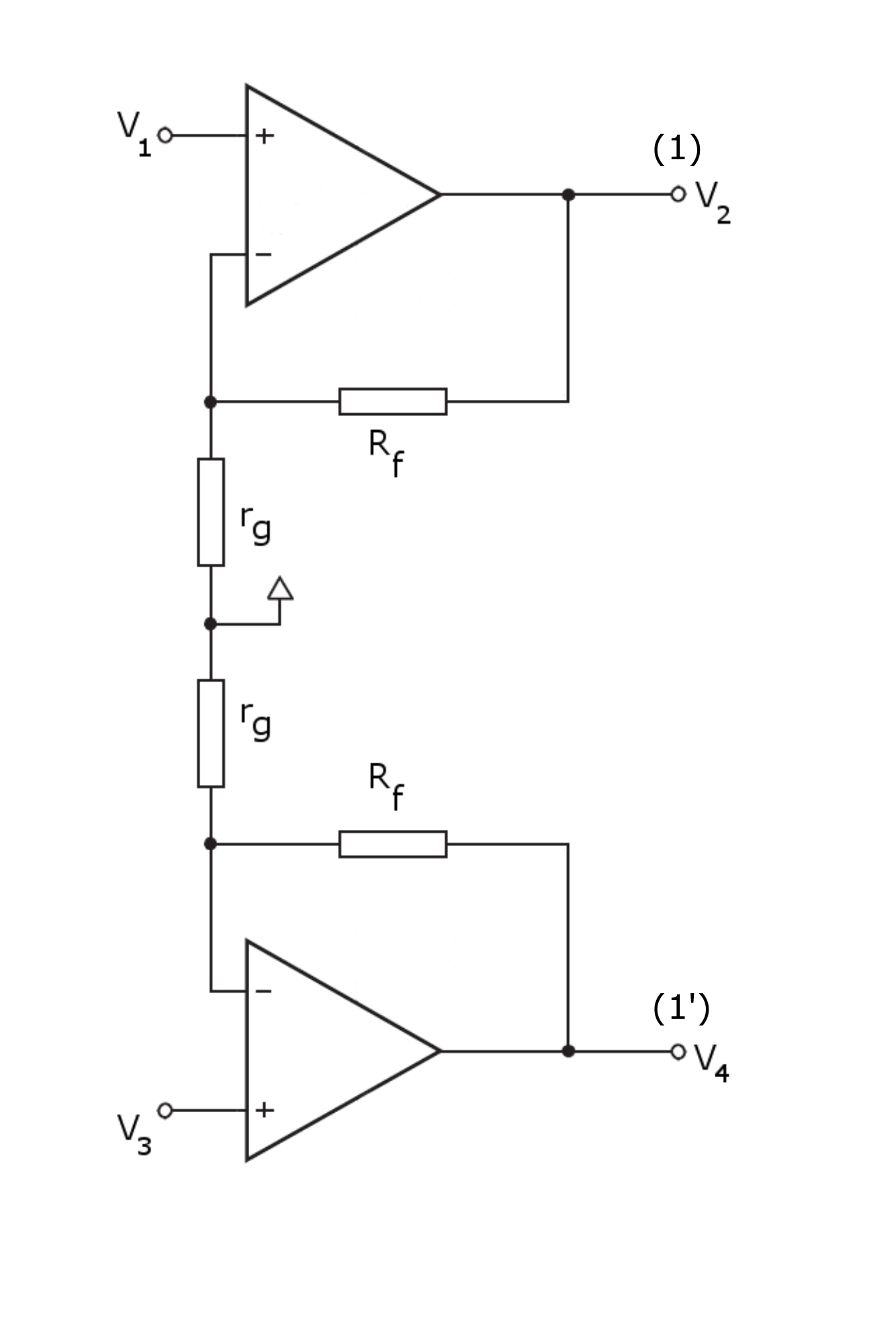}
\caption{Puffer}
\label{Buffer_d}
\end{minipage}
\begin{minipage}[t]{0.36\linewidth}
\includegraphics[scale=0.28]{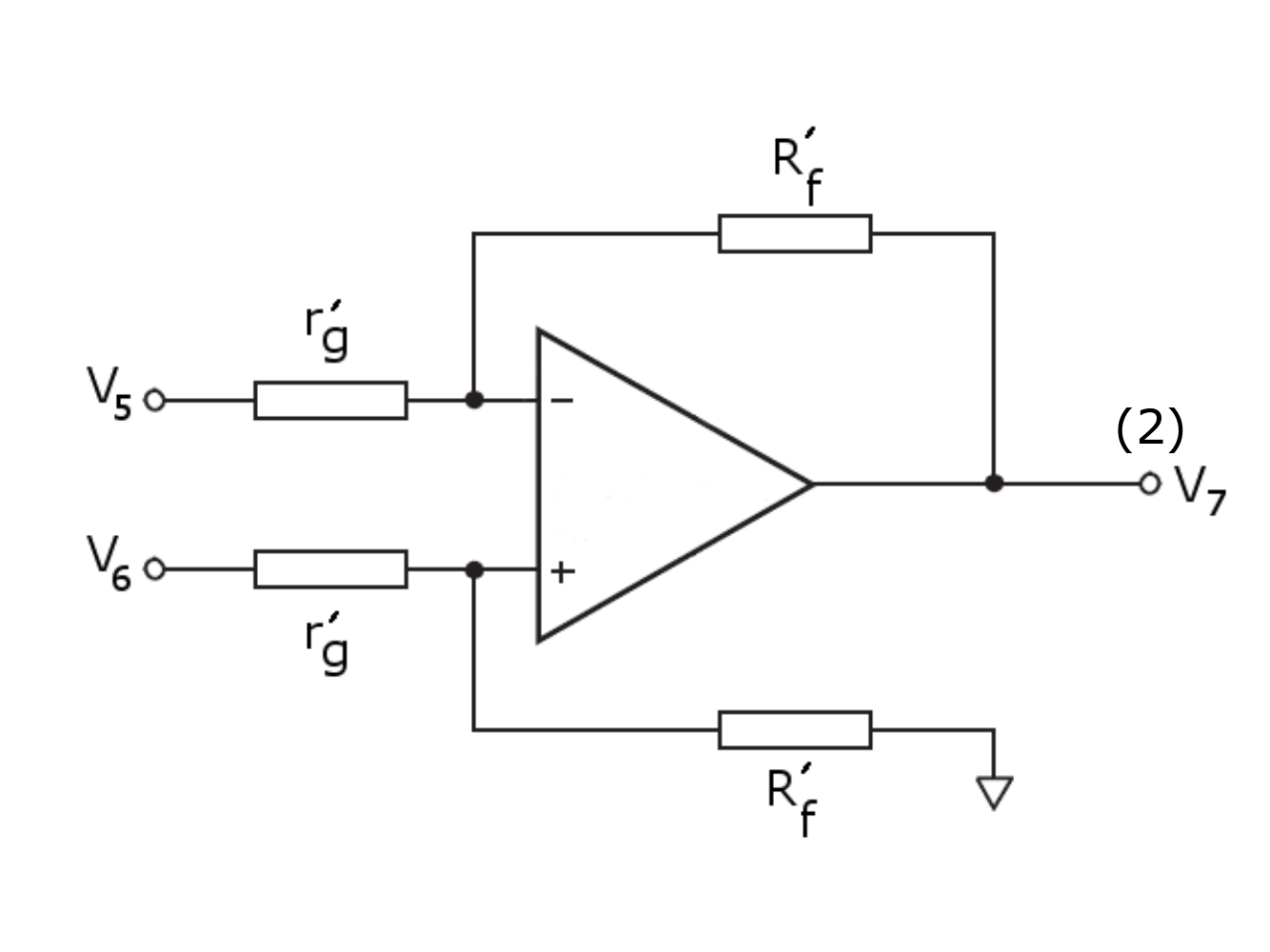}
\caption{Differenzverst\"arker}
\label{Differential amplifier_d}
\end{minipage}
\begin{minipage}[c]{0.4\linewidth}
\includegraphics[scale=0.28]{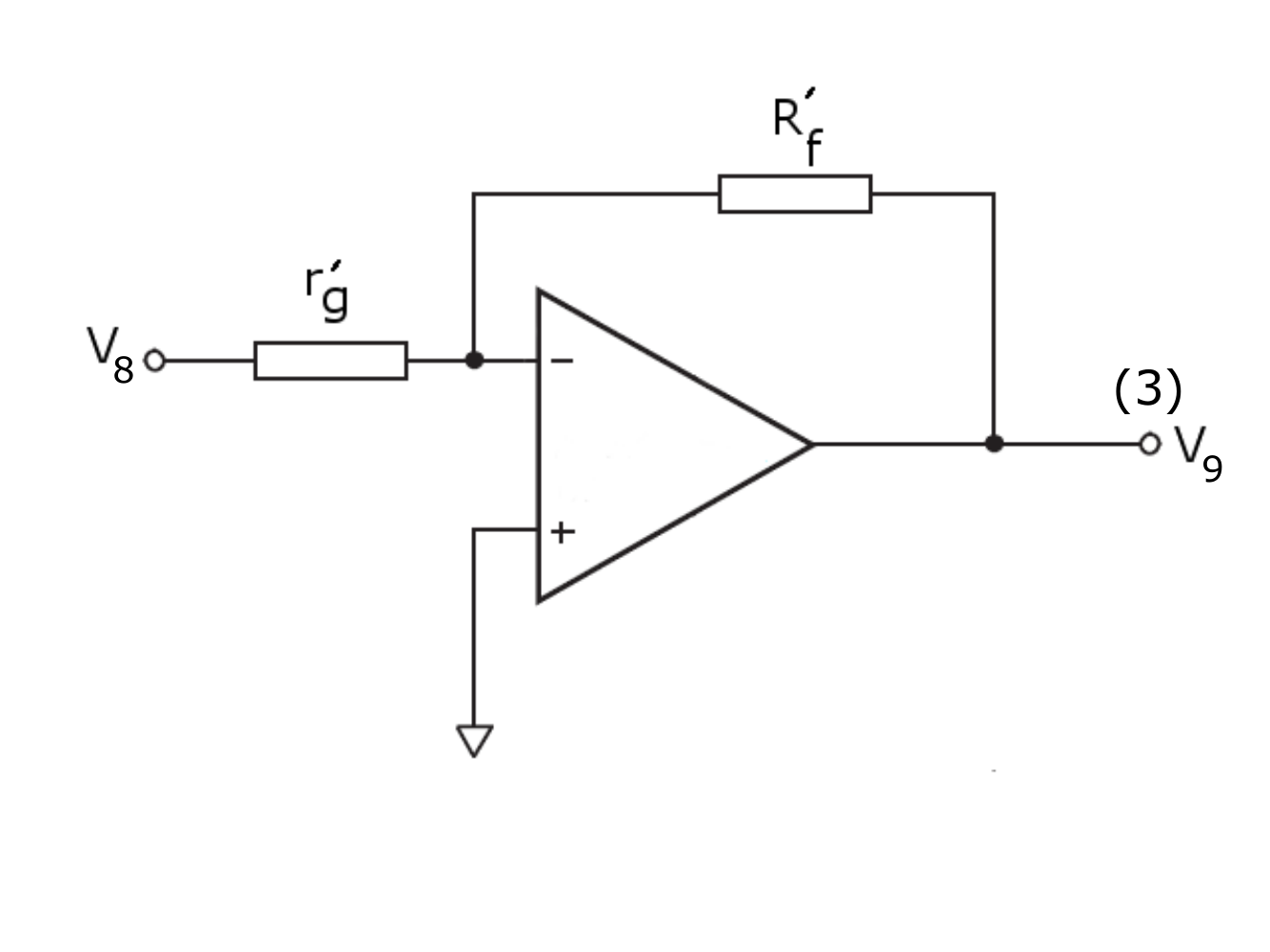}
\caption{Invertierender Verst\"arker}
\label{Inverting amplifier_d}
\end{minipage}
\begin{minipage}[c]{0.57\linewidth}
\includegraphics[scale=0.28]{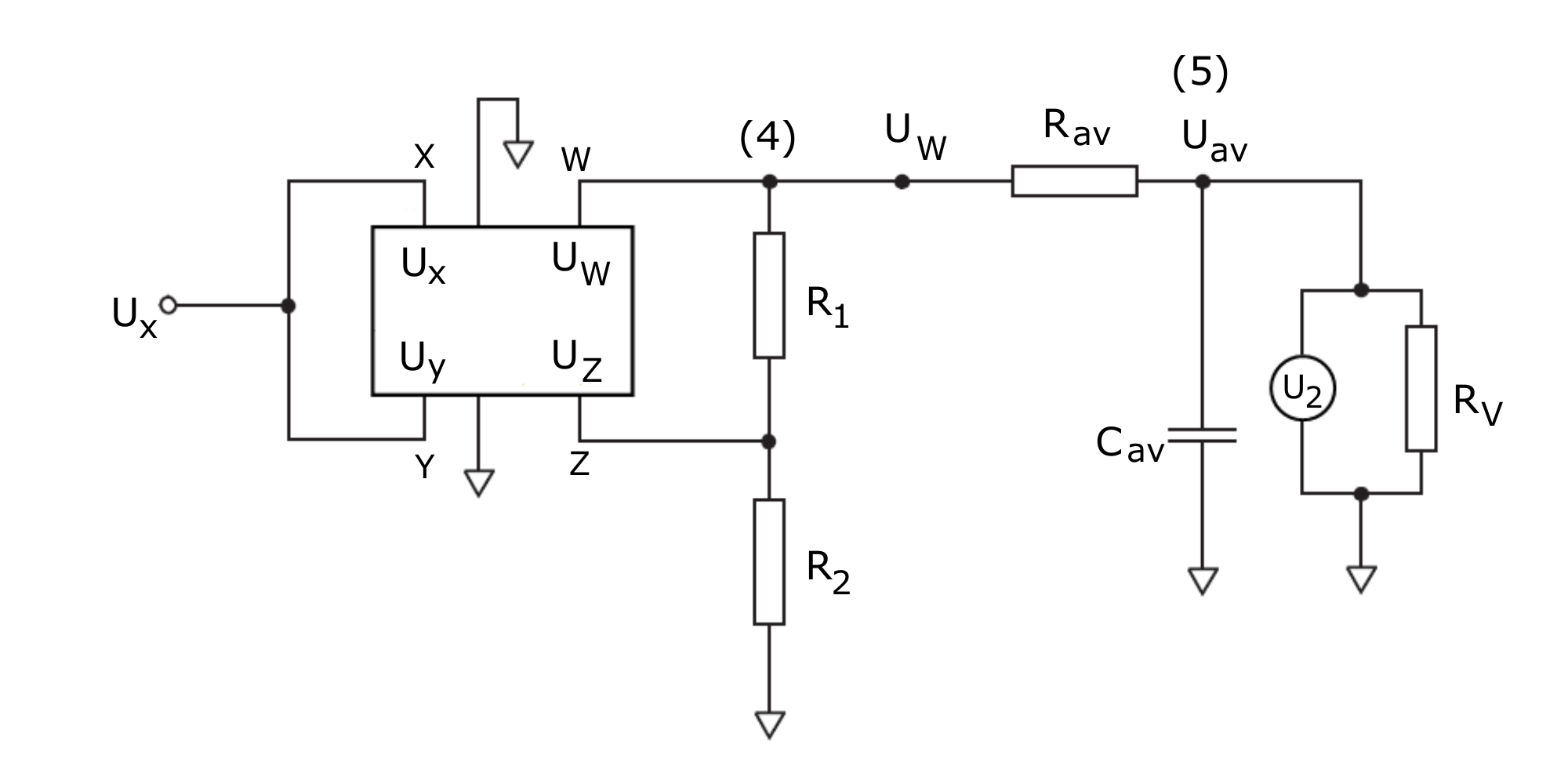}
\caption{Analoger Multiplizierer}
\label{Analog multiplier_d}
\end{minipage}
\end{figure}

Die L\"osung muss in der Nacht nach der Olympiade bis Sonnenaufgang an die E-Mail Adresse der Olympiade, epo@bgphsics.eu. geschickt werden. 
Dem Teilnehmer mit der besten L\"osung wird der Sommerfeldpreis \"ubergeben, ein Geldpreis in H\"ohe von  70EUR. 

\section{Weiterf\"uhrende Aufgaben}
Der experimentelle Aufbau ist tats\"achlich ein Lock-in-Verst\"arker, mit dem du eine Wechselspannung kleiner als $\mathrm{\mu V}$ messen kannst. Auf Anfrage werden wir dir sagen, wie du den ``NC``-Ausgang in diesem Fall benutzen kannst. 

\end{enumerate}

\clearpage


\renewcommand{\appendixname}{Appendix}
\appendix

\section{Unsere Messergebnisse}

\begin{enumerate}

\item{Aufgabe S: Einstieg}
\begin{table}[h]
\begin{tabular}{ c  c}
		\tableline
			& \\[-1em]
		Batterie  &\quad $U$  [V] \\ \tableline
			&  \\ [-1em]
			$B_1$	 	& 9,61 \\
			$B_2$	 	& 9,60\\
			$B_3$ 		& 9,55  \\
			$B_4$ 		& 9,56 \\ 
			$B'_1$	 & 1,59 \\
			$B'_2$	 & 1,59 \\
			$B'_3$	 & 1,59 \\
\tableline
\end{tabular}
\caption{Die gemessenen Spannungen der Batterien.}
\end{table}
Intervall der Spannung durch das Potentiometer  $ 0 V < U < 4,76 V $

\item{Aufgabe M: Multiplizierer}
\begin{table}[h]
\begin{tabular}{ c  r  r  r }
		\tableline \tableline
		&  \\ [-1em]
		i & $U_1$ [V] & \hspace{2.5pt} $U_2$ [V] & \hspace{5pt} $U_1^2$ [V$^2$] \\ \tableline 
			&  \\ [-1em] 
			1	&	-4,75	&	2,73	&	22,563\\
			2	&	-4,21	&	2,74	&	17,724\\
			3	&	-3,36	&	2,74	&	11,289\\
			4	&	-2,80	&	2,74	&	7,840	\\
			5	&	-2,34	&	2,15	&	5,476\\
			6	&	-1,08	&	0,46	&	1,166\\
			7	&	-0,44	&	0,08	&	0,194\\			
			8	&	0,17	&	0,01	&	0,028\\
			9	&	1,00	&	0,38	&	1,000\\
			10	&	1,25	&	0,61	&	1,563\\
			11	&	1,85	&	1,34	&	3,423	\\
			12	&	2,55	&	2,56	&	6,503\\
			13	&	3,60	&	2,74	&	12,960	\\
			14	&	4,75	&	2,74	&	22,563\\					
\tableline \tableline
\end{tabular}
\caption{Experimentelle Daten aus den Messungen von $U_1$, $U_2$ und berechnet $U_1^2.$}
\label{tab:M}
\end{table}
\begin{figure}[h]
\includegraphics[scale=0.5]{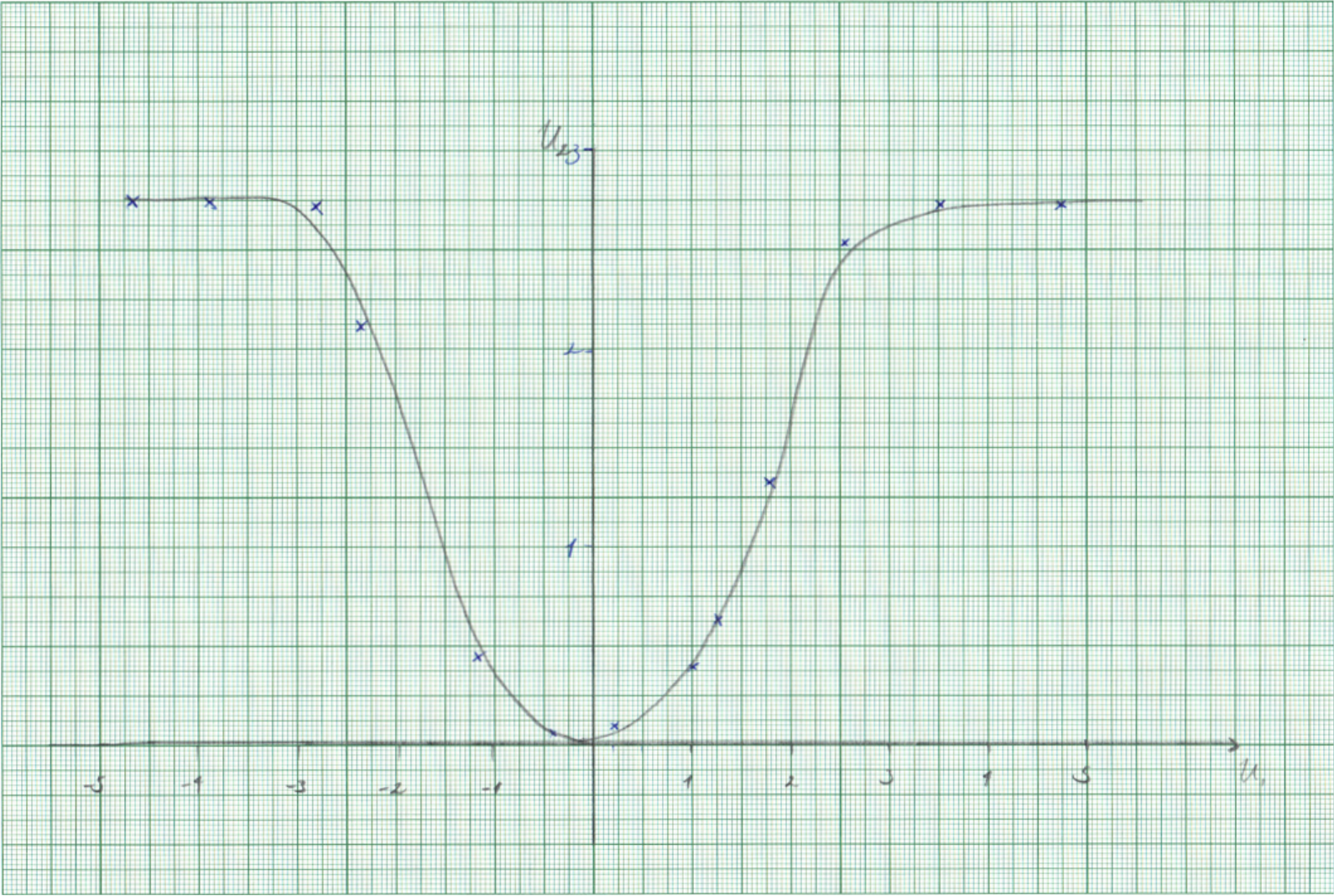}
\caption{Die grafisch dargestellte Abhängigkeit $U_2$ von $U_1$ aus Tabelle~\ref{tab:M}.}
\label{fig:u1u2}
\end{figure}
\begin{figure}[h]
\includegraphics[scale=0.5]{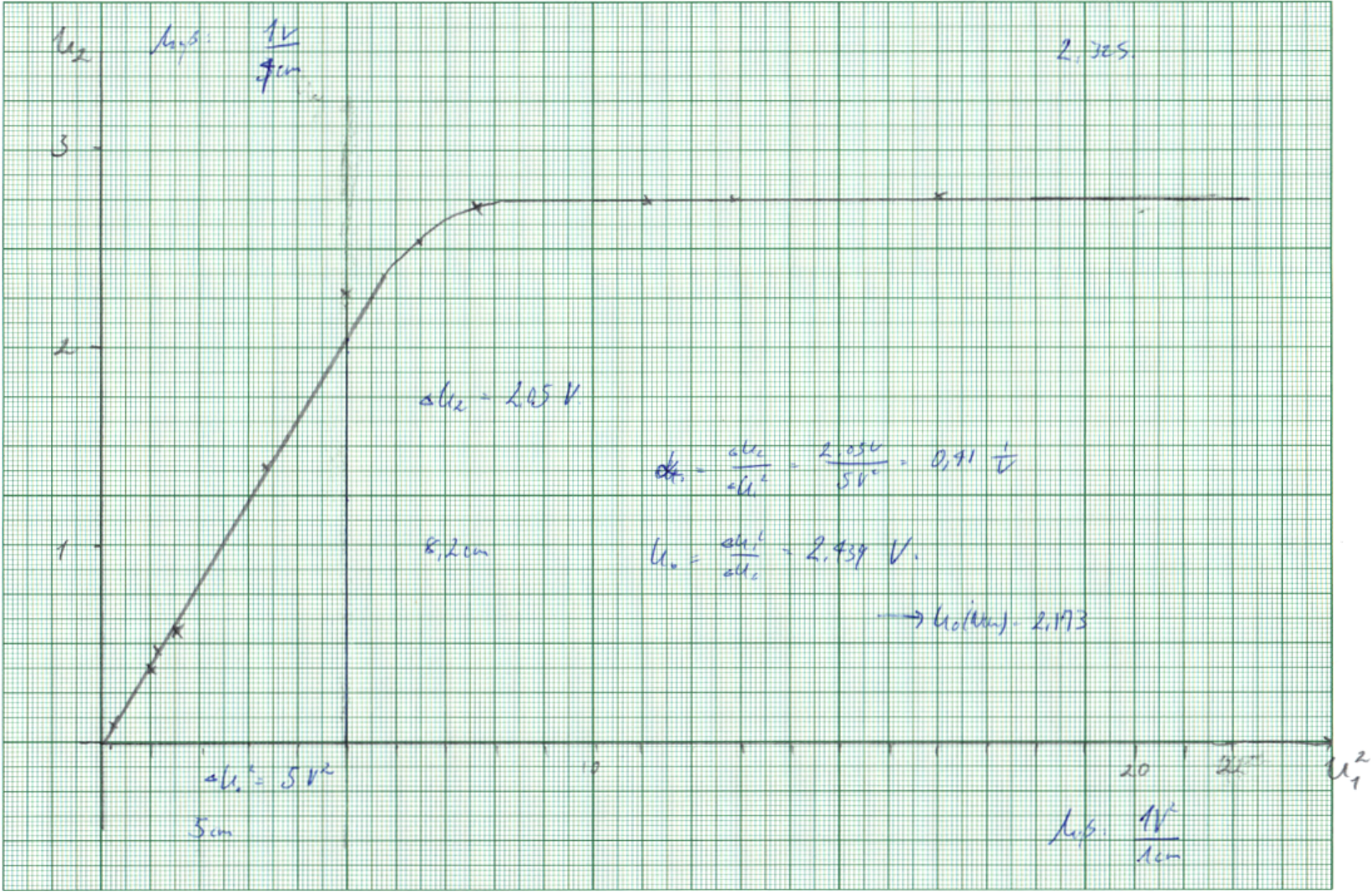}
\caption{Die grafisch dargestellte Abhängigkeit $U_2$ von $U_1^2$ aus Tabelle~\ref{tab:M}.}
\label{fig:u1u1u2}
\end{figure}

\item{Aufgabe L: Elektronenladung $q_e$ bestimmen}
\begin{table}[h]
\begin{tabular}{ c  r  r }
		\tableline \tableline
		&  \\ [-1em]
		i & $U_{\pm}$ [mV] & \hspace{2.5pt} $U_S$ [mV] \\
\tableline 
			&  \\ [-1em] 
			1	&	303		&	19,0\\
			2	&	400		&	19,2\\
			3	&	495		&	19,6\\
			4	&	612		&	20,0\\
			5	&	706		&	20,6\\
			6	&	807		&	21,2\\
			7	&	930		&	21,7\\			
			8	&	1070	&	22,2\\
			9	&	1158	&	22,7	\\
			10	&	1258	&	23,1\\
\tableline \tableline
\end{tabular}
\caption{Experimentelle Daten aus den Messungen von $U_\pm$ und $U_S$.}
\label{tab:L}
\end{table}
\begin{figure}[h]
\includegraphics[scale=0.5]{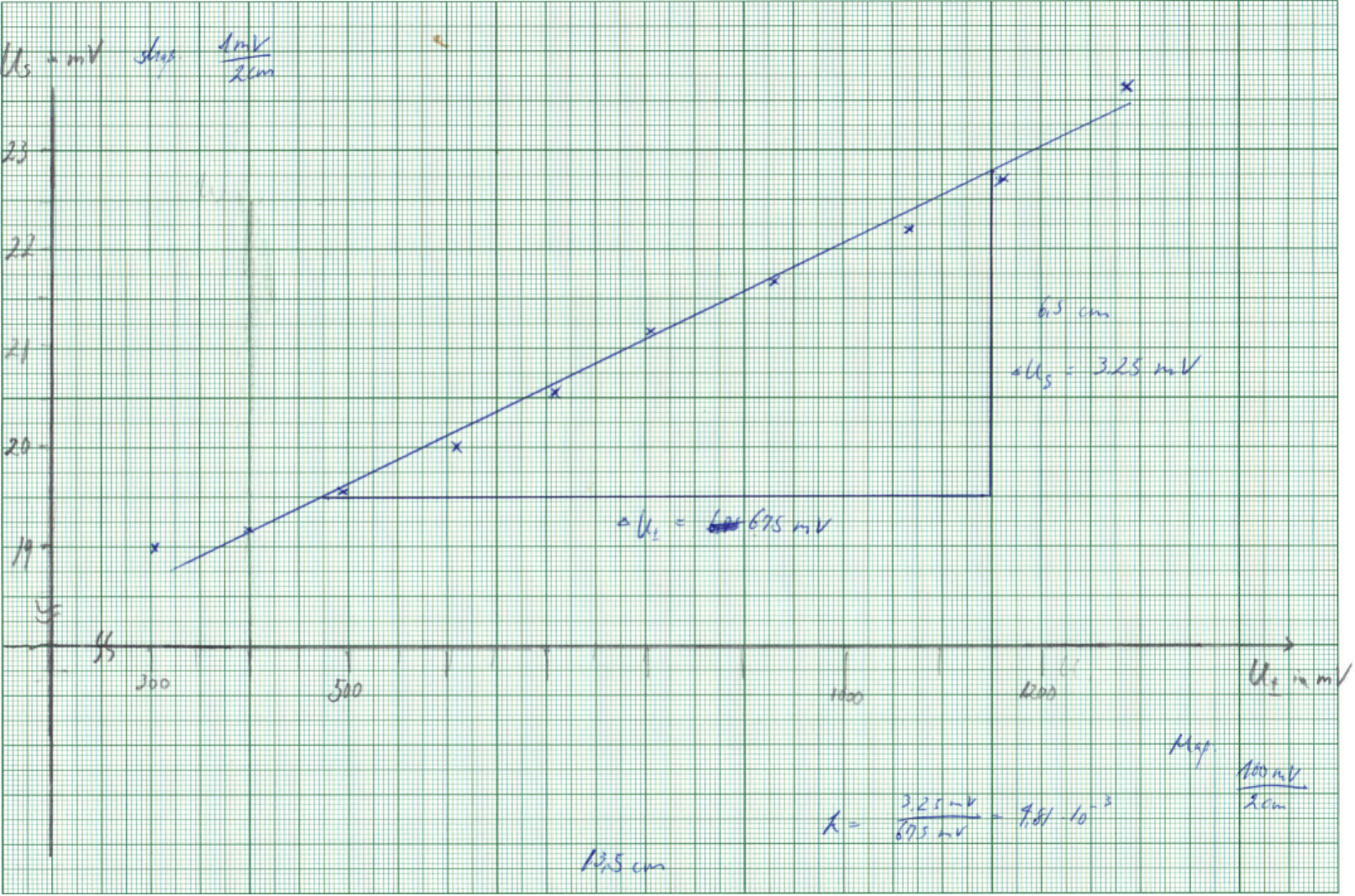}
\caption{Die grafisch dargestellte Abhängigkeit $U_S$ von $U_\pm$ aus Tabelle~\ref{tab:L}.}
\label{fig:k_d}
\end{figure}

Diese Werte best\"atigen unsere angenommene lineare Abh\"angigkeit zwischen $U_{\pm}$ und $U_S$, welche durch die lineare Regressionsgerade 
\[ U_S = kU_{\pm}+\mathrm{const.}, \qquad \qquad k = \frac{\Delta U_S}{\Delta U_{\pm}} = 0,00449. \]
dargestellt werden kann. Damit ergibt sich f\"ur die experimentell bestimmte Elementarladung nach der gegebenen Formel zu 
\[ q_e = 2\times    0,00449\times  \frac{101\times    510\Omega \times 44,2 \mathrm{nF}\times 2,439\mathrm{V}}{(1,01\times 10^{6})^2\times    200\Omega} = 2,444 \times 10^{-19}~\mathrm{C}.\]
Ber\"ucksichtigen wir noch weiter den Fehler des Kondensators $C_L$ mit $ \epsilon(C_L)|_{C_L = 44\mathrm{nF}}= 6,74\%$
\[ q_e = 2\times 0,00449 \times \frac{101\times 510\Omega \times 44,2 \mathrm{nF}\times 2,439\mathrm{V}}{(1-0,0674)(1,01\times 10^{6})^2\times 200\Omega} = 2,620 \times 10^{-19}~\mathrm{C}.\]
Abweichung zum Literaturwert ergibt sich zu  \; 
\[ \frac{q_{e \hspace{1pt}gemessen}}{q_{e \hspace{1pt}lit.}} = 1,635  
\qquad \Rightarrow 63,5\%  \mbox{ \"uber \; dem \; Literaturwert}.\]
\end{enumerate}


\section{Bestimmung der Fehlerquellen in Aufgabe L}
Jegliche Art von Stromrauschen, das nicht durch die Photodiode entstanden ist, stellt eine Fehlerquelle dar. Es ist sehr schwierig, alle st\"orenden Rauschquellen zu eliminieren, allerdings gibt es ein paar einfache Ma\ss nahmen, die man ergreifen kann:
\begin{itemize}
\item
Versuche das Glasfaserkabel, das das Licht zur Photodiode transportiert, so gut wie m\"oglich zu stabilisieren. Schon kleine Bewegungen des Kabels k\"onnen zu einer Ver\"anderung der Photospannung  $U_{\pm}$ f\"uhren. Diese Ver\"anderung l\"asst $U_S$ kurzfristig erheblich steigen und kann zu einer starken Ver\"anderung deiner Messwerte f\"uhren, wenn du nicht lange genug wartest, bis sich $U_S$ wieder eingependelt hat. Je n\"aher du mit dem Glasfaserkabel an die Photodiode herankommst, desto gr\"o\ss er werden die Spannungsfluktuationen $U_{\pm}$ durch kleine Bewegungen des Kabels und das System kann sich kaum mehr einpendeln.
\item
Ein starker Wechsel der \"au\ss eren Lichtverh\"altnisse kann zu einer Ver\"anderung der Photospannung  $U_{\pm}$ f\"uhren, was zu den oben beschriebenen Effekten f\"uhrt. Versuche also, wenn irgendwie m\"oglich, die \"au\ss eren Lichtverh\"altnisse m\"oglichst konstant zu halten.

\item
Die Photodiode sollte immer so gut wie m\"oglich gek\"uhlt sein. Je h\"oher die Temperatur der Photodiode, desto gr\"o\ss er ist deren thermisches Rauschen, was zu einer Verf\"alschung der Messergebnisse f\"uhrt.
\end{itemize}
Eine weitere Fehlerquelle, die man leider nicht aufheben kann, sind st\"orende elektromagnetische Wellen in der Luft zum Beispiel durch Radiosender, Mobilfunknetze, oder \"ahnliches. Die meisten dieser St\"orfrequenzen werden durch das Setup aufgehalten, allerdings nicht alle.\\
 Eine optimale M\"oglichkeit, sehr genaue Werte zu erziehlen, ist,  das Experiment in einem K\"uhlschrank durchzuf\"uhren. Man hat dort konstante Lichtverh\"altnisse, die gesamte Elektronik wird durchg\"angig gek\"uhlt und au\ss erdem schirmt ein K\"uhlschrank auch einige St\"orsignale gut ab. Es wurden schon Genauigkeiten von bis zu 1\% mit dieser Methode erreicht.